\documentclass[reprint,amsmath,amssymb,aps,prb,nofootinbib]{revtex4-2}

\usepackage[dvipdfmx]{graphicx}
\usepackage[FIGTOPCAP,nooneline]{subfigure}
\subcapraggedrighttrue
\usepackage{dcolumn}
\usepackage{bm}
\usepackage{braket}
\usepackage[dvipdfmx]{color}
\usepackage[title,titletoc]{appendix}
\usepackage{here}
\usepackage{mathtools}
\usepackage{tikz}
\usepackage{multirow}
\usepackage{appendix}
\usepackage{physics}
\usepackage{url}
\usepackage{hyperref}
\hypersetup{
    colorlinks=true,
    citecolor=blue,
    linkcolor=blue,
    urlcolor=blue,
}

\begin{document}

\preprint{APS/123-QED}

\title{Reciprocal and nonreciprocal paraconductivity in bilayer multiphase superconductors}

\author{Tsugumi Matsumoto}
 \email{matsumoto.tsugumi.78w@st.kyoto-u.ac.jp}
\author{Youichi Yanase}
\author{Akito Daido}
\affiliation{Department of Physics, Graduate School of Science, Kyoto University, Kyoto 606-8502, Japan}

\date{\today}

\begin{abstract}
Thin-film multiphase superconductors are attracting much attention, and it is important to propose how to detect them in experiments.
In this work, we study the reciprocal and nonreciprocal paraconductivity of a bilayer model with staggered Rashba-type spin-orbit coupling with and without the potential gradient and Zeeman field.
This model shows the Bardeen-Cooper-Schrieffer phase, the pair-density-wave phase, and the Fulde-Ferrell-Larkin-Ovchinnikov (FFLO) phase, and we focus on how their properties are encoded to the charge transport.
We show that the reciprocal paraconductivity has a peak associated with
the phase transitions between different superconducting states due
to the degeneracy of the transition temperatures as well as the paramagnetic depairing effect.
We also show that the FFLO superconducting state shows a sizable nonreciprocal paraconductivity once the degeneracy of Cooper pairs is lifted by applying the potential gradient.
Observation of the peaked reciprocal and nonreciprocal paraconductivity can be used as a probe of multiphase superconductivity.
\end{abstract}
\maketitle

\section{Introduction}
Multiphase superconductors, which have several superconducting states in a phase diagram, are platforms of many attractive and characteristic features and are
candidates for exotic superconducting states such as odd-parity, spin-triplet~\cite{Sigrist1991-lz}, and topological superconductivity~\cite{Qi2011-vx,Tanaka2012-vn,Sato2016-ab,Sato2017-nn}.
For instance, $\mathrm{CeRh_{2}As_{2}}$ is considered to have two superconducting phases at ambient pressure
and one of the superconducting phases is supposed to realize an odd-parity superconducting state
\cite{Khim2021-ml,Onishi2022-ak,Landaeta2022-gz,Ogata2023-ry,Siddiquee2023-wx,Mockli2021-mw,Schertenleib2021-sb,Skurativska2021-lg,Hazra2023-jy,Kimura2021-wx,Ishizuka2024-PRB,Nogaki2021-cv,Nogaki2022-pd,Lee2023-wp,Cavanagh2022-xe,Nogaki2024-PRB}.
$\mathrm{UTe_{2}}$ is known to realize various superconducting states by changing the pressure and magnetic field
and considered to be a candidate of spin-triplet superconductors and topological superconductors
\cite{Hazra2023-jy,Ran2019-yp,Aoki2022-zg,Ran2019-mh,Ran2020-hl,Miao2020-iw,Lin2020-so,Aoki2019-vq,Aoki2020-xg,Aoki2021-dz,Aoki2022-dj,Knebel2020-cr,Sakai2023-qx,Tokiwa2022-ji,Matsumura2023-ia,Kinjo2023-ri,Suetsugu2024-ua,Braithwaite2019-cr,Rosuel2023-st,Thomas2020-ah,Rosa2022-ud,Ishizuka2019-hv,Tei2023-nq,Tei2024-gv,Hakuno2024-cx}.
These materials
have been investigated intensively, and the superconducting multiphases have been identified mainly by using thermodynamic and magnetic measurements.

In addition to the bulk multiphase superconductors, atomically thin film superconductors have 
emerged as a new and important platform for multiphase superconductors after the technical development of thin-film fabrication.
For instance, $\mathrm{CeCoIn_{5}}$ thin films and superlattices~\cite{Naritsuka2021-ux} are considered to be candidates of multiphase superconductors which may have not only the Bardeen-Cooper-Schrieffer (BCS) state but also the pair-density wave (PDW) state and/or Fulde-Ferrell-Larkin-Ovchinnikov (FFLO) state
\cite{Radzihovsky2011-jx,She2012-lw,Yanase2022-bq,Asaba2024-jp}.
Although FFLO superconductivity was proposed a long time ago~\cite{Fulde1964-bx,Larkin1964-aa} and has been studied both in solid-state and cold-atom systems~\cite{Yin2014-wo,Xu2015-ql}, its presence is still elusive and new probes are awaited.
Moreover, $\mathrm{TaS_{2}}$~\cite{Nagata1992-bt,Navarro-Moratalla2016-mg,Yang2018-ck,Bekaert2020-ph,Almoalem2024-cm,Wan2024-wx}
intercalated with chiral molecules has the potential to realize the PDW state
\cite{Fischer2023-yq}.
The PDW state is an odd-parity superconducting state, which is rare in nature, and therefore, has attracted much attention.
Finite-momentum superconductivity beyond the conventional FFLO state has also been reported in multilayer NbSe$_2$~\cite{Wan2023-rh}.

Generally speaking, available experimental probes are limited for thin-film superconductors compared with bulk superconductors, since thermodynamic measurements are difficult to be performed. In this regard, the electric measurements, including reciprocal and nonreciprocal charge transport
\cite{Tokura2018-yp,Ideue2021-yo,Hoshino2018-qr,Nagaosa2024-uk}, may offer suitable probes of thin-film multiphase superconductors.
In this study, we focus on
the reciprocal and nonreciprocal paraconductivity~\cite{Varlamov1992-eu,Mishonov2002-ud,Puica2006-gp,larkin2005theory,Konschelle2007-me,Konschelle2009-mj,Wakatsuki2018-vd,Hoshino2018-qr,Daido2024-is,Nunchot2022-mc}
as a probe of two-dimensional multiphase superconductors~\cite{Yoshida2012-wa,Maruyama2012-cb,Sigrist2014-rc,Nakamura2017-vb,Mockli2018-ci}.
The paraconductivity refers to the excess conductivity by the thermally fluctuating Cooper pairs at temperature slightly larger than the mean-field transition temperature.
The reciprocal paraconductivity has been studied both in linear and nonlinear regimes~\cite{
Varlamov1992-eu,Mishonov2002-ud,Puica2006-gp}, while the nonreciprocal paraconductivity in noncentrosymmetric superconductors are attracting much attention recently~\cite{Wakatsuki2017-hw,Wakatsuki2018-vd,Hoshino2018-qr,Daido2024-is}.
In this paper, we focus on the leading contributions to the reciprocal and nonreciprocal paraconductivities, namely the fluctuation contributions to the linear and second-order nonlinear electric conductivities.

As minimal models of two-dimensional multiphase superconductors, we adopt bilayer models with a staggered Rashba-type spin-orbit coupling \cite{Yoshida2012-wa}.
Our model has the BCS and PDW phases in the perpendicular Zeeman fields~\cite{Yoshida2012-wa},
while has the BCS and FFLO phases in the in-plane Zeeman field~\cite{Yoshida2013}, allowing us to study the two representative superconducting multiphases.
Here, the PDW state is defined as an odd-parity state whose order parameter has an opposite sign on each layer, $(\Delta_1, \Delta_2)=(+\Delta,-\Delta)$. 
Note that this state should be distinguished from the state with finite in-plane momentum proposed e.g., for cuprate superconductors~\cite{Agterberg2020-gs}, which is also called a PDW state.
While the original model preserves the global inversion symmetry, we can also study the nonreciprocal paraconductivity by applying the potential gradient and breaking the inversion symmetry while applying the Zeeman field with a finite in-plane component.
We show that the degeneracy of the superconducting states in multiphase and/or finite-momentum superconductors give rise to characteristic features in reciprocal and nonreciprocal paraconductivity, and thereby argue that they can be used as a probe of the exotic superconducting states.

This paper is organized as follows.
In Sec.~\ref{sec:method}, we introduce the model Hamiltonian and the methods including the
general formulas to calculate the paraconductivity.
In Sec.~\ref{sec:linear}, we consider the system in the perpendicular Zeeman field.
First, we consider the system with inversion symmetry. 
We show the superconducting phase diagram which includes the first- and second-order superconducting transition lines.
We discuss the reciprocal paraconductivity with particular emphasis on its behavior near the first-order phase transition point between the BCS and PDW phases.
Then, we apply a potential gradient to break inversion symmetry
and discuss its impact on the features of the reciprocal paraconductivity.
In Sec.~\ref{sec:nonlinear}, we consider the system in the in-plane Zeeman field with the BCS and FFLO phases.
We show the reciprocal and nonreciprocal paraconductivity with and without the potential gradient and discuss their features from the perspective of the degeneracy of superconducting states and/or Cooper pair momenta.
We summarize the results in Sec.~\ref{sec:summary}.

\section{Setup and Formulation}
\label{sec:method}
In this section, we introduce the model and the general formulas to calculate the reciprocal and nonreciprocal paraconductivity.
\subsection{Bilayer model with staggered Rashba spin-orbit coupling}
We consider a bilayer model with staggered Rashba-type spin-orbit coupling \cite{Yoshida2012-wa},
\begin{equation}
  \hat{H} = \hat{H}_{\parallel} + \hat{H}_{\perp} + \hat{H}_{\mathrm{int}} + \hat{H}_{\mathrm{Rashba}} + \hat{H}_{\mathrm{Zeeman}},
\end{equation}
\begin{equation}
  \hat{H}_{\parallel} = \sum_{\bm{k},s,m}\xi(\bm{k})c^{\dagger}_{\bm{k}sm}c_{\bm{k}sm},
\end{equation}
\begin{equation}
  \hat{H}_{\perp} = t_{\perp}\sum_{\bm{k},s,\ev{m,m'}}c^{\dagger}_{\bm{k}sm}c_{\bm{k}sm'},
\end{equation}
\begin{align}
  \hat{H}_{\mathrm{int}} &= \sum_{\bm{k},\bm{k'},\bm{q},m}U(\bm{k},\bm{k'},\bm{q})c^{\dagger}_{\bm{k}+\bm{q}/2\uparrow m}c^{\dagger}_{-\bm{k}+\bm{q}/2\downarrow m}\nonumber\\
  &\qquad\qquad\qquad\qquad\times
  c_{-\bm{k'}+\bm{q}/2\downarrow m}c_{\bm{k'}+\bm{q}/2\uparrow m},
\end{align}
\begin{align}
  \hat{H}_{\mathrm{Rashba}} = \sum_{\bm{k},s,s',m}\alpha_{m}\bm{g}(\bm{k})\cdot\bm{\sigma}_{ss'}c^{\dagger}_{\bm{k}sm}c_{\bm{k}s'm},
\end{align}
\begin{equation}
  \hat{H}_{\mathrm{Zeeman}} = -\sum_{\bm{k},s,s',m}\mu_{\mathrm{B}}\bm{H}\cdot\bm{\sigma}_{ss'}c^{\dagger}_{\bm{k}sm}c_{\bm{k}s'm},
\end{equation}
where $m=1,2$ is the layer index and $s=\uparrow,\downarrow$ is the spin index.
Here, we define $\ev{m,m'}$ to be summation for nearest-neighbor layers, namely, $(m,m')=(1,2)$ and $(2,1)$.
$\xi(\bm{k}) = -2t(\cos k_{x}+ \cos k_{y}) - \mu $ represents intralayer hopping energy,
and $\bm{g}(\bm{k}) = (-\sin k_{y},\sin k_{x},0) $ represents the Rashba-type spin-orbit coupling.
The Zeeman field $\bm{H}=(H_x,H_y,H_z)$ is applied either in the perpendicular ($H_x=H_y=0$) or in the in-plane ($H_z=H_x=0$) directions.
Here we set $\alpha_{m}$ to be $(\alpha_{1}, \alpha_{2}) = (+\alpha,-\alpha)$, and thus the system has the global inversion symmetry.

In our model, the Zeeman field is adopted as the representative tuning knob of the superconducting multiphases.
This can be originated from the in-plane external magnetic field, the proximity to magnetic materials, and so on.
Our results may also be compared to the systems in the perpendicular external magnetic field, when the system has a large Maki parameter.

We assume isotropic $s$-wave superconductivity and apply the mean-field approximation to the Hamiltonian.
For simplicity, we assume $U(\bm{k},\bm{k'},\bm{q}) = -U$ with the on-site attractive interaction $U$.
In this study, the fluctuation of the superconducting order parameter will be treated in the Gaussian approximation of the  Ginzburg-Landau (GL) free energy.
The (nonlinear) gap equation for the order parameter, which is given by
\begin{align}
    \Delta_{m}(\bm{q}) = -U\sum_{\bm{k}}\ev{c_{\bm{-k+q/2}\downarrow m}c_{\bm{k+q/2}\uparrow m}},
    \label{eq:Gap_equation}
\end{align}
will be solved only to determine the first-order transition line within superconducting regions in the phase diagram, and it is not used to calculate the reciprocal and nonreciprocal paraconductivity.
Here, Eq.~\eqref{eq:Gap_equation} represents the spin-singlet $s$-wave Cooper pairs in each layer $m=1,2$.

In this study, we set parameters to $(t,t_{\perp},\mu,U) = (1.0,0.1,2.0,1.7)$ and adopt the unit $\mu_{\mathrm{B}} = k_{\rm B} = 1 $.
It is known that the superconducting phase diagram depends on the value of $\alpha$, and the superconducting multiphase including the BCS and PDW phases appears in the perpendicular Zeeman field when $\alpha/t_{\perp}$ takes a sufficiently large value \cite{Yoshida2012-wa}.
To study the superconducting multiphase, we set $\alpha=0.2$, and the
transition temperature calculated with these parameters is $T_{\rm c0} \sim 0.0255$, which is determined by the GL framework.

For the latter use, we define the normal-state Hamiltonian $H_{\mathrm{N}}(\bm{k})$,
\begin{widetext}
    \begin{align}
      \resizebox{\textwidth}{!}{
        $
        H_{\mathrm{N}}(\bm{k}) =
        \mqty(
        \xi(\bm{k})-\mu_{\mathrm{B}}H_z & \alpha(-\sin k_{y} -i\sin k_{x})+i\mu_{\mathrm{B}}H_y & t_{\perp} & 0 \\
        \alpha(-\sin k_{y} +i\sin k_{x})-i\mu_{\mathrm{B}}H_y & \xi(\bm{k})+\mu_{\mathrm{B}}H_z & 0 & t_{\perp} \\
        t_{\perp} & 0 & \xi(\bm{k})-\mu_{\mathrm{B}}H_z & \alpha(\sin k_{y} + i\sin k_{x})+i\mu_{\mathrm{B}}H_y \\
        0 & t_{\perp} & \alpha(\sin k_{y} -i\sin k_{x}) -i\mu_{\mathrm{B}}H_y& \xi(\bm{k})+\mu_{\mathrm{B}}H_z
        ),
        $
      }
    \end{align}
\end{widetext}
where $H_x=0$ is abbreviated.
For the centrosymmetric bilayers [Secs.~\ref{subsec:3A} and~\ref{subsec:4A}], we analyze the model introduced above. Furthermore, we will consider the effects of a potential gradient that breaks global inversion symmetry [Secs.~\ref{subsec:3B} and~\ref{subsec:4B}]. 
In the presence of the potential gradient $v$, the potential Hamiltonian $H_{v}$ is added to $H_{\mathrm{N}}(\bm{k})$ with
\begin{equation}
  H_{v} =
  \begin{pmatrix}
      v&0&0&0\\
      0&v&0&0\\
      0&0&-v&0\\
      0&0&0&-v
  \end{pmatrix}.
\end{equation}
\subsection{Ginzburg-Landau free energy}
We calculate the GL free energy with the Hamiltonian introduced in the previous section. By expanding the free energy in terms of the order parameter, we obtain
\begin{align}
    F
    &= -\frac{1}{\beta}\ln\tr e^{-\beta\hat{H}} \nonumber\\
    &= F_{0} + V\,\mqty(\Delta^{*}_{1} & \Delta^{*}_{2})\hat{\alpha}(\bm{q})\mqty(\Delta_{1} \\ \Delta_{2})+\mathcal{O}(\Delta^4).
\end{align}
Here, $\beta=1/k_{\rm B}T$ is the inverse temperature, and $\bm{q}$ corresponds to the Cooper-pair momentum as introduced in Eq.~$\eqref{eq:Gap_equation}$.
The matrix $\hat{\alpha}(\bm{q})$ corresponds to the GL coefficients, and its components are given by
\begin{align}
    \alpha_{ij}(\bm{q})=\frac{1}{U}\delta_{ij}-\frac{1}{2L_xL_y}\sum_{\bm{k}}\sum_{n,m}F_{nm}(\bm{k},\bm{q})
    [\hat{\alpha}_{nm}(\bm{k},\bm{q})]_{ij},
    \label{eq:alpha}
\end{align}
where $\alpha_{11}(\bm{q})$ and $\alpha_{12}(\bm{q})$, for instance, describe the Cooper-pair motion within the layer $1$ and between the layers $1,2$, respectively.
Here, $L_x$ and $L_y$ are the system dimensions in the $x$ and $y$ directions, respectively, and $F_{nm}(\bm{k},\bm{q})$ and $\hat{\alpha}_{nm}(\bm{k},\bm{q})$ are written as
\begin{equation}
  F_{nm}(\bm{k},\bm{q}) = \frac{f(\epsilon_{n}(\bm{k},\bm{q}))-f(-\epsilon_{m}(-\bm{k},\bm{q}))}{\epsilon_{n}(\bm{k},\bm{q})+\epsilon_{m}(-\bm{k},\bm{q})},
  \label{eq:F_nm}
\end{equation}
\begin{widetext}
  \begin{equation}
    \hat{\alpha}_{nm}(\bm{k},\bm{q}) =
    \mqty(
    \abs{\bra{u_{n}(\bm{k},\bm{q})}\varphi_{1}\ket{u_{m}^{*}(\bm{-k},\bm{q})}}^{2} & \bra{u_{m}^{*}(\bm{-k},\bm{q})}\varphi_{1}^{\dagger}
    \ket{u_{n}(\bm{k},\bm{q})}\bra{u_{n}(\bm{k},\bm{q})}\varphi_{2}\ket{u_{m}^{*}(\bm{-k},\bm{q})} \\
    \bra{u_{m}^{*}(\bm{-k},\bm{q})}\varphi_{2}^{\dagger}
    \ket{u_{n}(\bm{k},\bm{q})}\bra{u_{n}(\bm{k},\bm{q})}\varphi_{1}\ket{u_{m}^{*}(\bm{-k},\bm{q})}
    & \abs{\bra{u_{n}(\bm{k},\bm{q})}\varphi_{2}\ket{u_{m}^{*}(\bm{-k},\bm{q})}}^{2}
    ),
  \end{equation}
\end{widetext}
with the Fermi-Dirac function $f(\epsilon)=1/(e^{\beta\epsilon}+1)$, eignevalues and eigenstates of $H_{\mathrm{N}}(\bm{k},\bm{q}) = H_{\mathrm{N}}(\bm{k}+\bm{q}/2)$, $\epsilon_{n}(\bm{k},\bm{q})$,
and $\ket{u_{n}(\bm{k},\bm{q})}$, and form factors $\varphi_{1}$ and $\varphi_{2}$,
\begin{align}
  \varphi_{1} &=
  \mqty(
  \varphi & 0 \\
  0 & 0
  ), \qquad
  \varphi_{2} =
  \mqty(
  0 & 0 \\
  0 & \varphi
  ),
\end{align}
where $  \varphi =
  \mqty(
  0 & 1 \\
  -1 & 0
  )$ 
is the two-by-two matrix in the spin space(see Appendix.~\ref{app:GL_alpha}.)

The second-order mean-field transition line from the normal to the superconducting states can be determined by using $\hat{\alpha}(\bm{q})$.
Let us write its eigenequation as 
\begin{equation}
\hat{\alpha}(\bm{q})\ket{\mu(\bm{q})} = \alpha_{\mu}(\bm{q})\ket{\mu(\bm{q})}.
\end{equation}
Then, the superconducting transition occurs when the smallest eigenvalue $\alpha_\mu(\bm{q})$ among various $\bm{q}$ and $\mu$ vanishes upon lowering the temperature.
We note that $\hat{\alpha}(\bm{q})$ is calculated with the normal-state Hamiltonian given in the previous section, and therefore takes into account the microscopic information of the system beyond a phenomenology.
The transition line obtained in this way coincides with that determined by solving the gap equation~\eqref{eq:Gap_equation}.

In the presence of inversion symmetry, the state $\bm{q}=0$ gives an extremum of the free energy, and indeed gives a minimum when the Zeeman field is not so large. Then, the
eigenstates of $\hat{\alpha}(0)$ correspond to the BCS and the PDW states, which are expressed as
\begin{align}
  \ket{\mathrm{BCS}} &=\frac{1}{\sqrt{2}} \mqty(1 \\ 1), \\
  \ket{\mathrm{PDW}} &=\frac{1}{\sqrt{2}}\mqty(-1 \\ 1).
\end{align}
These eigenstates correspond to order parameters of the form $(\Delta_{1}, \Delta_{2}) = (+\Delta, +\Delta)$ for the BCS state, while $(\Delta_{1}, \Delta_{2}) = (+\Delta, -\Delta)$ for the PDW state.

Generally speaking, the BCS and PDW states are mixed with each other for a finite $\bm{q}$, but it can be shown that the eigenstates of $\hat{\alpha}(\bm{q})$ continue to be $\ket{\mathrm{BCS}}$ and $\ket{\mathrm{PDW}}$ for the system in the perpendicular Zeeman field (see Appendix~\ref{app:GL_alpha}).
Thus, in the absence of the in-plane component of the Zeeman field, we can identify the indices of the eigenstates as $\mu=\mathrm{BCS},\mathrm{PDW}$, and we write
\begin{subequations}
\begin{align}
\alpha_{\mathrm{BCS}}(\bm{q},H)&\equiv\bra{\mathrm{BCS}}\hat{\alpha}(\bm{q},H)\ket{\mathrm{BCS}},\\
\alpha_{\mathrm{PDW}}(\bm{q},H)&\equiv\bra{\mathrm{PDW}}\hat{\alpha}(\bm{q},H)\ket{\mathrm{PDW}}.
\end{align}\label{eqs:alphaBCSPDW}
\end{subequations}
Here, $H$ represents, if any, the perpendicular Zeeman field.
On the other hand, in the case of noncentrosymmetric systems (i.e., in the presence of the potential gradient $v$), the BCS and PDW states are generally admixed, and we can not identify $\mu$ as BCS or PDW.
\subsection{Time-dependent Ginzburg-Landau equation}
To calculate the reciprocal and nonreciprocal paraconductivities, we use the general formulas derived from the phenomenological time-dependent Ginzburg-Landau (TDGL) equation
in momentum space \cite{Hoshino2018-qr,Daido2024-is},
\begin{equation}
  \Gamma_{0}\frac{\partial\psi_{\bm{q}}(t)}{\partial t} = -\alpha_{\bm{q}}(t)\psi_{\bm{q}}(t)+\zeta_{\bm{q}}(t),
\end{equation}
where $\Gamma_{0}$ introduces the lifetime of Cooper pairs and $\zeta_{\bm{q}}(t)$ is the white noise,
\begin{equation}
  \ev{\zeta^{*}_{\bm{q}}(t)\zeta_{\bm{q'}}(t')} = \frac{2\Gamma_{0}}{\beta V}\delta(t-t')\delta_{\bm{q},\bm{q'}}.
\end{equation}
Here, we are focusing on the temperatures slightly above the mean-field transition line, and the fluctuating Cooper pairs are considered within the Gaussian approximation.
In this paper, we assume that $\Gamma_0$ is independent of the temperature and the Zeeman field, and we set $\Gamma_0$ to be unity, which does not affect the qualitative results as seen later.
We can obtain the paraconductivity by expanding the excess current contributed by Cooper pairs with respect to the electric field by using the solution of the TDGL equation,
\begin{equation}
  j_{i} = \sigma^{ij}_{1\mathrm{s}}E_{j} + \sigma^{ijk}_{2\mathrm{s}}E_{j}E_{k} + \order{E^{3}}.
\end{equation}
Here, the electric field is taken into account by introducing the vector potential
$\bm{A}(t) = -\bm{E}t$ by $\hat{\alpha}_{\bm{q}}(t) = \hat{\alpha}_{\bm{q}-2\bm{A}(t)}$.
The general formulas of the reciprocal and nonreciprocal paraconductivities are expressed as follows~\cite{Daido2024-is}:
\begin{align}
  \sigma^{ij}_{1\mathrm{s}} &= \frac{\Gamma_{0}}{\beta V}\sum_{\bm{q}}\sum_{\mu\nu}\frac{\Re\qty[\bra{\mu}j_{i}\ket{\nu}\bra{\nu}j_{j}\ket{\mu}]}
  {\alpha_{\mu}\alpha_{\nu}(\alpha_{\mu}+\alpha_{\nu})}, \label{eq:sigma_1s}\\
  \sigma^{ijk}_{2\mathrm{s}} &= \frac{2\Gamma^{2}_{0}}{\beta V}\sum_{\bm{q}}\sum_{\mu\nu\lambda}
  \frac{J^{ijk}_{\mu\nu\lambda}(\alpha_{\mu}+\alpha_{\nu}+2\alpha_{\lambda})}
  {\alpha_{\lambda}(\alpha_{\mu}+\alpha_{\nu})(\alpha_{\mu}+\alpha_{\lambda})^{2}(\alpha_{\nu}+\alpha_{\lambda})^{2}},\label{eq:sigma_2s}
  \nonumber\\
\end{align}
with the argument $\bm{q}$ of quantities suppressed. Here,
$J^{ijk}_{\mu\nu\lambda}=\Re\qty[\bra{\mu}j_{i}\ket{\nu}\bra{\nu}j_{j}\ket{\lambda}\bra{\lambda}j_{k}\ket{\mu}]$,
$j_{i}(\bm{q}) = -\partial_{A_{i}}\hat{\alpha}(\bm{q})$ and $V = L_{x}L_{y}$.
In actual calculations, the contribution from the momentum around $\bm{q}$, which gives the minimum eigenvalue of $\hat{\alpha}(\bm{q})$, is dominant.
For this reason, we introduce the cutoff $q_{\mathrm{c}}$ to the momentum integral rather than evaluating the contribution of every $\bm{q}$ in the whole Brillouin zone.
When the minimum eigenvalue of $\hat{\alpha}(\bm{q})$ is achieved at finite momenta $\bm{q}\neq0$, the system shows the FFLO states. Note that we do not have to distinguish the FF and LO states as long as the paraconductivity is concerned within the Gauss approximation, since their difference comes from the quartic term in the GL free energy.
We numerically evaluate $j_i$ by
\begin{equation}
  -\partial_{A_{i}} \hat{\alpha}_{\bm{q}} = 2\partial_{q_{i}}\hat{\alpha}_{\bm{q}}
  = \frac{\hat{\alpha}_{\bm{q}+dq_{i}\hat{x}_{i}}-\hat{\alpha}_{\bm{q}-dq_{i}\hat{x}_{i}}}{dq_{i}} + \order{dq_{i}^{2}},
\end{equation}
where $dq_{i} = 0.001$.
\section{Reciprocal paraconductivity in the perpendicular Zeeman field}
\label{sec:linear}
In this section, we investigate the superconducting phase diagram
and reciprocal paraconductivity in the perpendicular Zeeman field.
The system has the inversion symmetry
\begin{equation}
  \mathcal{I}\hat{H}_{\mathrm{N}}(\bm{k},\bm{q})\mathcal{I}^{-1} = \hat{H}_{\mathrm{N}}(\bm{-k},\bm{-q}),
\end{equation}
with inversion operator $\mathcal{I} = \sigma_{x}\otimes\sigma_{0}$,
when the potential gradient $v$ is absent. 
The system also has a fourfold rotation symmetry, and thus we can write the reciprocal paraconductivity as $\sigma_{1\rm s}^{ij}=\delta_{ij}\sigma_{1\rm s}$.
In the following, we first consider the inversion symmetric system with $v=0$ and subsequently consider the case with $v\neq0$.
\subsection{The system with inversion symmetry}
\label{subsec:3A}
We first consider the system with inversion symmetry $v=0$.
We calculate the superconducting mean-field transition line by using the condition that the minimum eigenvalue of the GL coefficient $\hat{\alpha}(\bm{q})$ vanishes, while we determine the first-order transition line between the BCS and PDW phases by evaluating the mean-field free energy along with solving the (nonlinear) gap equation \eqref{eq:Gap_equation}.
\begin{figure}[tbp]
    \centering
    \includegraphics[width=1\linewidth]{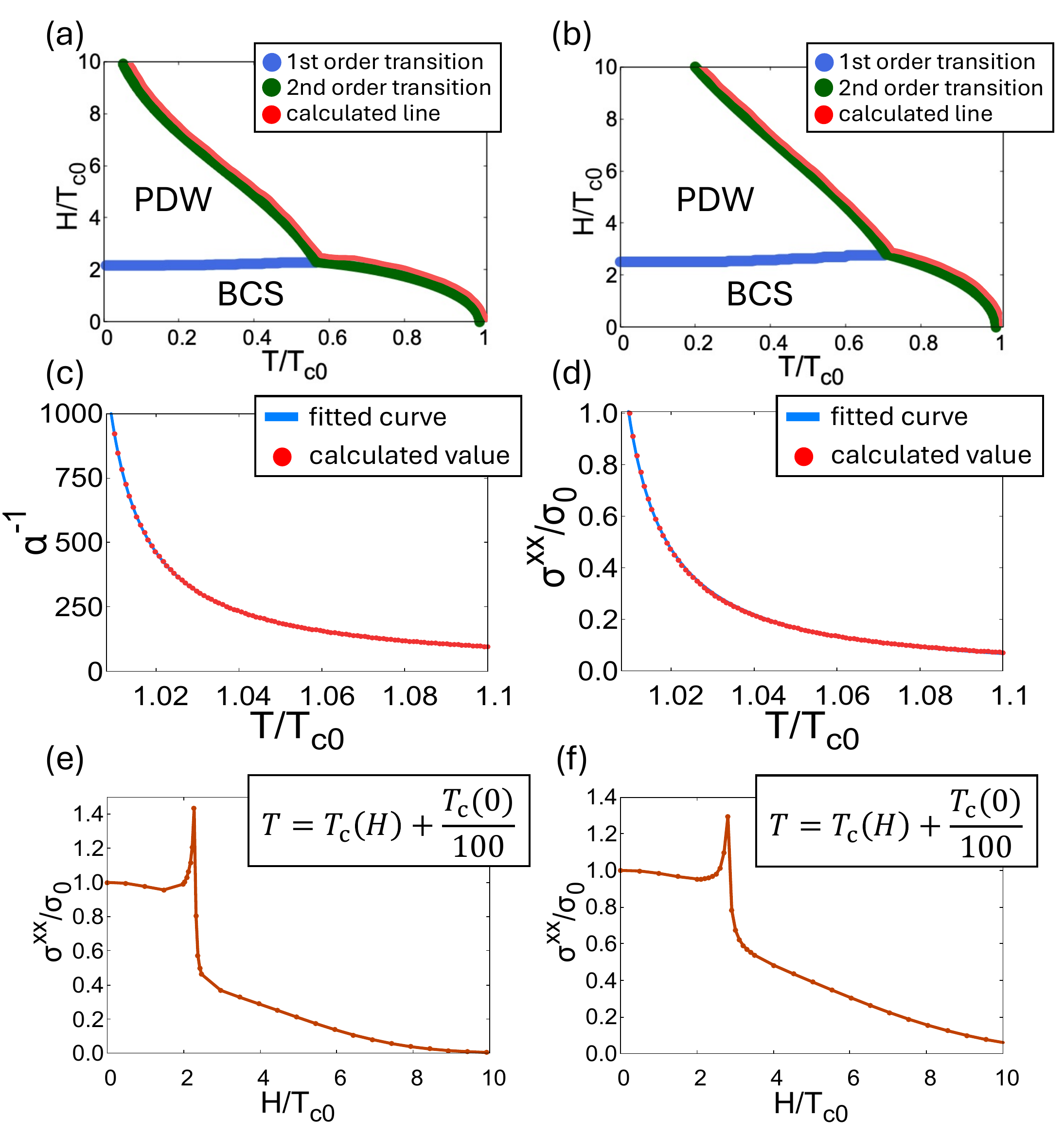}
    \caption{
    (a), (b) Mean-field phase diagram of the model. The Zeeman field and temperature are normalized with $T_{\rm c0} \equiv T_{\rm c}(H=0)$, the transition temperature in the absence of $H$. The blue line is the first-order transition line determined by solving the self-consistent equation \eqref{eq:Gap_equation}, the green line is the second-order transition line, and the red line
    indicates the line $(T,H)$ on which the paraconductivity is calculated.
    (c), (d) The numerical results of $\alpha^{-1}_{\mathrm{BCS}}$ and $\sigma_{1\rm s}$ at $H=0$ near the mean-field transition temperature (red points).
    The results are well fitted by Eqs.~\eqref{3:eq:alpha_T_fit} and \eqref{3:eq:sigma_fit}, with
    $(a_{1},b_{1}) \sim (9.173,4.543)$ and $(a_{2},b_{2}) \sim (0.010,-0.035)$ (blue lines).
    (e), (f) Reciprocal paraconductivity calculated along the red lines of (a) and (b), respectively. In this calculation, we adopted the temperature $T - T_{\rm c}(H) = T_{\rm c}(0)/100$ for each $H$ and normalized paraconductivity with $\sigma_{0} \equiv \sigma_{\rm 1s}(H = 0)$.
    Panels (a), (c), (d), and (e) are for $\alpha=0.2$, while panels (b) and (f) are for $\alpha=0.3$.}
    \label{3:fig:phase_diagram_and_linear}
\end{figure}
As shown in Figs.~\ref{3:fig:phase_diagram_and_linear}(a) and \ref{3:fig:phase_diagram_and_linear}(b),
wherein the different magnitudes of Rashba spin-orbit coupling $\alpha$ are adopted,
the BCS and PDW states are stabilized in the low and high Zeeman fields, respectively,
reproducing the known phase diagrams of the bilayer superconductor \cite{Yoshida2012-wa}. 
In this case, the states with zero-momentum $\bm{q}=0$ are stabilized in the whole phase diagram.
Thus, by using Eqs.~\eqref{eqs:alphaBCSPDW} and denoting the mean-field transition temperature in the Zeeman field $H$ as $T_{\rm c}(H)$, we should have
\begin{align}
\alpha_{\mathrm{BCS}}(H)\equiv\alpha_{\rm BCS}(0,H)\sim T-T_{\rm c}(H)
\label{eq:BCS_asym}
\end{align}
in the low Zeeman field and
\begin{align}
\alpha_{\mathrm{PDW}}(H)\equiv\alpha_{\rm PDW}(0,H)\sim T-T_{\rm c}(H)
\end{align}
in the high Zeeman field.
This behavior has been confirmed as shown in Fig.~\ref{3:fig:phase_diagram_and_linear} (c) in the case of $\alpha_{\mathrm{BCS}}(H=0)$ with the spin-orbit coupling $\alpha=0.2$, supporting the validity of our numerical calculations.
The calculated data (red points) are well fitted by the curve (blue line)
\begin{align}
\alpha^{-1}_{\mathrm{BCS}}(0) &= \frac{a_{1}}{[T-T_{\mathrm{c}}(0)]/T_{\mathrm{c}0}} + b_{1},\label{3:eq:alpha_T_fit} 
\end{align}
where the $b_1$ term represents the deviation from the asymptotic form Eq.~\eqref{eq:BCS_asym} due to $T-T_{\rm c}(0)\neq0$, namely the $O(T-T_{\rm c}(0))^2$ term in $\alpha_{\mathrm{BCS}}(0)$.

It is known that the reciprocal paraconductivity increases as $(T-T_{\rm c})^{-1}$ upon approaching the mean-field transition temperature in two-dimensional superconductors~\cite{larkin2005theory}.
To illustrate the validity of our calculations, we calculated the reciprocal paraconductivity $\sigma_{1\rm s}$ in the case of $H=0$ and $\alpha=0.2$ by using Eq.~\eqref{eq:sigma_1s}, and we show the temperature dependence of $\sigma_{1\mathrm{s}}$ in Fig. \ref{3:fig:phase_diagram_and_linear}(d).
The result indicates that $\sigma_{1\mathrm{s}}$ shows an expected asymptotic behavior.
Indeed, the calculated data (red points) are well fitted by the fitting curve (blue line)
\begin{align}
\frac{\sigma_{1\mathrm{s}}}{\sigma_{0}} &= \frac{a_{2}}{[T-T_{\mathrm{c}}(0)]/T_{\mathrm{c}0}} + b_{2}.
  \label{3:eq:sigma_fit}
\end{align}

We calculate the reciprocal paraconductivity along the transition line highlighted with green in Figs.~\ref{3:fig:phase_diagram_and_linear}(a) and \ref{3:fig:phase_diagram_and_linear}(b), by using Eq.~\eqref{eq:sigma_1s}.
Specifically, for each value of the Zeeman field $H$, we adopt the temperature $T$,
\begin{equation}
  \label{eq:tempTch}
  T=T_{\mathrm{c}}(H) +\frac{T_{\mathrm{c}}(0)}{100},
\end{equation}
which is slightly above the mean-field transition temperature $T_{\rm c}(H)$. For clarity, the lines where the paraconductivity is calculated are highlighted with red in Figs.~\ref{3:fig:phase_diagram_and_linear}(a) and \ref{3:fig:phase_diagram_and_linear}(b).
We adopt $(T,H)$ as given by Eq.~\eqref{eq:tempTch} for calculations of the reciprocal and nonreciprocal paraconductivity in this paper unless otherwise specified.

We see that the reciprocal paraconductivity shows a peak near the BCS-PDW transition point in both cases of $\alpha = 0.2$ and $\alpha = 0.3$ in Figs.~\ref{3:fig:phase_diagram_and_linear}(e) and \ref{3:fig:phase_diagram_and_linear}(f).
\begin{figure}[tbp]
    \centering
    \includegraphics[width=1.0\linewidth]{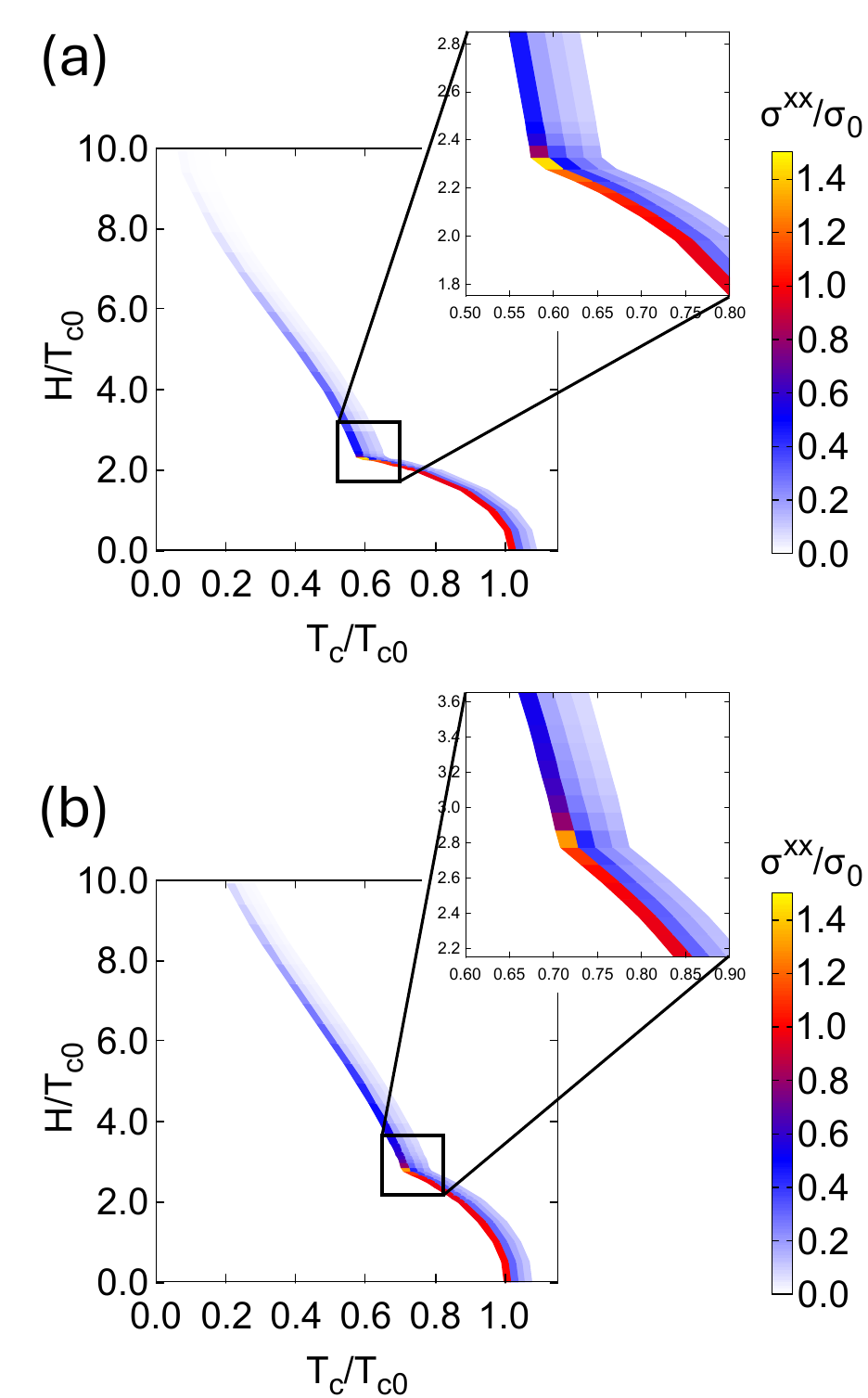}
    \caption{The color map of the reciprocal paraconductivity near the mean-field transition line.
    (a) is the case of $\alpha = 0.2$ and (b) is the case of $\alpha = 0.3$.
    The calculations are performed in the temperature interval $T_{\rm c0}/100\le T-T_{\rm c}(H)\le 9T_{\rm c0}/100$ for each $H$.}
    \label{fig:linear_color}
\end{figure}
The peak structures of the paraconductivity are manifest in Figs.~\ref{fig:linear_color}(a) and (b), where $\sigma_{\rm 1s}$ is shown by the color map near the mean-field transition line.
These peaks stem from the degeneracy of the superconducting states at the BCS-PDW transition point.
We can consider this
by separating the paraconductivity to contributions of the BCS and PDW fluctuation contributions as
\begin{equation}
  \sigma_{1\mathrm{s}} = \sigma_{\mathrm{BCS}} + \sigma_{\mathrm{PDW}} + \sigma_{\mathrm{BCS-PDW}}.
\end{equation}
Here, $\sigma_{\mathrm{BCS}}$ ($ \sigma_{\mathrm{PDW}}$) refers to the contribution with $\mu=\nu=\mathrm{BCS}$ (PDW) in Eq.$~\eqref{eq:sigma_1s}$,
and they are explicitly given by
\begin{align}
    \sigma_{\rm BCS} &= \frac{\Gamma_{0}}{\beta V}\sum_{\bm{q}}\sum_{\mu\nu}\frac{\Re[\bra{\rm BCS}j_{i}\ket{\rm BCS}\bra{\rm BCS}j_{j}\ket{\rm BCS}]}{2\alpha^{3}_{\rm BCS}},
    \label{eq:sigma_BCS} \\
    \sigma_{\rm PDW} &= \frac{\Gamma_{0}}{\beta V}\sum_{\bm{q}}\sum_{\mu\nu}\frac{\Re[\bra{\rm PDW}j_{i}\ket{\rm PDW}\bra{\rm PDW}j_{j}\ket{\rm PDW}]}{2\alpha^{3}_{\rm PDW}},
    \label{eq:sigma_PDW}
\end{align}
while $\sigma_{\mathrm{BCS-PDW}}$ represents the contribution with $\mu\neq \nu$ and vanishes in the perpendicular Zeeman field as shown in Appendix~\ref{app:GL_alpha}.
Here, arguments $\bm{q}$ and $H$ of the quantities $j_i(\bm{q},H)$, $\alpha_{\mathrm{BCS}}(\bm{q},H)$, and $\alpha_{\mathrm{PDW}}(\bm{q},H)$ are implicit.

\begin{figure}[tbp]
    \centering
    \includegraphics[width=1\linewidth]{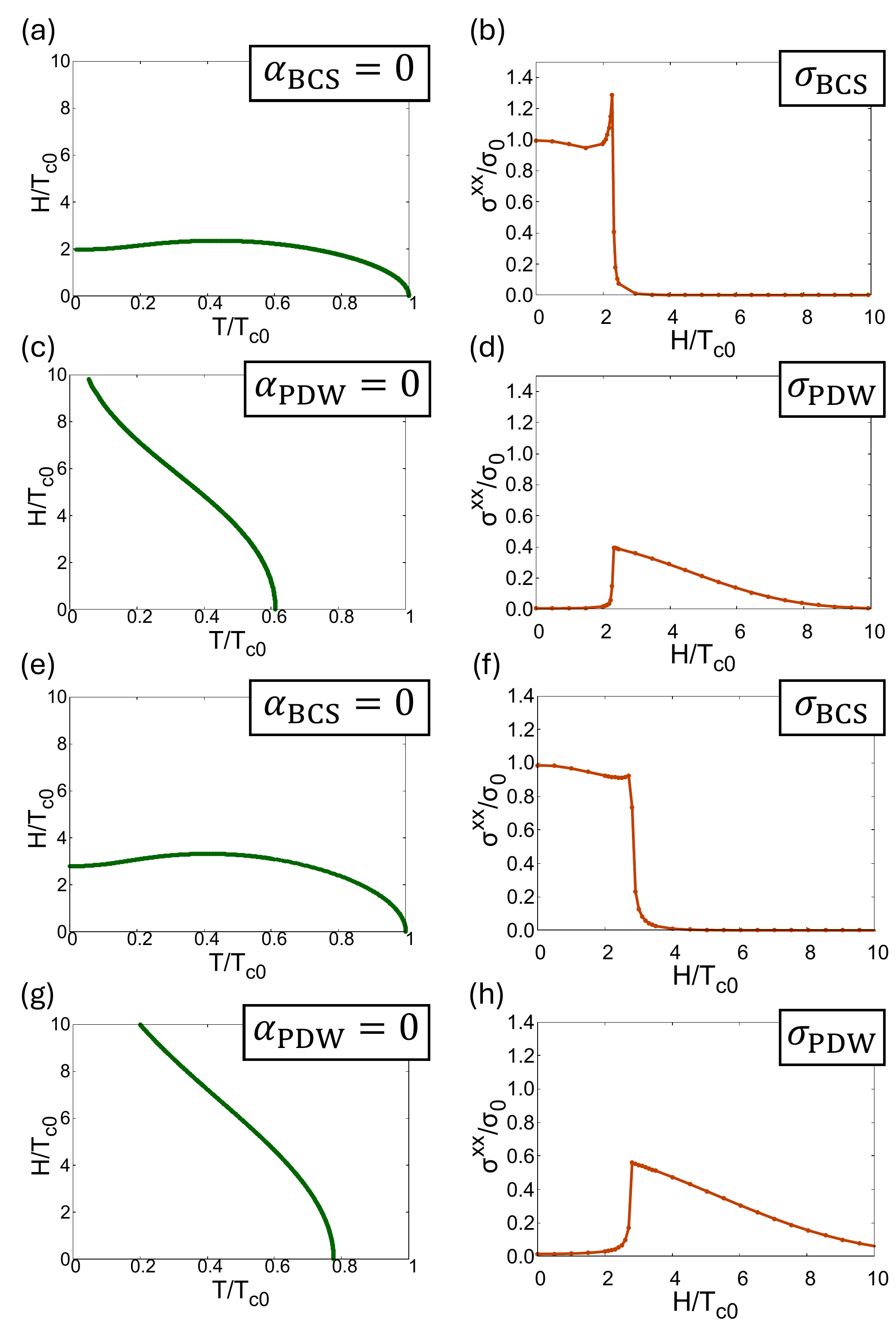}
    \caption{
    (a), (c), (e), (g) Putative transition lines of the BCS and PDW states, which are introduced to discuss the contributions of the BCS and PDW fluctuations to the reciprocal paraconductivity (see the main text).
    (a) and (e) show the line $\alpha_{\mathrm{BCS}} = 0$, and (c) and (g) show the line $\alpha_{\mathrm{PDW}} = 0$.
    (b), (d), (f), (h) Reciprocal paraconductivity contributed from the BCS and PDW fluctuations, that is, $\sigma_{\rm BCS}$ and $\sigma_{\rm PDW}$ defined in Eqs.~\eqref{eq:sigma_BCS} and \eqref{eq:sigma_PDW}, respectively.
    (b) and (f) show $\sigma_{\mathrm{BCS}}$, and (d) and (h) show $\sigma_{\mathrm{PDW}}$. Note that they are calculated along the transition line (red lines in Figs.~\ref{3:fig:phase_diagram_and_linear}(a) and (b)).
    Panels (a-d) are the results for $\alpha = 0.2$ while (e-h) are for $\alpha = 0.3$.}
    \label{3:fig:BCS-PDW}
\end{figure}

To discuss the BCS and PDW contributions,
it is convenient to 
consider the putative transition lines of the BCS and PDW states.
For instance, while the true mean-field transition temperature is determined by $\alpha_{\mathrm{BCS}}(T,H)=0$ in the low Zeeman field, we can define something like a PDW transition line by $\alpha_{\mathrm{PDW}}(T,H)=0$, which, if any, appears in the lower temperature region.
This is shown in Figs.~\ref{3:fig:BCS-PDW}(c) and (g) for $\alpha=0.2$ and $0.3$, respectively.
Note that the putative transition line is introduced only to discuss the paraconductivity, and does not generally coincide with the true first-order BCS-PDW transition line in Figs.~\ref{3:fig:phase_diagram_and_linear} (a) and (b) since the effects of the finite superconducting order parameter are not taken into account.
We can also define the putative BCS transition line in the same way, and they are shown in Figs.~\ref{3:fig:BCS-PDW} (a) and (e) for $\alpha=0.2$ and $0.3$.
The BCS (PDW) putative transition line coincides with the true mean-field transition line $T_{\rm c}(H)$ shown by green lines in Figs.~\ref{3:fig:phase_diagram_and_linear}(a) and (b) in the low (high) Zeeman field.

Since BCS and PDW contributions to the reciprocal paraconductivity Eqs.~\eqref{eq:sigma_BCS} and~\eqref{eq:sigma_PDW} are expected to be large for small $\alpha_{\mathrm{BCS}}(T,H)$ and $\alpha_{\mathrm{PDW}}(T,H)$, respectively, $\sigma_{\mathrm{BCS}}$ and $\sigma_{\mathrm{PDW}}$ would be sizable when the putative transition lines are close to the true mean-field transition line $T_{\mathrm c}(H)$. To clarify this point, we show $\sigma_{\mathrm{BCS}}$ ($\sigma_{\mathrm{PDW}}$) for $\alpha=0.2$ and $0.3$ in
Figs.~\ref{3:fig:BCS-PDW}(b) and \ref{3:fig:BCS-PDW}(f)  (Figs.~\ref{3:fig:BCS-PDW}(d) and \ref{3:fig:BCS-PDW}(h)), respectively.
They are calculated along the true mean-field transition line, and summing up Figs.~\ref{3:fig:BCS-PDW} (b) and (f) reproduces Fig.~\ref{3:fig:phase_diagram_and_linear} (e), for instance.
As expected, $\sigma_{\mathrm{BCS}}$ and $\sigma_{\mathrm{PDW}}$ mainly contributes in the low and high Zeeman field, respectively.
Importantly, around the BCS-PDW transition point,
both the BCS and PDW contributions take sizable values so that the reciprocal paraconductivity is enhanced.
Thus, the enhanced fluctuation due to the degeneracy of the BCS and PDW states is one of the main reasons for the peaked reciprocal paraconductivity.

\begin{figure}[tbp]
    \centering
    \includegraphics[width=1.0\linewidth]{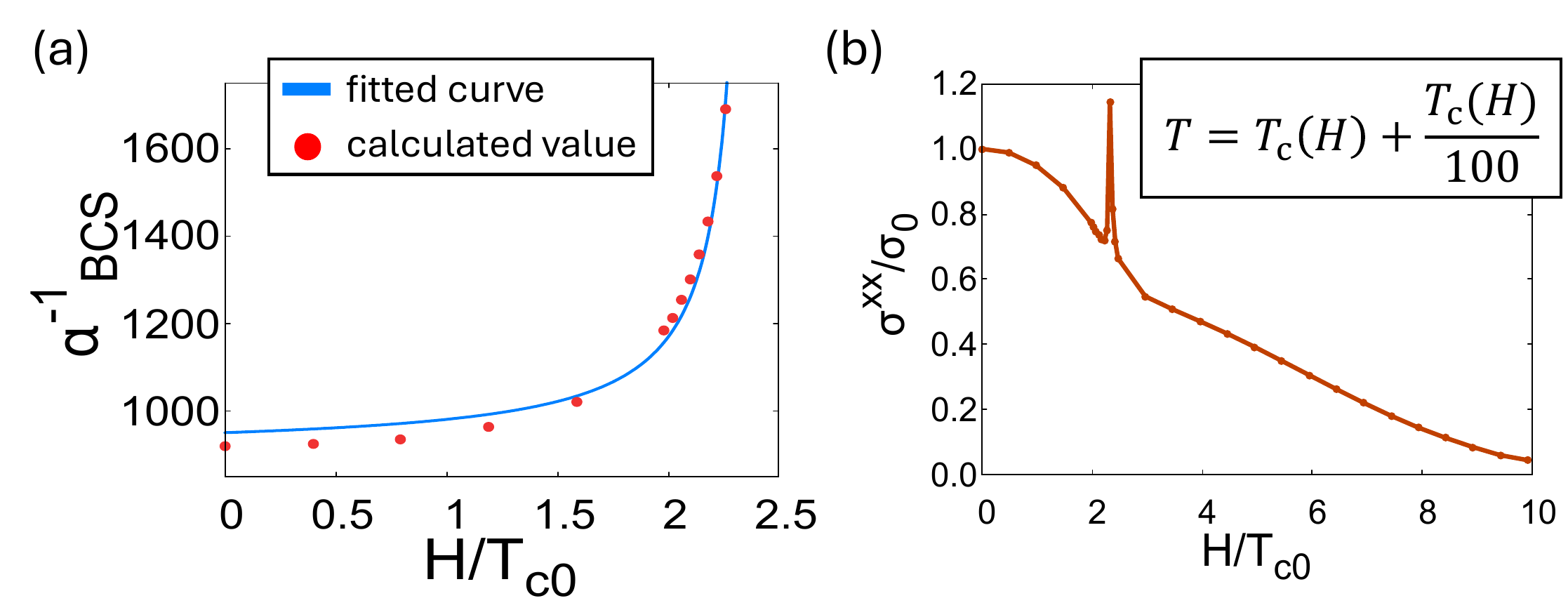}
    \caption{(a) The 
    Zeeman field dependence of $\alpha_{\mathrm{BCS}}^{-1}(H)$ at the temperature $T=T_{\rm c}(H)+T_{\rm c0}/100$ (red points).
    The numerical data shown by red points are well fitted by the form Eq.~\eqref{eq:alpha_h} (blue line).
    Fitting parameters are $(a_{3},b_{3}) \sim (99.9,909.8)$.
    (b) Reciprocal paraconductivity $\sigma_{1\rm s}$ calculated with $\Gamma_{0} = \tau_{0}N_{0}$ and $T-T_{\mathrm{c}}(H) = T_{\mathrm{c}}(H)/100$.}
    \label{3:fig:fit_tau}
\end{figure}

Interestingly, Fig.~\ref{3:fig:BCS-PDW}(b) shows that the BCS contribution in the case of $\alpha = 0.2$ by itself has a peak at the BCS-PDW transition point.
This provides another contribution to the peak structure of $\sigma_{1\mathrm{s}}$ in Fig.~\ref{3:fig:phase_diagram_and_linear}(e).
The peak of $\sigma_{\mathrm{BCS}}$ stems from the Pauli paramagnetic depairing effect, which can be understood as follows. The putative transition line in Fig.~\ref{3:fig:BCS-PDW}(a) can be expressed as the putative critical Zeeman field of the BCS state $H=H_{\mathrm{BCS}}(T)$, which is defined by $\alpha_{\mathrm{BCS}}(T,H_{\mathrm{BCS}}(T))=0$. 
It can also be expressed as the putative transition temperature by $\alpha_{\mathrm{BCS}}(T_{\mathrm{BCS}}(H),H)=0$.
The $H_{\mathrm{BCS}}(T)$ line has a maximum, and we denote this point in the phase diagram by $(T^*,H^*)$.
When we consider the region near $(T^*,H^*)$,  $\alpha_{\mathrm{BCS}}$ up to $\mathcal{O}(T-T_\mathrm{BCS}(H))$,
\begin{align}
  \alpha_{\mathrm{BCS}}(T,H)  \simeq&\left.\frac{\partial\alpha_{\mathrm{BCS}}}{\partial T}\right|_{T_{\mathrm{BCS}}(H),H}[T-T_{\mathrm{BCS}}(H)],
\end{align}
 is approximately proportional to $(H^{*}-H)$.
 This is because the coefficient satisfies
\begin{align}
\frac{\partial\alpha_{\mathrm{BCS}}}{\partial T}=-\frac{dH_{\mathrm{BCS}}}{dT}\frac{\partial\alpha_{\mathrm{BCS}}}{\partial H},
\end{align}
on the transition line and thus vanishes
at $(T^*,H^*)$, leading to the $H-H^*$ dependence when seen as a function of the Zeeman field $H$.
This behavior is confirmed by fitting the numerical results of $\alpha_{\mathrm{BCS}}^{-1}$ with the functional form,
\begin{align}
    &\alpha^{-1}_{\rm BCS}(T = T_{\rm c}(H)+T_{\rm c}(0)/100,H)
    \notag\\
    &= \frac{a_{3}}{(H^{*}-H)/T_{\rm c0}} + b_{3},
    \label{eq:alpha_h} 
\end{align}
and the fitting parameters $a_3$ and $b_3$, as shown in
Fig.~\ref{3:fig:fit_tau}(a).
Thus, $\alpha_{\rm BCS}(T,H)$ is small near $(T^*,H^*)$ and naturally enhances the reciprocal paraconductivity, giving rise to the peak structure.
The maximum $H^*$ of the putative BCS critical Zeeman field is typically obtained in the phase diagram of the FFLO superconductivity~\cite{Matsuda2007-xw}.
The decreasing tendency of $\partial\alpha_{\mathrm{BCS}}/\partial T\propto dH_{\mathrm{BCS}}/dT$
along the transition line is considered to be a general feature of superconductors suffering the Pauli paramagnetic depairing effect, which are characterized by the expression
$H_{\mathrm{BCS}}(T)=O( \sqrt{[T_{\mathrm{BCS}}(0)-T]/T_{\mathrm{BCS}}(0)})$ for $T\sim T_{\mathrm{BCS}}(0)$.
Thus, the Pauli paramagnetic depairing effect is identified as the origin of the peaked $\sigma_{\mathrm{BCS}}$, which is an origin of the peak in the paraconductivity.

In contrast to the case of $\alpha=0.2$ in Fig.~\ref{3:fig:BCS-PDW}(b),
$\sigma_{\mathrm{BCS}}$ of the case $\alpha=0.3$ in Fig.~\ref{3:fig:BCS-PDW}(f) does not show a pronounced peak structure by itself. This is because
the BCS-PDW transition point on the green line in Fig.~\ref{3:fig:phase_diagram_and_linear}(b) for $\alpha = 0.3$ is farther from $(T^*,H^*)$,
so that the Pauli paramagnetic depairing effect is less significant in Fig.~\ref{3:fig:BCS-PDW}(f).
As for the PDW contributions $\sigma_{\mathrm{PDW}}$,
they are least affected by the Pauli paramagnetic depairing effect. Indeed, we can see in Figs.~\ref{3:fig:BCS-PDW}(c), \ref{3:fig:BCS-PDW}(d),
\ref{3:fig:BCS-PDW}(g), and \ref{3:fig:BCS-PDW}(h) that
its putative critical Zeeman field does not have an extremum, and thus there is no sharp peak in $\sigma_{\mathrm{PDW}}$ like that in Fig.~\ref{3:fig:BCS-PDW}(b) [Figs.~\ref{3:fig:BCS-PDW}(d) and \ref{3:fig:BCS-PDW}(h)].
Physically, this is attributed to the robustness of the PDW state against the Pauli depairing effect~\cite{Yoshida2012-wa,Fischer2023-kd}, as is also manifest from its transition line extending in the high-Zeeman field region compared with that of the BCS state.

We comment on the decaying behavior of $\sigma_{\mathrm{PDW}}$ in Figs.~\ref{3:fig:BCS-PDW}(d) and (h) in the high Zeeman field:
This can be understood as follows. When we write $\alpha_{\mathrm{PDW}}(\bm{q})=N_{\mathrm{PDW}}(\epsilon+\xi_{\mathrm{PDW}}^2q^2)+O(q^4)$, the reciprocal paraconductivity in the high Zeeman field can be written as $\sigma_{1\rm s}\simeq \sigma_{\mathrm{PDW}}\simeq\frac{\Gamma_0T}{2\pi\epsilon N_{\mathrm{PDW}}}$~\cite{Daido2024-is,larkin2005theory}.
It turns out that $T/N_{\mathrm{PDW}}$ decreases with increasing $H$ (data not shown), due to the decreasing $T_{\mathrm{c}}(H)$.
Note also that the reduced temperature $\epsilon=(T-T_{\mathrm{c}}(H))/T_{\mathrm{c}}(H)$ increases since $T-T_{\mathrm{c}}(H)$ is fixed to $T_{\mathrm{c}}(0)/100$.
Thus, these two factors are responsible for the decreasing $\sigma_{1\mathrm{s}}$ in the high Zeeman field.

So far, we have discussed the reciprocal paraconductivity by adopting the phenomenological relaxation parameter $\Gamma_0=1$.
Another possible choice would be $\Gamma_0=N_0\tau_{0}$, where $\tau_0=\pi/8T$ is the GL relaxation time~\cite{larkin2005theory} and it would be feasible to choose $N_0$ as $\partial_\epsilon\alpha_\mu(\bm{q})$ for the dominant superconducting state.
We show in Fig.~\ref{3:fig:fit_tau}(b) the reciprocal paraconductivity by assuming $\Gamma_0=N_0\tau_0$.
The results follow the formula $\sigma_{1\rm s}\simeq 1/16\epsilon$~\cite{larkin2005theory} except near the BCS-PDW transition points. We can see that the qualitative results remain unchanged, accompanying the peak structure of $\sigma_{1\mathrm{s}}$ coming from the degenerate superconducting states.~\footnote{Strictly speaking, the peak in $\sigma_{1\mathrm{s}}$ from the Pauli depairing effect would tend to cancel out with the contribution from $N_0$.
The other properties seem to be unaffected by the choice of $\Gamma_0$.
Note also that $N_0$ of the present choice is discontinuous at the BCS-PDW transition.}
We adopt $\Gamma_0=1$ in the rest of the paper.

The results obtained in this section suggest that we can experimentally detect the BCS-PDW phase transition by observing the enhancement of the reciprocal paraconductivity, as shown in Fig.~\ref{fig:linear_color}.
This would generally work as strong evidence for the degeneracy of the two different superconducting states not limited to the BCS and PDW states.

\subsection{The system without inversion symmetry}
\label{subsec:3B}
Next, we discuss the system without inversion symmetry,
which is broken by the application of the potential gradient $v$.
We focus on how the behavior of the paraconductivity changes while changing the magnitude of $v$.
To see this, we adopt two strengths of the potential gradient, $v=0.01$ and $v=0.05$.
Considering that the transition temperature in the absence of both $v$ and $H$ is $T_{\mathrm{c}0} \sim 0.0255$,
these two cases correspond to weakly and strongly noncentrosymmetric superconductivity, respectively.
The following discussion proceeds with $\alpha = 0.2$.

\begin{figure}[tbp]
    \centering
    \includegraphics[width=1.0\linewidth]{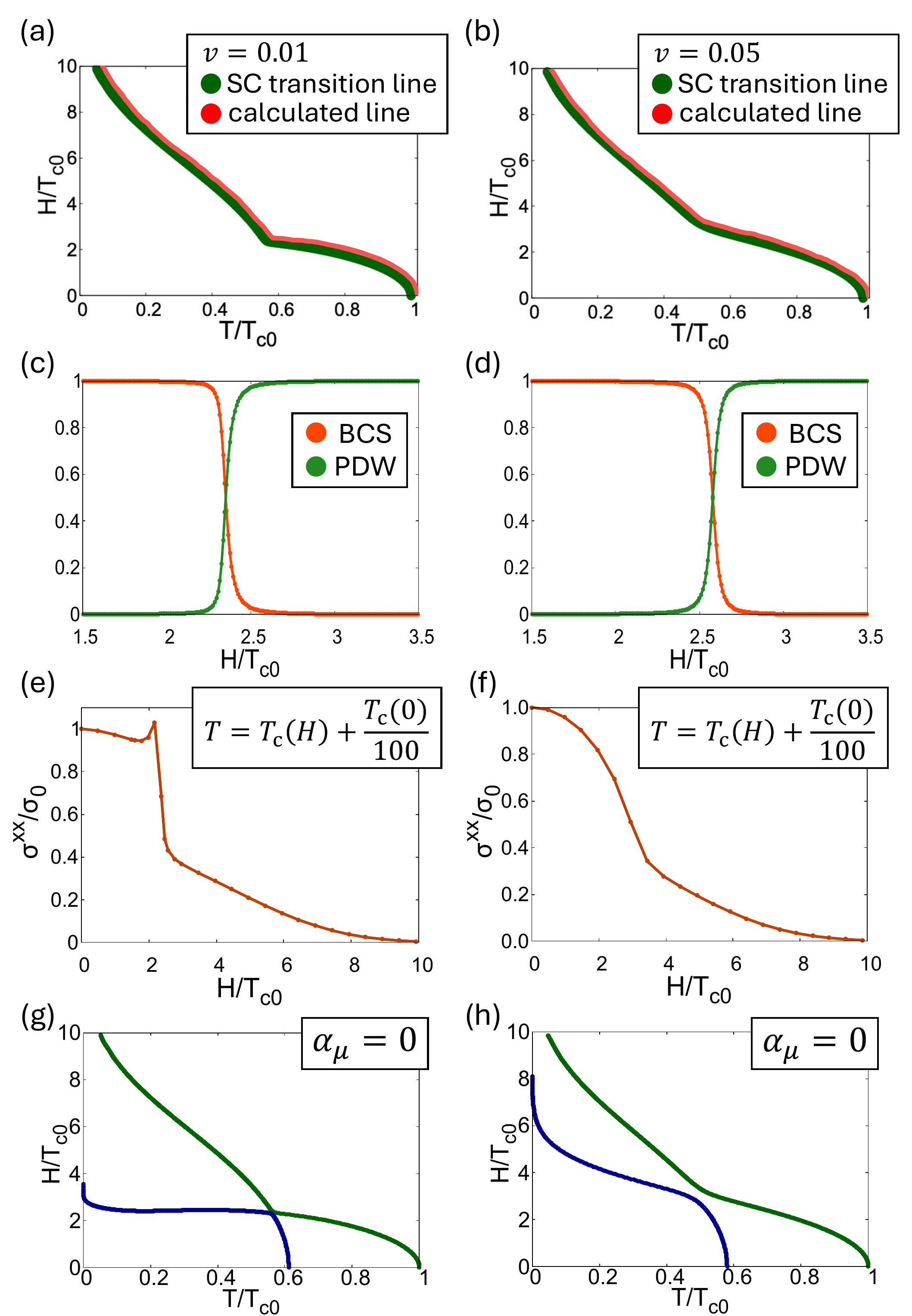}
    \caption{(a,b) Superconducting transition lines in the presence of potential gradient (a) $v = 0.01$ and (b) $v = 0.05$ (green lines).
    The red line indicates the temperature and the Zeeman field we adopted to calculate the paraconductivity.
    (c,d) $H$-dependence of the BCS and PDW components of the order parameter $|\gamma|^2$ and $|\delta|^2$ defined in Eq.~\eqref{4:eq:eigenstate}.
    The BCS-PDW crossover occurs at $H/T_{\mathrm{c}0} \sim 2.4$ in panel (c) ($v = 0.01$) while at $H/T_{\mathrm{c}0} \sim 2.6$ in panel (d) ($v = 0.05$).
    (e, f) Reciprocal paraconductivity calculated along the red line of the panels (a) and (b) for (e) $v = 0.01$ and (f) $v = 0.05$, respectively.
    (g-h) The putative transition lines defined by $\alpha_{\mu} = 0$ with $\mu=1,2$ for
    (g) $v = 0.01$ and (h) $v = 0.05$.}
    \label{3:fig:ISB}
\end{figure}

In the beginning, we calculate the superconducting transition line, and we show the results in
Figs.~\ref{3:fig:ISB}(a) and \ref{3:fig:ISB}(b).
Due to the application of the potential gradient and the broken inversion symmetry, the first-order phase transition between the BCS and PDW states changes to the BCS-PDW crossover.
This can be verified by writing the eigenstates for $\bm{q}=0$ as
\begin{align}
  \ket{\mu} = \gamma\ket{\mathrm{BCS}}+\delta\ket{\mathrm{PDW}},
  \label{4:eq:eigenstate}
\end{align}
with $\mu=1,2$, and looking at the squared values of its coefficients $\gamma$ and $\delta$.
We can see the crossover in Figs.~\ref{3:fig:ISB}(c) and \ref{3:fig:ISB}(d), where the evolution of the BCS component $|\gamma|^2$ and the PDW component $|\delta|^2$ along the mean-field transition lines is shown by red and green, respectively. We
find that the onset of the crossover regime in Fig.~\ref{3:fig:ISB}(d) is shifted to the higher Zeeman field than Fig.~\ref{3:fig:ISB}(c) by increasing the potential gradient.
By comparing the cases of $v=0.01$ and $v=0.05$, we can see that in the former case the kink point of $T_{\mathrm{c}}(H)$ remains near the BCS-PDW crossover area,
while in the latter case the kink changes to a smooth upturn.
The difference between the two cases is attributed to the extent to which the degeneracy of the BCS and PDW eigenvalues is lifted.

We show the reciprocal paraconductivity of the system with the potential gradient in
Figs.~\ref{3:fig:ISB}(e) and \ref{3:fig:ISB}(f).
It is shown that the reciprocal paraconductivity has a peak accompanied by the BCS-PDW crossover for $v = 0.01$.
Thus, the reciprocal paraconductivity of the weakly noncentrosymmetric system resembles that of the centrosymmetric system, as expected.
On the other hand, the peak of the paraconductivity disappears in the strongly noncentrosymmetric system $v = 0.05$.
This seems to be correlated with the smoother transition line observed in Fig.~\ref{3:fig:ISB}(h) compared with that in Fig.~\ref{3:fig:ISB}(g).
We show in Figs.~\ref{3:fig:ISB}(g) and \ref{3:fig:ISB}(h)
by green and blue lines the zero contour lines of the two eigenvalues of the GL coefficient $\hat{\alpha}$, which correspond to the putative transition lines of the mixed BCS and PDW states.
The mixing of the BCS and PDW states, or in other words the avoided crossing of the two putative transition lines,
clearly explains the presence and the absence of the approximate kink structure in the weak and strong potential gradients, respectively.
The lifted degeneracy naturally suppresses the fluctuation in the sub-leading superconducting channel, resulting in the absence of the peak in reciprocal conductivity in the strong potential gradient.

\begin{figure}[tbp]
    \centering
    \includegraphics[height=5cm]{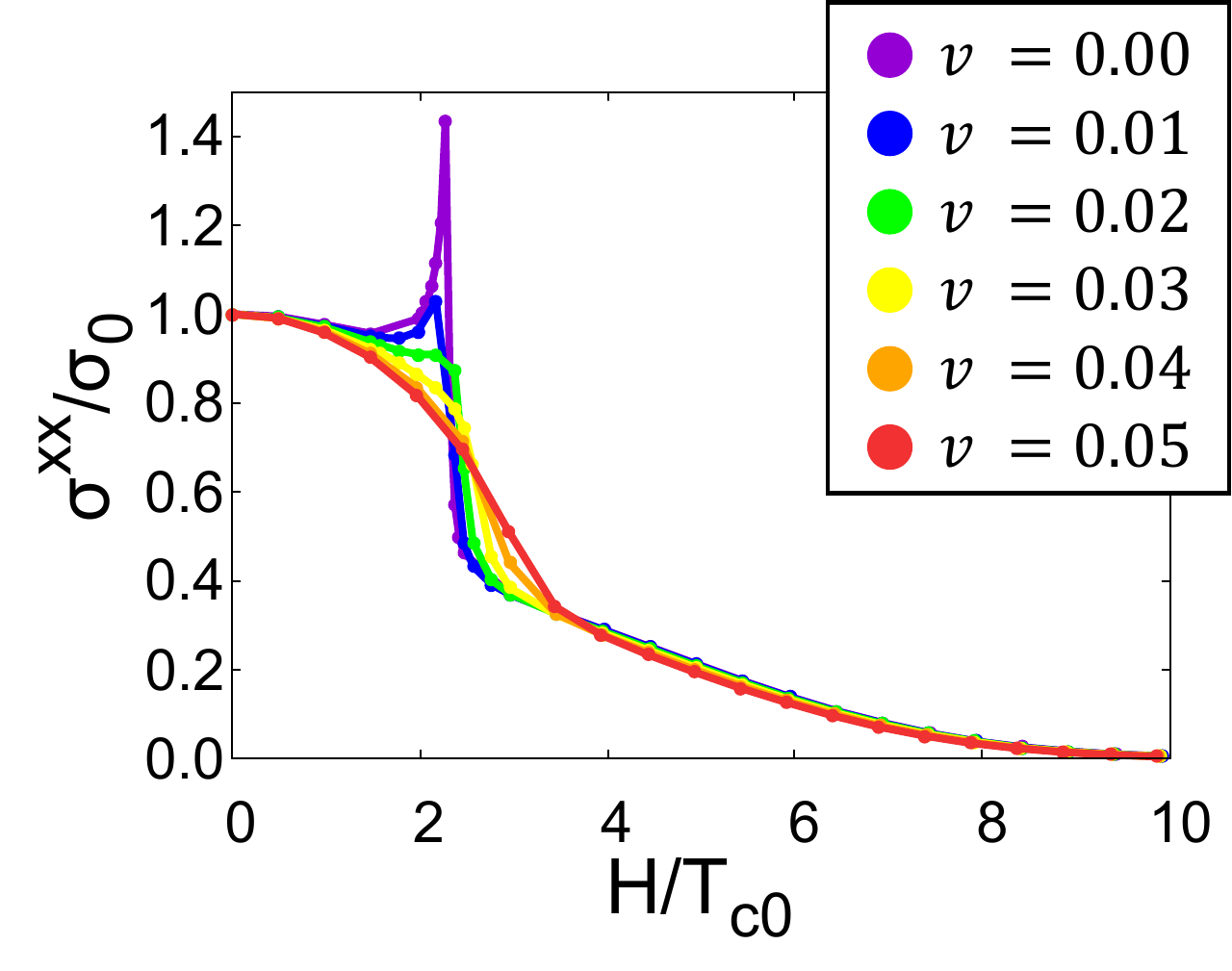}
    \caption{Reciprocal paraconductivity $\sigma_{1\rm s}$ for various values of the potential gradient.
    It can be seen that the peaks are gradually suppressed as increasing the potential gradient $v$.
    }
    \label{3:fig:linear_E_dep}
\end{figure}

Lastly, we discuss the potential gradient dependence of the reciprocal paraconductivity as shown in Fig.~\ref{3:fig:linear_E_dep}.
We can see that the peak height is monotonically decreasing upon increasing the strength of the potential gradient.
The peak vanishes near
\begin{equation}
  v \sim T_{\mathrm{c}0}.
\end{equation}
The reciprocal paraconductivity in the low and high Zeeman fields is almost unaffected by the potential gradient,
due to the dominance of the BCS and PDW states, respectively.
Experimental observation of the peaked conductivity and its decaying tendency under the application of the potential gradient will strongly support the superconducting multiphase.
\section{Reciprocal and nonreciprocal paraconductivity in the in-plane Zeeman field}
\label{sec:nonlinear}
In this section, we study the reciprocal and nonreciprocal paraconductivity of the system in the in-plane Zeeman field with and without the potential gradient.

\subsection{The system with inversion symmetry}
\label{subsec:4A}
In the beginning, we consider the system without the potential gradient. In this case, the system has inversion symmetry, and therefore the nonreciprocal paraconductivity vanishes.
We calculate the superconducting transition line and the reciprocal paraconductivity in the same way as Sec.~\ref{sec:linear}.
\begin{figure}[tbp]
    \centering
    \includegraphics[width=1\linewidth]{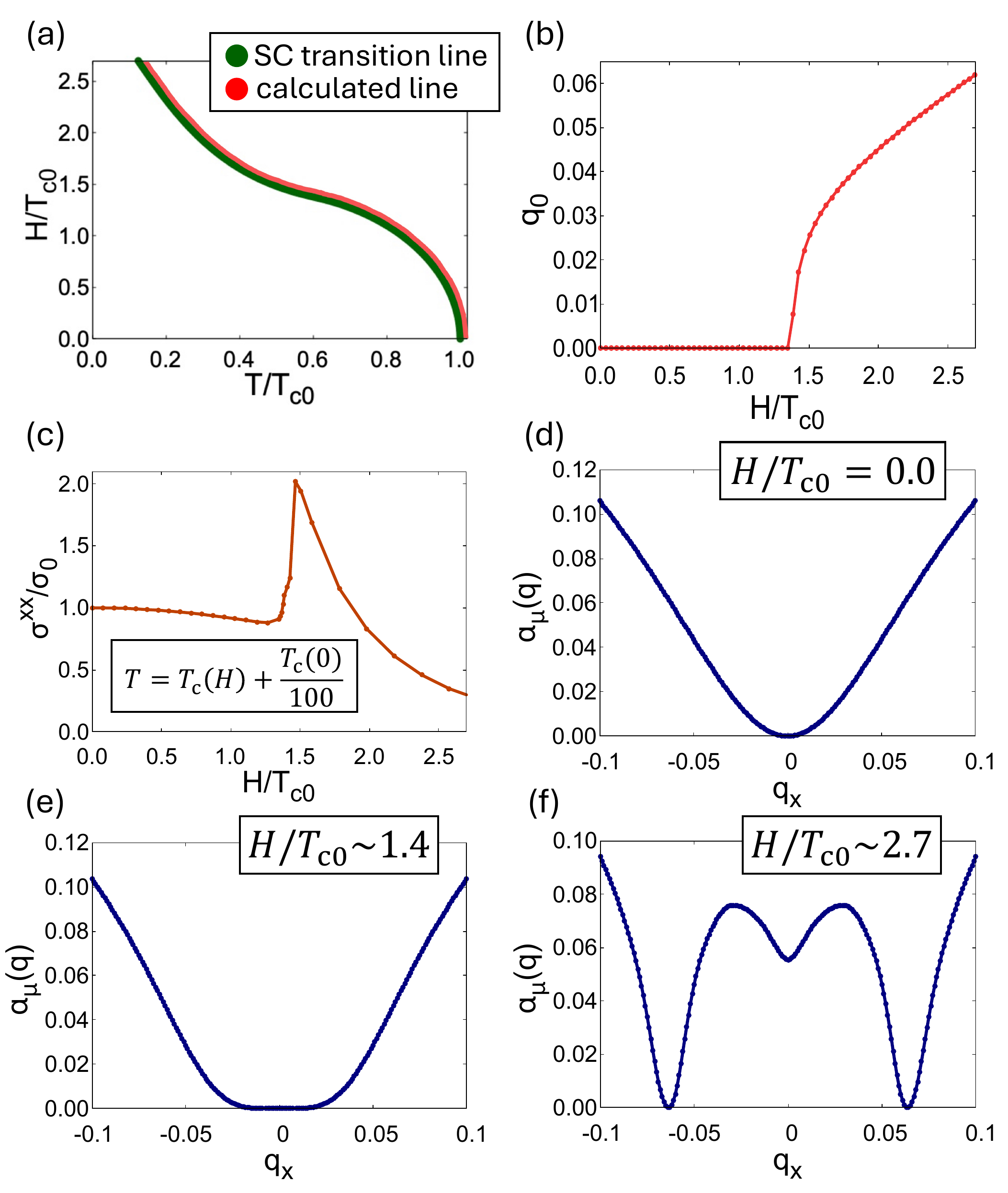}
    \caption{(a) The mean-field transition line of the system in the in-plane Zeeman field (green line).
    The red line indicates the temperature and Zeeman field adopted to calculate the paraconductivity.
    (b) The Zeeman field dependence of the equilibrium Cooper-pair momentum $q_0$ along the transition line (the green line in panel (a)).
    Superconducting state changes from the BCS state to the FFLO state and $q_0$ becomes finite
    near $H/T_{\mathrm{c}0} \sim 1.4$.
    (c) The reciprocal paraconductivity calculated along the red line in panel (a).
    (d-f) Momentum dependence of $\alpha_{\mu}$ for $q_{y} = 0$. 
    (d), (e), and (f) correspond to $H/T_{\rm c0} =0.0$, $1.4,$ and $2.7$, respectively.}
    \label{4:fig:IS}
\end{figure}

We show the mean-field superconducting transition line and the Zeeman-field dependence of the equilibrium Cooper-pair momentum along the transition line in Figs.~\ref{4:fig:IS}(a) and (b). Here, we determine the equilibrium Cooper-pair momentum $\bm{q}_0=(q_0,0)$ to minimize the eigenvalue of $\hat{\alpha}(\bm{q})$ among all $\bm{q}$, or in other words, to satisfy $\bra{\mu(\bm{q})}j_i(\bm{q})\ket{\mu(\bm{q})}=0$
~\cite{Bohm1949-pd}.
As understood from Figs.~\ref{4:fig:IS}(a) and \ref{4:fig:IS}(b), the superconducting phase diagram has the BCS phase in the low Zeeman field and the FFLO phase in the high Zeeman field~\cite{Yoshida2013}: The equilibrium Cooper-pair momentum $q_0$ becomes finite for the Zeeman field $H\gtrsim1.4T_{\rm c0}$.

We show in Figs.~\ref{4:fig:IS}(c) the reciprocal paraconductivity $\sigma_{1\rm s}^{xx}$ calculated along the red curve in Fig.~\ref{4:fig:IS}(a).
The reciprocal paraconductivity $\sigma_{1\rm s}^{xx}$ has a peak near the tricritical point, which is attributed to the shape of $\alpha_{\mu}(\bm{q})$.
We show the momentum dependence of $\alpha_{\mu}(\bm{q})$ for the system in the BCS phase ($H/T_{\rm c0} = 0.00$) and the FFLO phase ($H/T_{\rm c0} \sim 2.7$) and near the tricritical point ($H/T_{\rm c0} = 1.4$),
in Figs.~\ref{4:fig:IS}(d), \ref{4:fig:IS}(f) and \ref{4:fig:IS}(e), respectively.
Here and hereafter,  $\alpha_{\mu}(\bm{q})$ specifies the smaller of the two eigenvalues $\alpha_{\mu=1,2}(\bm{q})$.
It should be noted that the contribution to the paraconductivity formula Eq.~\eqref{eq:sigma_1s} mainly comes from the region where $\alpha_\mu(\bm{q})$ is vanishingly small. Near the tricritical point, we can see from Fig.~\ref{4:fig:IS}(e) that the area of small $\alpha_\mu(\bm{q})$ is much larger than that in Figs.~\ref{4:fig:IS}(d) and \ref{4:fig:IS}(f).
This naturally leads to the peak of reciprocal paraconductivity.
In other words, the peaked paraconductivity near the tricritical point may be interpreted as coming from the degeneracy of the BCS and FFLO states, in analogy with the case of the degenerate BCS and PDW phases in the perpendicular Zeeman field.
The result obtained here is consistent with a previous study that pointed out the enhanced reciprocal paraconductivity with the anomalous temperature scaling near the tricritical point~\cite{Konschelle2007-me,Konschelle2009-mj}.
This result implies that the observation of the reciprocal paraconductivity can be used as a probe of FFLO superconductors.
\subsection{The system without inversion symmetry}
\label{subsec:4B}
In this section, we apply the potential gradients $v = 0.01$ and $v = 0.10$, which are smaller and larger than the transition temperature $T_{\rm c0}\sim0.0255$, and correspond to weakly and strongly noncentrosymmetric superconductors, respectively.
We consider not only the reciprocal paraconductivity but also the nonreciprocal paraconductivity, which is allowed due to the broken inversion symmetry and $C_{2z}$ rotation symmetry.
As an indicator of the nonreciprocity, in addition to $\sigma_{2\rm s}^{xxx}$, we focus on the quantity
\begin{equation}
  \eta^{xxx} = \frac{\sigma^{xxx}_{2\mathrm{s}}}{(\sigma^{xx}_{1\mathrm{s}})^{2}},
  \label{4:eq:nonreciprocity}
\end{equation}
to study the nonreciprocal charge transport.
This quantity is convergent as it approaches the transition temperature, in contrast to $\sigma_{1\rm s}$ and $\sigma_{2\rm s}$. It has the dimension of the inverse of the current density and, therefore, represents the typical value of the inverse current density where nonreciprocity becomes visible~\cite{Daido2024-is}.

\begin{figure*}
    \centering
    \includegraphics[height=8.5cm]{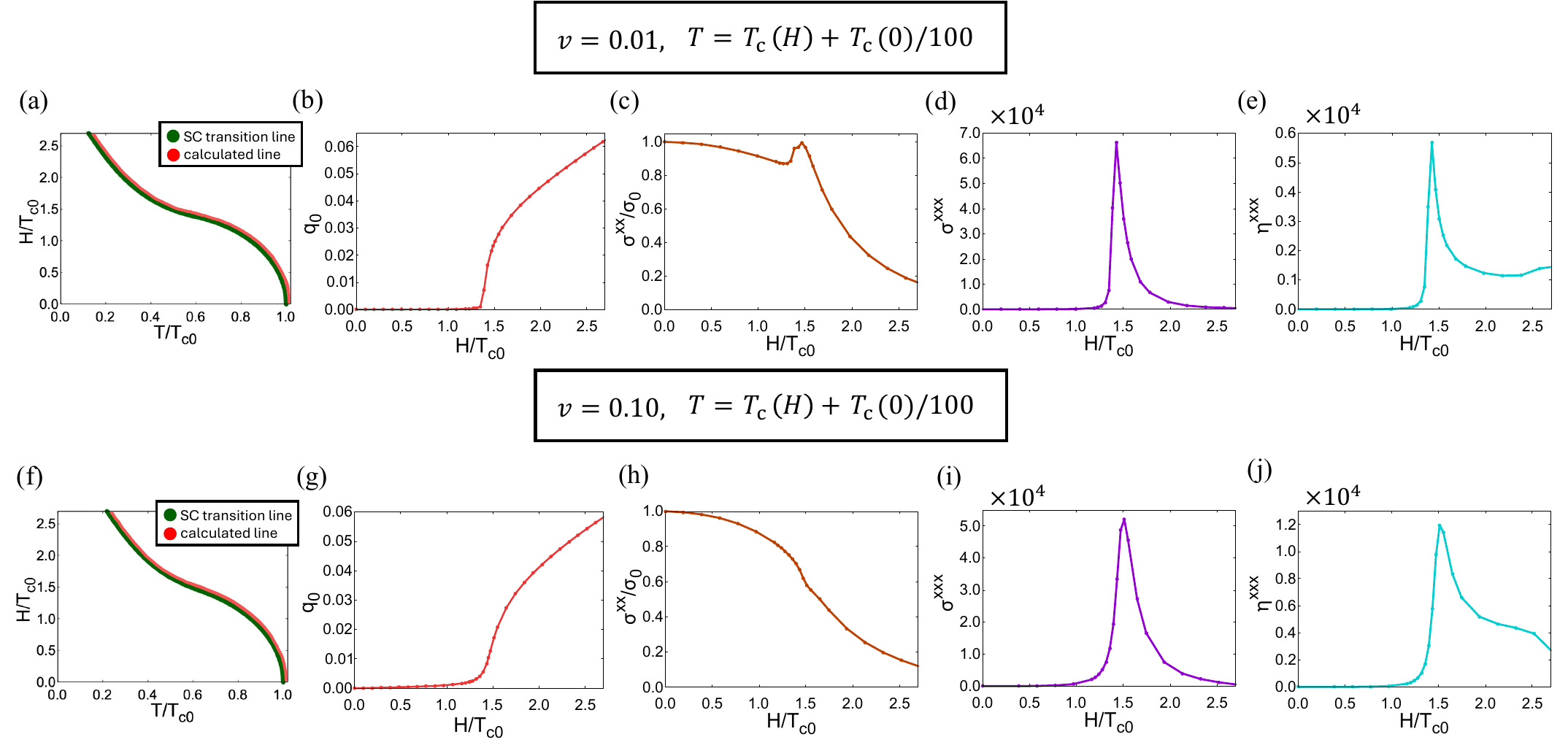}
    \caption{The mean-field transition line (green) and the line along which the paraconductivity is calculated (red) [(a), (f)],
    equilibrium Cooper-pair momentum $q_0$ [(b), (g)] along the transition line (red lines in panel (a) and (f)), reciprocal [(c), (h)]
    and nonreciprocal [(d), (i)] paraconductivities,
    and $\eta^{xxx}$ defined in Eq.~\eqref{4:eq:nonreciprocity} [(e), (j)] of the system in the in-plane Zeeman field and potential gradient.
    Panels (a-e) and (f-j) show the results for the potential gradient $v = 0.01$, $0.10$, respectively.}
    \label{4:fig:ISB}
\end{figure*}

We show the calculation results in Fig.~\ref{4:fig:ISB}.
First, we discuss the difference of the superconducting phase diagrams in the presence and absence of inversion symmetry, by comparing Figs.~\ref{4:fig:IS} and \ref{4:fig:ISB}.
By switching on the potential gradient $v$, due to the inversion symmetry breaking, the BCS-FFLO transition in Fig.~\ref{4:fig:IS} begins to change to the crossover of the so-called helical superconductivity
\cite{Barzykin2002-fo,Agterberg2003-jx,Dimitrova2003-vp,Kaur2005-vk,Agterberg2007-zi,Dimitrova2007-fx,Yanase2008-mk,Samokhin2008-oa,Michaeli2012-ks,Houzet2015-ie,Daido2024-is,bauer2012non}.
Indeed, the equilibrium Cooper-pair momentum $q_0$ takes finite values in the low-field regime before it takes large values in the high Zeeman field, as shown in Figs.~\ref{4:fig:ISB}(b) and \ref{4:fig:ISB}(g).
For clarity, we call the crossover of $q_0$ upon increasing the in-plane Zeeman field in weakly and strongly noncentrosymmetric cases as the BCS-FFLO-like crossover and the helical crossover, respectively.

When the inversion-symmetry breaking is weak ($v=0.01$), the behavior of the reciprocal paraconductivity in the moderate Zeeman field is similar to that near the BCS-PDW transition.
Indeed, we can see that the reciprocal paraconductivity has a peak in the BCS-FFLO-like crossover region [Fig.~\ref{4:fig:ISB}(c)] as well as near the BCS-FFLO transition point [Fig.~\ref{4:fig:IS}(c)].
By contrast, Fig.~\ref{4:fig:ISB}(h) shows that the peak vanishes in the strongly noncentrosymmetric system, i.e., in the helical crossover regime.
These are natural results because the peaked reciprocal paraconductivity signals the degenerate superconducting states and the degeneracy is sufficiently lifted in the strongly noncentrosymmetric case in Fig.~\ref{4:fig:ISB}(h).

\begin{figure}[tbp]
    \centering
    \includegraphics[width=1.0\linewidth]{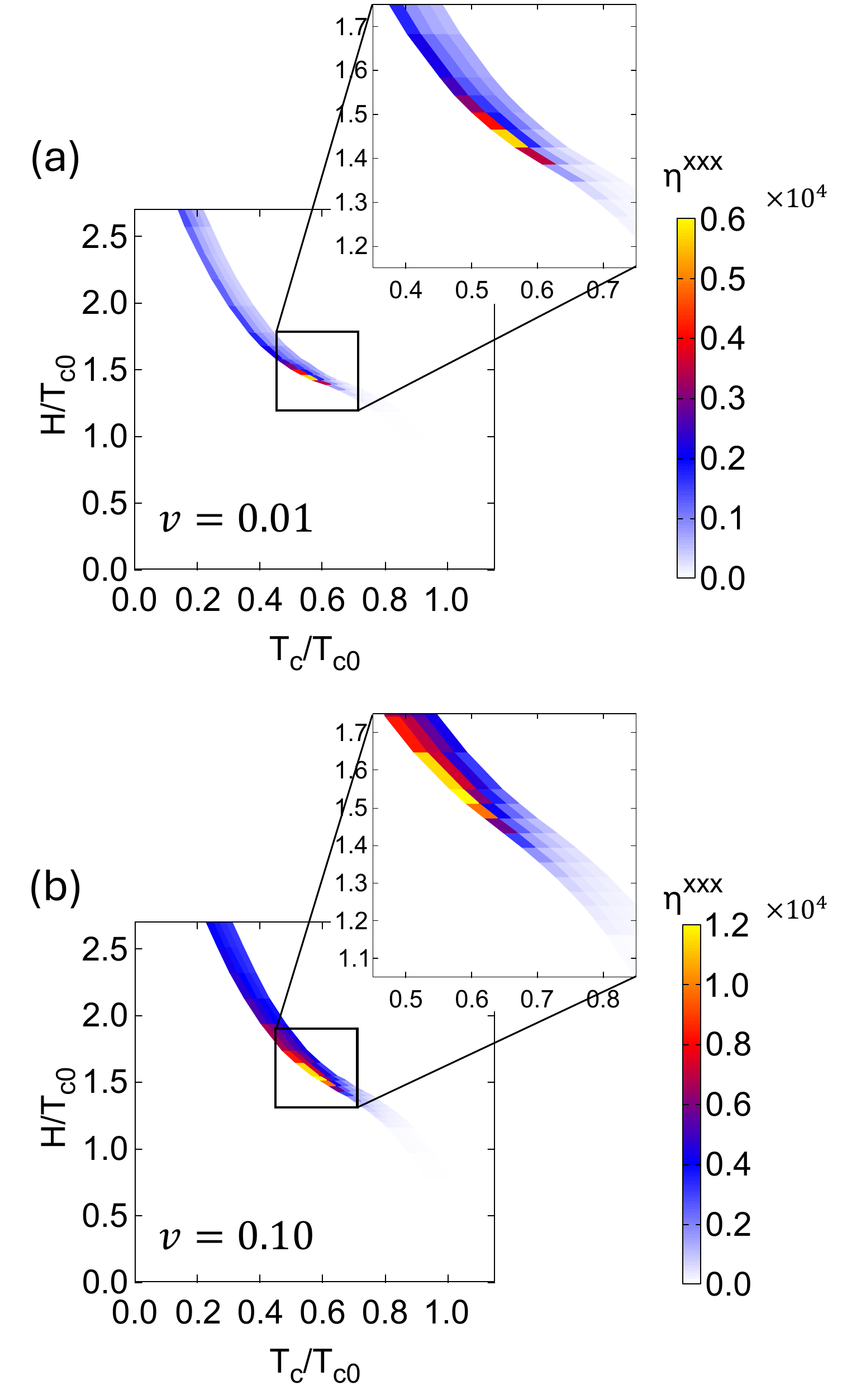}
    \caption{Temperature and Zeeman field dependence of $\eta^{xxx}$ defined in Eq.~\eqref{4:eq:nonreciprocity}, shown by the color plot.
    Panels (a) and (b) correspond to $v=0.01$ and $0.1$, respectively.
    The calculations are performed in the temperature interval $T_{\rm c0}/100\le T-T_{\rm c}(H)\le 9T_{\rm c0}/100$ for each $H$.}
    \label{fig:nonreciprocity_color}
\end{figure}

Interestingly, the behavior of the nonreciprocal paraconductivity is significantly different from the reciprocal paraconductivity.
We can see in Figs.~\ref{4:fig:ISB}(d) and \ref{4:fig:ISB}(i) that there is a peak in the nonreciprocal paraconductivity in both weakly and strongly noncentrosymmetric systems with comparable strengths of $\sigma_{2s}^{xxx}$ and $\eta^{xxx}$.
These peak structures are also manifest in the color plot of $\eta^{xxx}$ shown in Figs.~\ref{fig:nonreciprocity_color}(a) and (b) for $v=0.01$ and $0.1$, respectively.
The obtained peak in strongly noncentrosymmetric superconductors is consistent with the results of Ref.~\cite{Daido2024-is}, which studies the enhancement of nonreciprocal paraconductivity associated with the helical crossover in a strongly noncentrosymmetric mono-layer Rashba model.
Our results indicate that the enhanced nonreciprocal paraconductivity associated with the evolution of finite-momentum superconductivity also exists in the weakly noncentrosymmetric regime.

\begin{figure}[tbp]
    \centering
    \includegraphics[width=1.0\linewidth]{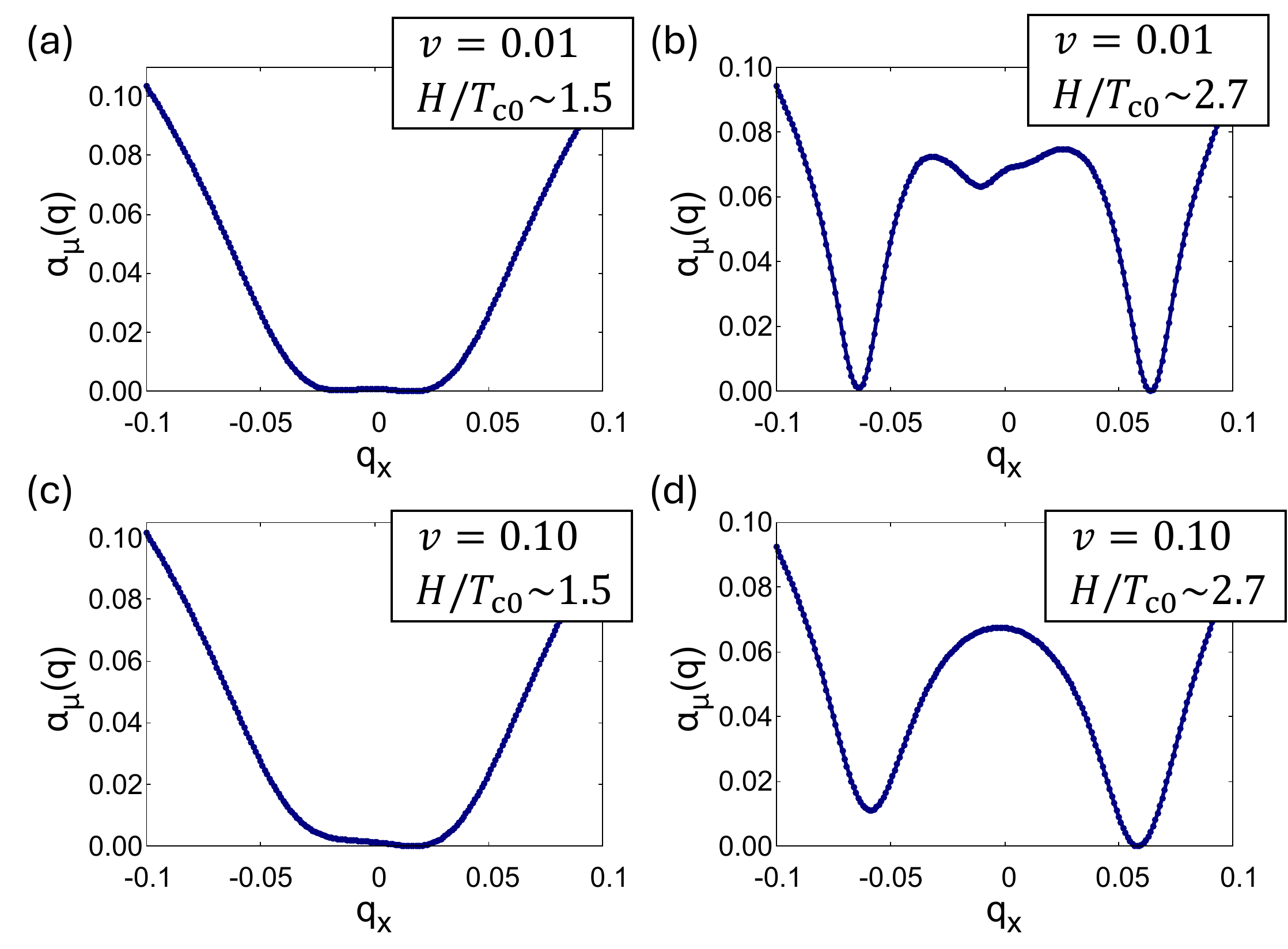}
    \caption{The momentum dependence of $\alpha_\mu(\bm{q})$.
    (a) and (b) are the case of $v = 0.01$, while (c) and (d) are the case of $v = 0.10$.
    (a) and (c) are for the moderate Zeeman field $H/T_{\rm c0} \sim 1.5$ where the nonreciprocal paraconductivity shows the peak structures.
    (b) and (d) are obtained in the high Zeeman field, $H/T_{\rm c0} \sim 2.7$.}
    \label{4:fig:alpha}
\end{figure}

\begin{figure}[tbp]
    \centering
    \includegraphics[width=1.0\linewidth]{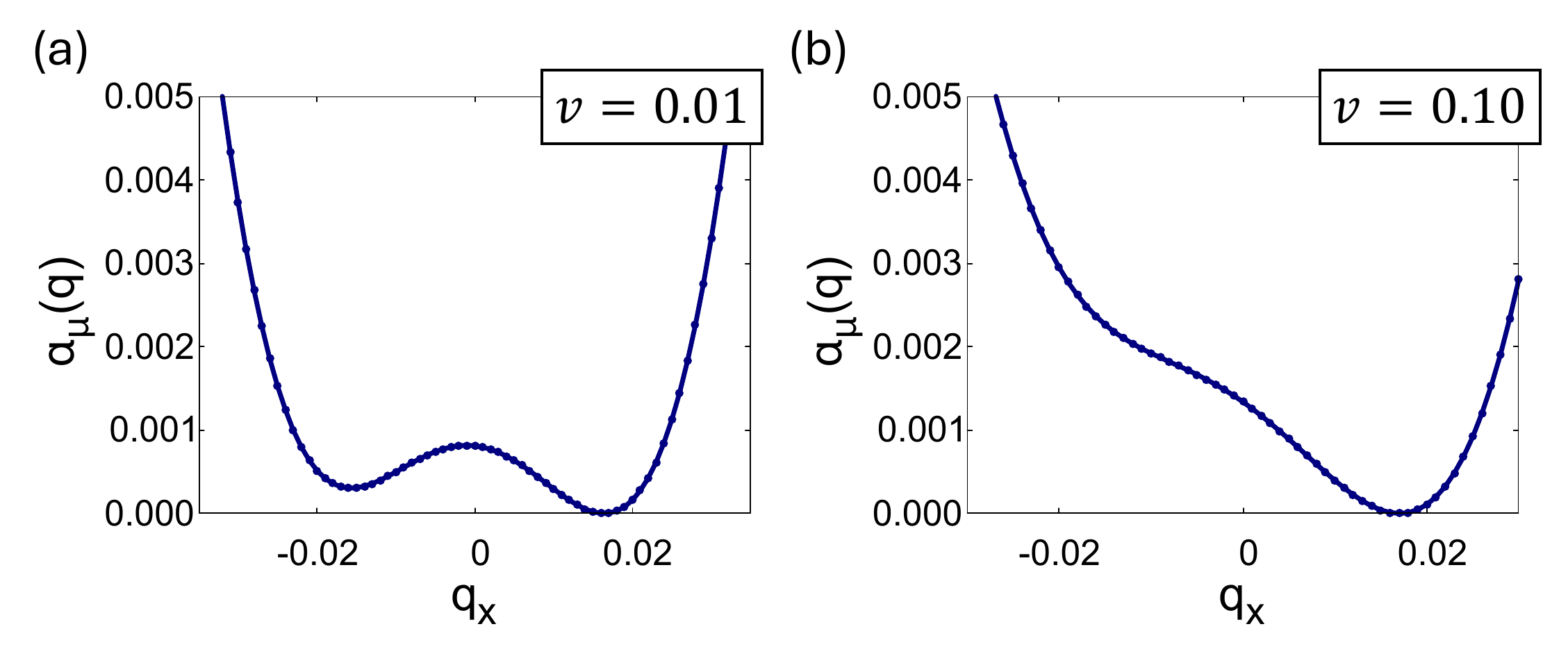}
    \caption{Enlarged figures of Figs.~\ref{4:fig:alpha}(a) and \ref{4:fig:alpha}(c) for small $\alpha_\mu(\bm{q})$.
    The panels (a) and (b) correspond to Figs.~\ref{4:fig:alpha}(a) and
    \ref{4:fig:alpha}(c), respectively.}
    \label{4:fig:alpha_enlarged}
\end{figure}

To understand the behavior of the nonreciprocal paraconductivity, we show in Figs.~\ref{4:fig:alpha} and ~\ref{4:fig:alpha_enlarged} the momentum dependence of $\alpha_\mu(\bm{q})$ with $\bm{q}=(q_x,0)$ in the moderate and strong Zeeman fields.
Here, $\mu$ specifies the minimum eigenvalue of $\hat{\alpha}(\bm{q})$.
These figures clearly show the realization of finite-momentum superconductivity in the in-plane field.
To understand the behavior of the nonreciprocal paraconductivity, we show in Figs.~\ref{4:fig:alpha} and ~\ref{4:fig:alpha_enlarged} the momentum dependence of $\alpha_\mu(\bm{q})$ with $\bm{q}=(q_x,0)$ in the moderate and strong Zeeman fields.
The results of $\alpha_\mu(\bm{q})$ for $v=0.01$ in Figs.~\ref{4:fig:alpha}(a) and \ref{4:fig:alpha}(b) are similar to those of the FFLO state shown in Figs.~\ref{4:fig:IS}(e) and \ref{4:fig:IS}(f), reflecting the weak inversion breaking ($v\ll T_{\rm c0}$), while the strongly noncentrosymmetric case ($v\gg T_{\rm c0}$) in Figs.~\ref{4:fig:alpha}(c) and \ref{4:fig:alpha}(d) shows a large deviation.
We also show in Figs.~\ref{4:fig:alpha_enlarged}(a) and \ref{4:fig:alpha_enlarged}(b) the enlarged figures of Figs.~\ref{4:fig:alpha}(a) and \ref{4:fig:alpha}(c), respectively, to consider the main contribution to the paraconductivity coming from the region $\alpha_\mu(\bm{q})\sim 0$. 
When seen in this energy scale, the structure of $\alpha_\mu(\bm{q})$ for $v=0.01$ is essentially the same as that for $v=0.1$.
Actually, the values of $\alpha_\mu(\bm{q})$ at the local minima $q\sim\pm0.02$, dubbed $\alpha_\pm$, are not degenerate due to the inversion symmetry breaking $v=0.01\neq0$, and they satisfy $\alpha_--\alpha_+\gg\alpha_+\gtrsim0$.
This means that the $\alpha_+$ valley mainly contributes to the nonreciprocal paraconductivity in the same way as the strongly noncentrosymmetric case.
Thus, at the temperatures considered here, 
the system is sufficiently noncentrosymmetric for the nonreciprocal paraconductivity even though $v\ll T_{\rm c0}$.
This explains the appearance of the peaked nonreciprocal paraconductivity in Fig.~\ref{4:fig:ISB}(d). 
 We expect that the peak width will decrease 
 when $v$ is decreased from $v=0.01$ with fixing the reduced temperature, 
 and eventually will vanish when $\alpha_--\alpha_+\sim \alpha_+$ is always satisfied along the transition line. This is consistent with $\sigma_{2s}^{xxx}=0$ at $v=0$ as enforced by the inversion symmetry.

The important point is that a small potential gradient $v\ll T_{\rm c0}$ is sufficient to obtain the sizable nonreciprocal paraconductivity.
Such a small $v$ does not usually modify superconducting properties.
In this case, on the other hand, the left and right valleys of the FFLO state shown e.g., in Fig.~\ref{4:fig:IS}(f), have intrinsic nonreciprocity even without $v$: The $\alpha_\mu(\bm{q})$ dispersion has a (cubic) asymmetry with respect to each local minimum, which 
has been identified as the origin of the nonreciprocal charge transport in noncentrosymmetric superconductors~\cite{Wakatsuki2017-hw,Wakatsuki2018-vd,Hoshino2018-qr,Daido2024-is}.
The application of the potential gradient lifts the degeneracy of the two valleys, and one of the valleys selectively contributes to the nonreciprocal paraconductivity, picking up the intrinsic nonreciprocity of finite-momentum Cooper pairs. 
A small potential gradient inevitably exists in most thin film superconductors due to substrate effects. 
Since the other contributions to the nonreciprocal paraconductivity unrelated to the FFLO state, if any, would be small when $v\ll T_{\rm c0}$, the observation of the peaked nonreciprocal paraconductivity in intrinsically centrosymmetric superconductors under a small potential gradient serves as strong evidence for the FFLO state.
In particular, as is clear in Fig.~\ref{fig:nonreciprocity_color}(a), a sizable $\eta^{xxx}$ is widely obtained in the high-field region, and therefore, enhanced $\eta^{xxx}$, not limited to the peaked structure, may offer a feasible experimental probe of the FFLO states.
\section{Summary and Conclusion}
\label{sec:summary}
In this study, we have investigated reciprocal and nonreciprocal paraconductivity in a bilayer system under perpendicular and in-plane Zeeman fields.
We have revealed that the reciprocal paraconductivity has a peak accompanied by the BCS-PDW transition in the system with a perpendicular Zeeman field.
The peak of reciprocal paraconductivity tends to vanish by applying a potential gradient, signaling the lifted degeneracy of the BCS and PDW states.
Similar results of the reciprocal paraconductivity are obtained for the in-plane Zeeman field.
In this case, the degeneracy of the BCS and FFLO states gives rise to the peak structure of the reciprocal paraconductivity, which is also smeared out by applying the strong potential gradient.
We have also studied the nonreciprocal paraconductivity of the system in the in-plane Zeeman field.
In contrast to the reciprocal paraconductivity, a peak appears robustly in both the weakly and strongly noncentrosymmetric cases, i.e., both near the BCS-FFLO-like crossover and the helical crossover regions.

The Zeeman field considered in our study can be realized by an external magnetic field, its proximity to magnetic materials, and so on.
When the external magnetic field is chosen as the source of the perpendicular Zeeman field, since
the orbital magnetic field is not taken into account in this paper, our results
correspond to superconductors with large Maki parameters.
Even when the orbital magnetic field can not be neglected, the qualitative results, such as the peaked reciprocal paraconductivity, are expected to continue to exsit due to the degeneracy of the superconducting phases. 
The results for the in-plane Zeeman field can be directly compared with the transport measurements of two-dimensional superconductors in an in-plane magnetic field.
Even when the in-plane orbital magnetic field can not be neglected due to the finite thickness of the thin films,
the enhanced  nonreciprocal paraconductivity in finite-momentum superconductors is also expected to survive to some extent~\cite{Daido2024-is}, leaving a quantitative study as a future issue.

Based on the results obtained in this paper, we propose to use reciprocal and nonreciprocal paraconductivity to probe multiphase and/or finite-momentum superconductors.
In particular, this proposal is expected to be useful in thin film superconductors because the thermodynamic measurements are challenging there. 
The reciprocal and nonreciprocal paraconductivity has different advantages and disadvantages.
The reciprocal paraconductivity can generally be used to probe degenerate superconducting states, through the detection of the peak structure, while it seems not to be sensitive to the detailed nature of the superconducting states.
The nonreciprocal paraconductivity can only be used by breaking the inversion symmetry. However, once the symmetry allows a finite response, it can provide strong evidence for the finite-momentum superconductivity, including (nearly-)FFLO and helical superconducting states.
A sizable response is obtained by lifting the degeneracy of the finite-momentum Cooper pairs.
It is known that there are a variety of finite-momentum superconducting states, such as the layer-decoupled state or the orbital FFLO state~\cite{Yuan2023-eg,Wan2023-rh,Xie2023-gs,Nakamura2024-qe,Xie2023-gs}, and the pair-density-wave state in strongly correlated superconductors~\cite{Agterberg2020-gs} (accompanying the in-plane modulation of the order parameter,
referring to a different state from the PDW state studied in this paper). It is an interesting future issue to quantitatively study the nonreciprocal charge transport of such finite momentum superconductors.
\begin{acknowledgments}
T.M. gratefully acknowledges fruitful discussion with Sho Ishikawa.
This work was supported by JSPS KAKENHI
No.~JP21K13880, No.~JP22H01181, No.~JP22H04933, No.~JP22H04476, No.~JP23K17353, No.~JP23K22452, No.~JP24H00007, No.~JP24K21530, and No.~JP24H01662. 
\end{acknowledgments}
\bibliography{ref.bib}

\begin{thebibliography}{100}%
\makeatletter
\providecommand \@ifxundefined [1]{%
 \@ifx{#1\undefined}
}%
\providecommand \@ifnum [1]{%
 \ifnum #1\expandafter \@firstoftwo
 \else \expandafter \@secondoftwo
 \fi
}%
\providecommand \@ifx [1]{%
 \ifx #1\expandafter \@firstoftwo
 \else \expandafter \@secondoftwo
 \fi
}%
\providecommand \natexlab [1]{#1}%
\providecommand \enquote  [1]{``#1''}%
\providecommand \bibnamefont  [1]{#1}%
\providecommand \bibfnamefont [1]{#1}%
\providecommand \citenamefont [1]{#1}%
\providecommand \href@noop [0]{\@secondoftwo}%
\providecommand \href [0]{\begingroup \@sanitize@url \@href}%
\providecommand \@href[1]{\@@startlink{#1}\@@href}%
\providecommand \@@href[1]{\endgroup#1\@@endlink}%
\providecommand \@sanitize@url [0]{\catcode `\\12\catcode `\$12\catcode `\&12\catcode `\#12\catcode `\^12\catcode `\_12\catcode `\%12\relax}%
\providecommand \@@startlink[1]{}%
\providecommand \@@endlink[0]{}%
\providecommand \url  [0]{\begingroup\@sanitize@url \@url }%
\providecommand \@url [1]{\endgroup\@href {#1}{\urlprefix }}%
\providecommand \urlprefix  [0]{URL }%
\providecommand \Eprint [0]{\href }%
\providecommand \doibase [0]{https://doi.org/}%
\providecommand \selectlanguage [0]{\@gobble}%
\providecommand \bibinfo  [0]{\@secondoftwo}%
\providecommand \bibfield  [0]{\@secondoftwo}%
\providecommand \translation [1]{[#1]}%
\providecommand \BibitemOpen [0]{}%
\providecommand \bibitemStop [0]{}%
\providecommand \bibitemNoStop [0]{.\EOS\space}%
\providecommand \EOS [0]{\spacefactor3000\relax}%
\providecommand \BibitemShut  [1]{\csname bibitem#1\endcsname}%
\let\auto@bib@innerbib\@empty
\bibitem [{\citenamefont {Sigrist}\ and\ \citenamefont {Ueda}(1991)}]{Sigrist1991-lz}%
  \BibitemOpen
  \bibfield  {author} {\bibinfo {author} {\bibfnamefont {M.}~\bibnamefont {Sigrist}}\ and\ \bibinfo {author} {\bibfnamefont {K.}~\bibnamefont {Ueda}},\ }\bibfield  {title} {\bibinfo {title} {Phenomenological theory of unconventional superconductivity},\ }\href {https://doi.org/10.1103/RevModPhys.63.239} {\bibfield  {journal} {\bibinfo  {journal} {Rev. Mod. Phys.}\ }\textbf {\bibinfo {volume} {63}},\ \bibinfo {pages} {239} (\bibinfo {year} {1991})}\BibitemShut {NoStop}%
\bibitem [{\citenamefont {Qi}\ and\ \citenamefont {Zhang}(2011)}]{Qi2011-vx}%
  \BibitemOpen
  \bibfield  {author} {\bibinfo {author} {\bibfnamefont {X.-L.}\ \bibnamefont {Qi}}\ and\ \bibinfo {author} {\bibfnamefont {S.-C.}\ \bibnamefont {Zhang}},\ }\bibfield  {title} {\bibinfo {title} {Topological insulators and superconductors},\ }\href {https://doi.org/10.1103/RevModPhys.83.1057} {\bibfield  {journal} {\bibinfo  {journal} {Rev. Mod. Phys.}\ }\textbf {\bibinfo {volume} {83}},\ \bibinfo {pages} {1057} (\bibinfo {year} {2011})}\BibitemShut {NoStop}%
\bibitem [{\citenamefont {Tanaka}\ \emph {et~al.}(2012)\citenamefont {Tanaka}, \citenamefont {Sato},\ and\ \citenamefont {Nagaosa}}]{Tanaka2012-vn}%
  \BibitemOpen
  \bibfield  {author} {\bibinfo {author} {\bibfnamefont {Y.}~\bibnamefont {Tanaka}}, \bibinfo {author} {\bibfnamefont {M.}~\bibnamefont {Sato}},\ and\ \bibinfo {author} {\bibfnamefont {N.}~\bibnamefont {Nagaosa}},\ }\bibfield  {title} {\bibinfo {title} {Symmetry and topology in superconductors {--Odd-Frequency} pairing and edge states--},\ }\href {https://doi.org/10.1143/JPSJ.81.011013} {\bibfield  {journal} {\bibinfo  {journal} {J. Phys. Soc. Jpn.}\ }\textbf {\bibinfo {volume} {81}},\ \bibinfo {pages} {011013} (\bibinfo {year} {2012})}\BibitemShut {NoStop}%
\bibitem [{\citenamefont {Sato}\ and\ \citenamefont {Fujimoto}(2016)}]{Sato2016-ab}%
  \BibitemOpen
  \bibfield  {author} {\bibinfo {author} {\bibfnamefont {M.}~\bibnamefont {Sato}}\ and\ \bibinfo {author} {\bibfnamefont {S.}~\bibnamefont {Fujimoto}},\ }\bibfield  {title} {\bibinfo {title} {Majorana fermions and topology in superconductors},\ }\href {https://doi.org/10.7566/JPSJ.85.072001} {\bibfield  {journal} {\bibinfo  {journal} {J. Phys. Soc. Jpn.}\ }\textbf {\bibinfo {volume} {85}},\ \bibinfo {pages} {072001} (\bibinfo {year} {2016})}\BibitemShut {NoStop}%
\bibitem [{\citenamefont {Sato}\ and\ \citenamefont {Ando}(2017)}]{Sato2017-nn}%
  \BibitemOpen
  \bibfield  {author} {\bibinfo {author} {\bibfnamefont {M.}~\bibnamefont {Sato}}\ and\ \bibinfo {author} {\bibfnamefont {Y.}~\bibnamefont {Ando}},\ }\bibfield  {title} {\bibinfo {title} {Topological superconductors: a review},\ }\href {https://doi.org/10.1088/1361-6633/aa6ac7} {\bibfield  {journal} {\bibinfo  {journal} {Rep. Prog. Phys.}\ }\textbf {\bibinfo {volume} {80}},\ \bibinfo {pages} {076501} (\bibinfo {year} {2017})}\BibitemShut {NoStop}%
\bibitem [{\citenamefont {Khim}\ \emph {et~al.}(2021)\citenamefont {Khim}, \citenamefont {Landaeta}, \citenamefont {Banda}, \citenamefont {Bannor}, \citenamefont {Brando}, \citenamefont {Brydon}, \citenamefont {Hafner}, \citenamefont {K{\"u}chler}, \citenamefont {Cardoso-Gil}, \citenamefont {Stockert}, \citenamefont {Mackenzie}, \citenamefont {Agterberg}, \citenamefont {Geibel},\ and\ \citenamefont {Hassinger}}]{Khim2021-ml}%
  \BibitemOpen
  \bibfield  {author} {\bibinfo {author} {\bibfnamefont {S.}~\bibnamefont {Khim}}, \bibinfo {author} {\bibfnamefont {J.~F.}\ \bibnamefont {Landaeta}}, \bibinfo {author} {\bibfnamefont {J.}~\bibnamefont {Banda}}, \bibinfo {author} {\bibfnamefont {N.}~\bibnamefont {Bannor}}, \bibinfo {author} {\bibfnamefont {M.}~\bibnamefont {Brando}}, \bibinfo {author} {\bibfnamefont {P.~M.~R.}\ \bibnamefont {Brydon}}, \bibinfo {author} {\bibfnamefont {D.}~\bibnamefont {Hafner}}, \bibinfo {author} {\bibfnamefont {R.}~\bibnamefont {K{\"u}chler}}, \bibinfo {author} {\bibfnamefont {R.}~\bibnamefont {Cardoso-Gil}}, \bibinfo {author} {\bibfnamefont {U.}~\bibnamefont {Stockert}}, \bibinfo {author} {\bibfnamefont {A.~P.}\ \bibnamefont {Mackenzie}}, \bibinfo {author} {\bibfnamefont {D.~F.}\ \bibnamefont {Agterberg}}, \bibinfo {author} {\bibfnamefont {C.}~\bibnamefont {Geibel}},\ and\ \bibinfo {author} {\bibfnamefont {E.}~\bibnamefont {Hassinger}},\ }\bibfield  {title} {\bibinfo {title} {Field-induced transition within the
  superconducting state of {CeRh$_{2}$As$_{2}$}},\ }\href {https://doi.org/10.1126/science.abe7518} {\bibfield  {journal} {\bibinfo  {journal} {Science}\ }\textbf {\bibinfo {volume} {373}},\ \bibinfo {pages} {1012} (\bibinfo {year} {2021})}\BibitemShut {NoStop}%
\bibitem [{\citenamefont {Onishi}\ \emph {et~al.}(2022)\citenamefont {Onishi}, \citenamefont {Stockert}, \citenamefont {Khim}, \citenamefont {Banda}, \citenamefont {Brando},\ and\ \citenamefont {Hassinger}}]{Onishi2022-ak}%
  \BibitemOpen
  \bibfield  {author} {\bibinfo {author} {\bibfnamefont {S.}~\bibnamefont {Onishi}}, \bibinfo {author} {\bibfnamefont {U.}~\bibnamefont {Stockert}}, \bibinfo {author} {\bibfnamefont {S.}~\bibnamefont {Khim}}, \bibinfo {author} {\bibfnamefont {J.}~\bibnamefont {Banda}}, \bibinfo {author} {\bibfnamefont {M.}~\bibnamefont {Brando}},\ and\ \bibinfo {author} {\bibfnamefont {E.}~\bibnamefont {Hassinger}},\ }\bibfield  {title} {\bibinfo {title} {{Low-Temperature} thermal conductivity of the {Two-Phase} superconductor {CeRh$_{2}$As$_{2}$}},\ }\bibfield  {journal} {\bibinfo  {journal} {Frontiers in Electronic Materials}\ }\textbf {\bibinfo {volume} {2}},\ \href {https://doi.org/10.3389/femat.2022.880579} {10.3389/femat.2022.880579} (\bibinfo {year} {2022})\BibitemShut {NoStop}%
\bibitem [{\citenamefont {Landaeta}\ \emph {et~al.}(2022)\citenamefont {Landaeta}, \citenamefont {Khanenko}, \citenamefont {Cavanagh}, \citenamefont {Geibel}, \citenamefont {Khim}, \citenamefont {Mishra}, \citenamefont {Sheikin}, \citenamefont {Brydon}, \citenamefont {Agterberg}, \citenamefont {Brando},\ and\ \citenamefont {Hassinger}}]{Landaeta2022-gz}%
  \BibitemOpen
  \bibfield  {author} {\bibinfo {author} {\bibfnamefont {J.~F.}\ \bibnamefont {Landaeta}}, \bibinfo {author} {\bibfnamefont {P.}~\bibnamefont {Khanenko}}, \bibinfo {author} {\bibfnamefont {D.~C.}\ \bibnamefont {Cavanagh}}, \bibinfo {author} {\bibfnamefont {C.}~\bibnamefont {Geibel}}, \bibinfo {author} {\bibfnamefont {S.}~\bibnamefont {Khim}}, \bibinfo {author} {\bibfnamefont {S.}~\bibnamefont {Mishra}}, \bibinfo {author} {\bibfnamefont {I.}~\bibnamefont {Sheikin}}, \bibinfo {author} {\bibfnamefont {P.~M.~R.}\ \bibnamefont {Brydon}}, \bibinfo {author} {\bibfnamefont {D.~F.}\ \bibnamefont {Agterberg}}, \bibinfo {author} {\bibfnamefont {M.}~\bibnamefont {Brando}},\ and\ \bibinfo {author} {\bibfnamefont {E.}~\bibnamefont {Hassinger}},\ }\bibfield  {title} {\bibinfo {title} {{Field-Angle} dependence reveals {Odd-Parity} superconductivity in {CeRh$_{2}$As$_{2}$}},\ }\href {https://doi.org/10.1103/PhysRevX.12.031001} {\bibfield  {journal} {\bibinfo  {journal} {Phys. Rev. X}\ }\textbf {\bibinfo {volume} {12}},\
  \bibinfo {pages} {031001} (\bibinfo {year} {2022})}\BibitemShut {NoStop}%
\bibitem [{\citenamefont {Ogata}\ \emph {et~al.}(2023)\citenamefont {Ogata}, \citenamefont {Kitagawa}, \citenamefont {Kinjo}, \citenamefont {Ishida}, \citenamefont {Brando}, \citenamefont {Hassinger}, \citenamefont {Geibel},\ and\ \citenamefont {Khim}}]{Ogata2023-ry}%
  \BibitemOpen
  \bibfield  {author} {\bibinfo {author} {\bibfnamefont {S.}~\bibnamefont {Ogata}}, \bibinfo {author} {\bibfnamefont {S.}~\bibnamefont {Kitagawa}}, \bibinfo {author} {\bibfnamefont {K.}~\bibnamefont {Kinjo}}, \bibinfo {author} {\bibfnamefont {K.}~\bibnamefont {Ishida}}, \bibinfo {author} {\bibfnamefont {M.}~\bibnamefont {Brando}}, \bibinfo {author} {\bibfnamefont {E.}~\bibnamefont {Hassinger}}, \bibinfo {author} {\bibfnamefont {C.}~\bibnamefont {Geibel}},\ and\ \bibinfo {author} {\bibfnamefont {S.}~\bibnamefont {Khim}},\ }\bibfield  {title} {\bibinfo {title} {Parity transition of {Spin-Singlet} superconductivity using sublattice degrees of freedom},\ }\href {https://doi.org/10.1103/PhysRevLett.130.166001} {\bibfield  {journal} {\bibinfo  {journal} {Phys. Rev. Lett.}\ }\textbf {\bibinfo {volume} {130}},\ \bibinfo {pages} {166001} (\bibinfo {year} {2023})}\BibitemShut {NoStop}%
\bibitem [{\citenamefont {Siddiquee}\ \emph {et~al.}(2023)\citenamefont {Siddiquee}, \citenamefont {Rehfuss}, \citenamefont {Broyles},\ and\ \citenamefont {Ran}}]{Siddiquee2023-wx}%
  \BibitemOpen
  \bibfield  {author} {\bibinfo {author} {\bibfnamefont {H.}~\bibnamefont {Siddiquee}}, \bibinfo {author} {\bibfnamefont {Z.}~\bibnamefont {Rehfuss}}, \bibinfo {author} {\bibfnamefont {C.}~\bibnamefont {Broyles}},\ and\ \bibinfo {author} {\bibfnamefont {S.}~\bibnamefont {Ran}},\ }\bibfield  {title} {\bibinfo {title} {Pressure dependence of superconductivity in {CeRh$_{2}$As$_{2}$}},\ }\href {https://doi.org/10.1103/PhysRevB.108.L020504} {\bibfield  {journal} {\bibinfo  {journal} {Phys. Rev. B Condens. Matter}\ }\textbf {\bibinfo {volume} {108}},\ \bibinfo {pages} {L020504} (\bibinfo {year} {2023})}\BibitemShut {NoStop}%
\bibitem [{\citenamefont {M{\"o}ckli}\ and\ \citenamefont {Ramires}(2021)}]{Mockli2021-mw}%
  \BibitemOpen
  \bibfield  {author} {\bibinfo {author} {\bibfnamefont {D.}~\bibnamefont {M{\"o}ckli}}\ and\ \bibinfo {author} {\bibfnamefont {A.}~\bibnamefont {Ramires}},\ }\bibfield  {title} {\bibinfo {title} {Two scenarios for superconductivity in {CeRh$_{2}$As$_{2}$}},\ }\href {https://doi.org/10.1103/PhysRevResearch.3.023204} {\bibfield  {journal} {\bibinfo  {journal} {Phys. Rev. Res.}\ }\textbf {\bibinfo {volume} {3}},\ \bibinfo {pages} {023204} (\bibinfo {year} {2021})}\BibitemShut {NoStop}%
\bibitem [{\citenamefont {Schertenleib}\ \emph {et~al.}(2021)\citenamefont {Schertenleib}, \citenamefont {Fischer},\ and\ \citenamefont {Sigrist}}]{Schertenleib2021-sb}%
  \BibitemOpen
  \bibfield  {author} {\bibinfo {author} {\bibfnamefont {E.~G.}\ \bibnamefont {Schertenleib}}, \bibinfo {author} {\bibfnamefont {M.~H.}\ \bibnamefont {Fischer}},\ and\ \bibinfo {author} {\bibfnamefont {M.}~\bibnamefont {Sigrist}},\ }\bibfield  {title} {\bibinfo {title} {Unusual - phase diagram of {CeRh$_{2}$As$_{2}$}: The role of staggered noncentrosymmetricity},\ }\href {https://doi.org/10.1103/PhysRevResearch.3.023179} {\bibfield  {journal} {\bibinfo  {journal} {Phys. Rev. Res.}\ }\textbf {\bibinfo {volume} {3}},\ \bibinfo {pages} {023179} (\bibinfo {year} {2021})}\BibitemShut {NoStop}%
\bibitem [{\citenamefont {Skurativska}\ \emph {et~al.}(2021)\citenamefont {Skurativska}, \citenamefont {Sigrist},\ and\ \citenamefont {Fischer}}]{Skurativska2021-lg}%
  \BibitemOpen
  \bibfield  {author} {\bibinfo {author} {\bibfnamefont {A.}~\bibnamefont {Skurativska}}, \bibinfo {author} {\bibfnamefont {M.}~\bibnamefont {Sigrist}},\ and\ \bibinfo {author} {\bibfnamefont {M.~H.}\ \bibnamefont {Fischer}},\ }\bibfield  {title} {\bibinfo {title} {Spin response and topology of a staggered-rashba superconductor},\ }\href {https://doi.org/10.1103/PhysRevResearch.3.033133} {\bibfield  {journal} {\bibinfo  {journal} {Phys. Rev. Res.}\ }\textbf {\bibinfo {volume} {3}},\ \bibinfo {pages} {033133} (\bibinfo {year} {2021})}\BibitemShut {NoStop}%
\bibitem [{\citenamefont {Hazra}\ and\ \citenamefont {Coleman}(2023)}]{Hazra2023-jy}%
  \BibitemOpen
  \bibfield  {author} {\bibinfo {author} {\bibfnamefont {T.}~\bibnamefont {Hazra}}\ and\ \bibinfo {author} {\bibfnamefont {P.}~\bibnamefont {Coleman}},\ }\bibfield  {title} {\bibinfo {title} {Triplet pairing mechanisms from {Hund's-Kondo} models: Applications to {UTe$_{2}$} and {CeRh$_{2}$As$_{2}$}},\ }\href {https://doi.org/10.1103/PhysRevLett.130.136002} {\bibfield  {journal} {\bibinfo  {journal} {Phys. Rev. Lett.}\ }\textbf {\bibinfo {volume} {130}},\ \bibinfo {pages} {136002} (\bibinfo {year} {2023})}\BibitemShut {NoStop}%
\bibitem [{\citenamefont {Kimura}\ \emph {et~al.}(2021)\citenamefont {Kimura}, \citenamefont {Sichelschmidt},\ and\ \citenamefont {Khim}}]{Kimura2021-wx}%
  \BibitemOpen
  \bibfield  {author} {\bibinfo {author} {\bibfnamefont {S.-I.}\ \bibnamefont {Kimura}}, \bibinfo {author} {\bibfnamefont {J.}~\bibnamefont {Sichelschmidt}},\ and\ \bibinfo {author} {\bibfnamefont {S.}~\bibnamefont {Khim}},\ }\bibfield  {title} {\bibinfo {title} {Optical study of the electronic structure of locally noncentrosymmetric {CeRh$_{2}$As$_{2}$}},\ }\href {https://doi.org/10.1103/PhysRevB.104.245116} {\bibfield  {journal} {\bibinfo  {journal} {Phys. Rev. B Condens. Matter}\ }\textbf {\bibinfo {volume} {104}},\ \bibinfo {pages} {245116} (\bibinfo {year} {2021})}\BibitemShut {NoStop}%
\bibitem [{\citenamefont {Ishizuka}\ \emph {et~al.}(2024)\citenamefont {Ishizuka}, \citenamefont {Nogaki}, \citenamefont {Sigrist},\ and\ \citenamefont {Yanase}}]{Ishizuka2024-PRB}%
  \BibitemOpen
  \bibfield  {author} {\bibinfo {author} {\bibfnamefont {J.}~\bibnamefont {Ishizuka}}, \bibinfo {author} {\bibfnamefont {K.}~\bibnamefont {Nogaki}}, \bibinfo {author} {\bibfnamefont {M.}~\bibnamefont {Sigrist}},\ and\ \bibinfo {author} {\bibfnamefont {Y.}~\bibnamefont {Yanase}},\ }\bibfield  {title} {\bibinfo {title} {Correlation-induced fermi surface evolution and topological crystalline superconductivity in ${\mathrm{cerh}}_{2}{\mathrm{as}}_{2}$},\ }\href {https://doi.org/10.1103/PhysRevB.110.L140505} {\bibfield  {journal} {\bibinfo  {journal} {Phys. Rev. B}\ }\textbf {\bibinfo {volume} {110}},\ \bibinfo {pages} {L140505} (\bibinfo {year} {2024})}\BibitemShut {NoStop}%
\bibitem [{\citenamefont {Nogaki}\ \emph {et~al.}(2021)\citenamefont {Nogaki}, \citenamefont {Daido}, \citenamefont {Ishizuka},\ and\ \citenamefont {Yanase}}]{Nogaki2021-cv}%
  \BibitemOpen
  \bibfield  {author} {\bibinfo {author} {\bibfnamefont {K.}~\bibnamefont {Nogaki}}, \bibinfo {author} {\bibfnamefont {A.}~\bibnamefont {Daido}}, \bibinfo {author} {\bibfnamefont {J.}~\bibnamefont {Ishizuka}},\ and\ \bibinfo {author} {\bibfnamefont {Y.}~\bibnamefont {Yanase}},\ }\bibfield  {title} {\bibinfo {title} {Topological crystalline superconductivity in locally noncentrosymmetric {CeRh$_{2}$As$_{2}$}},\ }\href {https://doi.org/10.1103/PhysRevResearch.3.L032071} {\bibfield  {journal} {\bibinfo  {journal} {Phys. Rev. Res.}\ }\textbf {\bibinfo {volume} {3}},\ \bibinfo {pages} {L032071} (\bibinfo {year} {2021})}\BibitemShut {NoStop}%
\bibitem [{\citenamefont {Nogaki}\ and\ \citenamefont {Yanase}(2022)}]{Nogaki2022-pd}%
  \BibitemOpen
  \bibfield  {author} {\bibinfo {author} {\bibfnamefont {K.}~\bibnamefont {Nogaki}}\ and\ \bibinfo {author} {\bibfnamefont {Y.}~\bibnamefont {Yanase}},\ }\bibfield  {title} {\bibinfo {title} {Even-odd parity transition in strongly correlated locally noncentrosymmetric superconductors: Application to {CeRh$_{2}$As$_{2}$}},\ }\href {https://doi.org/10.1103/PhysRevB.106.L100504} {\bibfield  {journal} {\bibinfo  {journal} {Phys. Rev. B Condens. Matter}\ }\textbf {\bibinfo {volume} {106}},\ \bibinfo {pages} {L100504} (\bibinfo {year} {2022})}\BibitemShut {NoStop}%
\bibitem [{\citenamefont {Lee}\ and\ \citenamefont {Chung}(2023)}]{Lee2023-wp}%
  \BibitemOpen
  \bibfield  {author} {\bibinfo {author} {\bibfnamefont {C.}~\bibnamefont {Lee}}\ and\ \bibinfo {author} {\bibfnamefont {S.~B.}\ \bibnamefont {Chung}},\ }\bibfield  {title} {\bibinfo {title} {Linear optical response from the odd-parity {Bardasis-Schrieffer} mode in locally non-centrosymmetric superconductors},\ }\href {https://doi.org/10.1038/s42005-023-01421-8} {\bibfield  {journal} {\bibinfo  {journal} {Communications Physics}\ }\textbf {\bibinfo {volume} {6}},\ \bibinfo {pages} {1} (\bibinfo {year} {2023})}\BibitemShut {NoStop}%
\bibitem [{\citenamefont {Cavanagh}\ \emph {et~al.}(2022)\citenamefont {Cavanagh}, \citenamefont {Shishidou}, \citenamefont {Weinert}, \citenamefont {Brydon},\ and\ \citenamefont {Agterberg}}]{Cavanagh2022-xe}%
  \BibitemOpen
  \bibfield  {author} {\bibinfo {author} {\bibfnamefont {D.~C.}\ \bibnamefont {Cavanagh}}, \bibinfo {author} {\bibfnamefont {T.}~\bibnamefont {Shishidou}}, \bibinfo {author} {\bibfnamefont {M.}~\bibnamefont {Weinert}}, \bibinfo {author} {\bibfnamefont {P.~M.~R.}\ \bibnamefont {Brydon}},\ and\ \bibinfo {author} {\bibfnamefont {D.~F.}\ \bibnamefont {Agterberg}},\ }\bibfield  {title} {\bibinfo {title} {Nonsymmorphic symmetry and field-driven odd-parity pairing in {CeRh$_{2}$As$_{2}$}},\ }\href {https://doi.org/10.1103/PhysRevB.105.L020505} {\bibfield  {journal} {\bibinfo  {journal} {Phys. Rev. B Condens. Matter}\ }\textbf {\bibinfo {volume} {105}},\ \bibinfo {pages} {L020505} (\bibinfo {year} {2022})}\BibitemShut {NoStop}%
\bibitem [{\citenamefont {Nogaki}\ and\ \citenamefont {Yanase}(2024)}]{Nogaki2024-PRB}%
  \BibitemOpen
  \bibfield  {author} {\bibinfo {author} {\bibfnamefont {K.}~\bibnamefont {Nogaki}}\ and\ \bibinfo {author} {\bibfnamefont {Y.}~\bibnamefont {Yanase}},\ }\bibfield  {title} {\bibinfo {title} {Field-induced superconductivity mediated by odd-parity multipole fluctuation},\ }\href {https://doi.org/10.1103/PhysRevB.110.184501} {\bibfield  {journal} {\bibinfo  {journal} {Phys. Rev. B}\ }\textbf {\bibinfo {volume} {110}},\ \bibinfo {pages} {184501} (\bibinfo {year} {2024})}\BibitemShut {NoStop}%
\bibitem [{\citenamefont {Ran}\ \emph {et~al.}(2019{\natexlab{a}})\citenamefont {Ran}, \citenamefont {Eckberg}, \citenamefont {Ding}, \citenamefont {Furukawa}, \citenamefont {Metz}, \citenamefont {Saha}, \citenamefont {Liu}, \citenamefont {Zic}, \citenamefont {Kim}, \citenamefont {Paglione},\ and\ \citenamefont {Butch}}]{Ran2019-yp}%
  \BibitemOpen
  \bibfield  {author} {\bibinfo {author} {\bibfnamefont {S.}~\bibnamefont {Ran}}, \bibinfo {author} {\bibfnamefont {C.}~\bibnamefont {Eckberg}}, \bibinfo {author} {\bibfnamefont {Q.-P.}\ \bibnamefont {Ding}}, \bibinfo {author} {\bibfnamefont {Y.}~\bibnamefont {Furukawa}}, \bibinfo {author} {\bibfnamefont {T.}~\bibnamefont {Metz}}, \bibinfo {author} {\bibfnamefont {S.~R.}\ \bibnamefont {Saha}}, \bibinfo {author} {\bibfnamefont {I.-L.}\ \bibnamefont {Liu}}, \bibinfo {author} {\bibfnamefont {M.}~\bibnamefont {Zic}}, \bibinfo {author} {\bibfnamefont {H.}~\bibnamefont {Kim}}, \bibinfo {author} {\bibfnamefont {J.}~\bibnamefont {Paglione}},\ and\ \bibinfo {author} {\bibfnamefont {N.~P.}\ \bibnamefont {Butch}},\ }\bibfield  {title} {\bibinfo {title} {Nearly ferromagnetic spin-triplet superconductivity},\ }\href {https://doi.org/10.1126/science.aav8645} {\bibfield  {journal} {\bibinfo  {journal} {Science}\ }\textbf {\bibinfo {volume} {365}},\ \bibinfo {pages} {684} (\bibinfo {year} {2019}{\natexlab{a}})}\BibitemShut
  {NoStop}%
\bibitem [{\citenamefont {Aoki}\ \emph {et~al.}(2022{\natexlab{a}})\citenamefont {Aoki}, \citenamefont {Brison}, \citenamefont {Flouquet}, \citenamefont {Ishida}, \citenamefont {Knebel}, \citenamefont {Tokunaga},\ and\ \citenamefont {Yanase}}]{Aoki2022-zg}%
  \BibitemOpen
  \bibfield  {author} {\bibinfo {author} {\bibfnamefont {D.}~\bibnamefont {Aoki}}, \bibinfo {author} {\bibfnamefont {J.-P.}\ \bibnamefont {Brison}}, \bibinfo {author} {\bibfnamefont {J.}~\bibnamefont {Flouquet}}, \bibinfo {author} {\bibfnamefont {K.}~\bibnamefont {Ishida}}, \bibinfo {author} {\bibfnamefont {G.}~\bibnamefont {Knebel}}, \bibinfo {author} {\bibfnamefont {Y.}~\bibnamefont {Tokunaga}},\ and\ \bibinfo {author} {\bibfnamefont {Y.}~\bibnamefont {Yanase}},\ }\bibfield  {title} {\bibinfo {title} {Unconventional superconductivity in {UTe$_{2}$}},\ }\href {https://doi.org/10.1088/1361-648X/ac5863} {\bibfield  {journal} {\bibinfo  {journal} {Journal of Physics: Condensed Matter}\ }\textbf {\bibinfo {volume} {34}},\ \bibinfo {pages} {243002} (\bibinfo {year} {2022}{\natexlab{a}})}\BibitemShut {NoStop}%
\bibitem [{\citenamefont {Ran}\ \emph {et~al.}(2019{\natexlab{b}})\citenamefont {Ran}, \citenamefont {Liu}, \citenamefont {Eo}, \citenamefont {Campbell}, \citenamefont {Neves}, \citenamefont {Fuhrman}, \citenamefont {Saha}, \citenamefont {Eckberg}, \citenamefont {Kim}, \citenamefont {Graf}, \citenamefont {Balakirev}, \citenamefont {Singleton}, \citenamefont {Paglione},\ and\ \citenamefont {Butch}}]{Ran2019-mh}%
  \BibitemOpen
  \bibfield  {author} {\bibinfo {author} {\bibfnamefont {S.}~\bibnamefont {Ran}}, \bibinfo {author} {\bibfnamefont {I.-L.}\ \bibnamefont {Liu}}, \bibinfo {author} {\bibfnamefont {Y.~S.}\ \bibnamefont {Eo}}, \bibinfo {author} {\bibfnamefont {D.~J.}\ \bibnamefont {Campbell}}, \bibinfo {author} {\bibfnamefont {P.~M.}\ \bibnamefont {Neves}}, \bibinfo {author} {\bibfnamefont {W.~T.}\ \bibnamefont {Fuhrman}}, \bibinfo {author} {\bibfnamefont {S.~R.}\ \bibnamefont {Saha}}, \bibinfo {author} {\bibfnamefont {C.}~\bibnamefont {Eckberg}}, \bibinfo {author} {\bibfnamefont {H.}~\bibnamefont {Kim}}, \bibinfo {author} {\bibfnamefont {D.}~\bibnamefont {Graf}}, \bibinfo {author} {\bibfnamefont {F.}~\bibnamefont {Balakirev}}, \bibinfo {author} {\bibfnamefont {J.}~\bibnamefont {Singleton}}, \bibinfo {author} {\bibfnamefont {J.}~\bibnamefont {Paglione}},\ and\ \bibinfo {author} {\bibfnamefont {N.~P.}\ \bibnamefont {Butch}},\ }\bibfield  {title} {\bibinfo {title} {Extreme magnetic field-boosted superconductivity},\ }\href
  {https://doi.org/10.1038/s41567-019-0670-x} {\bibfield  {journal} {\bibinfo  {journal} {Nat. Phys.}\ }\textbf {\bibinfo {volume} {15}},\ \bibinfo {pages} {1250} (\bibinfo {year} {2019}{\natexlab{b}})}\BibitemShut {NoStop}%
\bibitem [{\citenamefont {Ran}\ \emph {et~al.}(2020)\citenamefont {Ran}, \citenamefont {Kim}, \citenamefont {Liu}, \citenamefont {Saha}, \citenamefont {Hayes}, \citenamefont {Metz}, \citenamefont {Eo}, \citenamefont {Paglione},\ and\ \citenamefont {Butch}}]{Ran2020-hl}%
  \BibitemOpen
  \bibfield  {author} {\bibinfo {author} {\bibfnamefont {S.}~\bibnamefont {Ran}}, \bibinfo {author} {\bibfnamefont {H.}~\bibnamefont {Kim}}, \bibinfo {author} {\bibfnamefont {I.-L.}\ \bibnamefont {Liu}}, \bibinfo {author} {\bibfnamefont {S.~R.}\ \bibnamefont {Saha}}, \bibinfo {author} {\bibfnamefont {I.}~\bibnamefont {Hayes}}, \bibinfo {author} {\bibfnamefont {T.}~\bibnamefont {Metz}}, \bibinfo {author} {\bibfnamefont {Y.~S.}\ \bibnamefont {Eo}}, \bibinfo {author} {\bibfnamefont {J.}~\bibnamefont {Paglione}},\ and\ \bibinfo {author} {\bibfnamefont {N.~P.}\ \bibnamefont {Butch}},\ }\bibfield  {title} {\bibinfo {title} {Enhancement and reentrance of spin triplet superconductivity in ${\mathrm{ute}}_{2}$ under pressure},\ }\href {https://doi.org/10.1103/PhysRevB.101.140503} {\bibfield  {journal} {\bibinfo  {journal} {Phys. Rev. B}\ }\textbf {\bibinfo {volume} {101}},\ \bibinfo {pages} {140503(R)} (\bibinfo {year} {2020})}\BibitemShut {NoStop}%
\bibitem [{\citenamefont {Miao}\ \emph {et~al.}(2020)\citenamefont {Miao}, \citenamefont {Liu}, \citenamefont {Xu}, \citenamefont {Kotta}, \citenamefont {Kang}, \citenamefont {Ran}, \citenamefont {Paglione}, \citenamefont {Kotliar}, \citenamefont {Butch}, \citenamefont {Denlinger},\ and\ \citenamefont {Wray}}]{Miao2020-iw}%
  \BibitemOpen
  \bibfield  {author} {\bibinfo {author} {\bibfnamefont {L.}~\bibnamefont {Miao}}, \bibinfo {author} {\bibfnamefont {S.}~\bibnamefont {Liu}}, \bibinfo {author} {\bibfnamefont {Y.}~\bibnamefont {Xu}}, \bibinfo {author} {\bibfnamefont {E.~C.}\ \bibnamefont {Kotta}}, \bibinfo {author} {\bibfnamefont {C.-J.}\ \bibnamefont {Kang}}, \bibinfo {author} {\bibfnamefont {S.}~\bibnamefont {Ran}}, \bibinfo {author} {\bibfnamefont {J.}~\bibnamefont {Paglione}}, \bibinfo {author} {\bibfnamefont {G.}~\bibnamefont {Kotliar}}, \bibinfo {author} {\bibfnamefont {N.~P.}\ \bibnamefont {Butch}}, \bibinfo {author} {\bibfnamefont {J.~D.}\ \bibnamefont {Denlinger}},\ and\ \bibinfo {author} {\bibfnamefont {L.~A.}\ \bibnamefont {Wray}},\ }\bibfield  {title} {\bibinfo {title} {Low energy band structure and symmetries of {UTe$_{2}$} from {Angle-Resolved} photoemission spectroscopy},\ }\href {https://doi.org/10.1103/PhysRevLett.124.076401} {\bibfield  {journal} {\bibinfo  {journal} {Phys. Rev. Lett.}\ }\textbf {\bibinfo {volume} {124}},\
  \bibinfo {pages} {076401} (\bibinfo {year} {2020})}\BibitemShut {NoStop}%
\bibitem [{\citenamefont {Lin}\ \emph {et~al.}(2020)\citenamefont {Lin}, \citenamefont {Campbell}, \citenamefont {Ran}, \citenamefont {Liu}, \citenamefont {Kim}, \citenamefont {Nevidomskyy}, \citenamefont {Graf}, \citenamefont {Butch},\ and\ \citenamefont {Paglione}}]{Lin2020-so}%
  \BibitemOpen
  \bibfield  {author} {\bibinfo {author} {\bibfnamefont {W.-C.}\ \bibnamefont {Lin}}, \bibinfo {author} {\bibfnamefont {D.~J.}\ \bibnamefont {Campbell}}, \bibinfo {author} {\bibfnamefont {S.}~\bibnamefont {Ran}}, \bibinfo {author} {\bibfnamefont {I.-L.}\ \bibnamefont {Liu}}, \bibinfo {author} {\bibfnamefont {H.}~\bibnamefont {Kim}}, \bibinfo {author} {\bibfnamefont {A.~H.}\ \bibnamefont {Nevidomskyy}}, \bibinfo {author} {\bibfnamefont {D.}~\bibnamefont {Graf}}, \bibinfo {author} {\bibfnamefont {N.~P.}\ \bibnamefont {Butch}},\ and\ \bibinfo {author} {\bibfnamefont {J.}~\bibnamefont {Paglione}},\ }\bibfield  {title} {\bibinfo {title} {Tuning magnetic confinement of spin-triplet superconductivity},\ }\href {https://doi.org/10.1038/s41535-020-00270-w} {\bibfield  {journal} {\bibinfo  {journal} {npj Quantum Materials}\ }\textbf {\bibinfo {volume} {5}},\ \bibinfo {pages} {1} (\bibinfo {year} {2020})}\BibitemShut {NoStop}%
\bibitem [{\citenamefont {Aoki}\ \emph {et~al.}(2019)\citenamefont {Aoki}, \citenamefont {Nakamura}, \citenamefont {Honda}, \citenamefont {Li}, \citenamefont {Homma}, \citenamefont {Shimizu}, \citenamefont {Sato}, \citenamefont {Knebel}, \citenamefont {Brison}, \citenamefont {Pourret}, \citenamefont {Braithwaite}, \citenamefont {Lapertot}, \citenamefont {Niu}, \citenamefont {Vali{\v s}ka}, \citenamefont {Harima},\ and\ \citenamefont {Flouquet}}]{Aoki2019-vq}%
  \BibitemOpen
  \bibfield  {author} {\bibinfo {author} {\bibfnamefont {D.}~\bibnamefont {Aoki}}, \bibinfo {author} {\bibfnamefont {A.}~\bibnamefont {Nakamura}}, \bibinfo {author} {\bibfnamefont {F.}~\bibnamefont {Honda}}, \bibinfo {author} {\bibfnamefont {D.}~\bibnamefont {Li}}, \bibinfo {author} {\bibfnamefont {Y.}~\bibnamefont {Homma}}, \bibinfo {author} {\bibfnamefont {Y.}~\bibnamefont {Shimizu}}, \bibinfo {author} {\bibfnamefont {Y.~J.}\ \bibnamefont {Sato}}, \bibinfo {author} {\bibfnamefont {G.}~\bibnamefont {Knebel}}, \bibinfo {author} {\bibfnamefont {J.-P.}\ \bibnamefont {Brison}}, \bibinfo {author} {\bibfnamefont {A.}~\bibnamefont {Pourret}}, \bibinfo {author} {\bibfnamefont {D.}~\bibnamefont {Braithwaite}}, \bibinfo {author} {\bibfnamefont {G.}~\bibnamefont {Lapertot}}, \bibinfo {author} {\bibfnamefont {Q.}~\bibnamefont {Niu}}, \bibinfo {author} {\bibfnamefont {M.}~\bibnamefont {Vali{\v s}ka}}, \bibinfo {author} {\bibfnamefont {H.}~\bibnamefont {Harima}},\ and\ \bibinfo {author} {\bibfnamefont {J.}~\bibnamefont
  {Flouquet}},\ }\bibfield  {title} {\bibinfo {title} {Unconventional superconductivity in heavy fermion {UTe$_{2}$}},\ }\href {https://doi.org/10.7566/JPSJ.88.043702} {\bibfield  {journal} {\bibinfo  {journal} {J. Phys. Soc. Jpn.}\ }\textbf {\bibinfo {volume} {88}},\ \bibinfo {pages} {043702} (\bibinfo {year} {2019})}\BibitemShut {NoStop}%
\bibitem [{\citenamefont {Aoki}\ \emph {et~al.}(2020)\citenamefont {Aoki}, \citenamefont {Honda}, \citenamefont {Knebel}, \citenamefont {Braithwaite}, \citenamefont {Nakamura}, \citenamefont {Li}, \citenamefont {Homma}, \citenamefont {Shimizu}, \citenamefont {Sato}, \citenamefont {Brison},\ and\ \citenamefont {Flouquet}}]{Aoki2020-xg}%
  \BibitemOpen
  \bibfield  {author} {\bibinfo {author} {\bibfnamefont {D.}~\bibnamefont {Aoki}}, \bibinfo {author} {\bibfnamefont {F.}~\bibnamefont {Honda}}, \bibinfo {author} {\bibfnamefont {G.}~\bibnamefont {Knebel}}, \bibinfo {author} {\bibfnamefont {D.}~\bibnamefont {Braithwaite}}, \bibinfo {author} {\bibfnamefont {A.}~\bibnamefont {Nakamura}}, \bibinfo {author} {\bibfnamefont {D.}~\bibnamefont {Li}}, \bibinfo {author} {\bibfnamefont {Y.}~\bibnamefont {Homma}}, \bibinfo {author} {\bibfnamefont {Y.}~\bibnamefont {Shimizu}}, \bibinfo {author} {\bibfnamefont {Y.~J.}\ \bibnamefont {Sato}}, \bibinfo {author} {\bibfnamefont {J.-P.}\ \bibnamefont {Brison}},\ and\ \bibinfo {author} {\bibfnamefont {J.}~\bibnamefont {Flouquet}},\ }\bibfield  {title} {\bibinfo {title} {Multiple superconducting phases and unusual enhancement of the upper critical field in {UTe$_{2}$}},\ }\href {https://doi.org/10.7566/JPSJ.89.053705} {\bibfield  {journal} {\bibinfo  {journal} {J. Phys. Soc. Jpn.}\ }\textbf {\bibinfo {volume} {89}},\ \bibinfo
  {pages} {053705} (\bibinfo {year} {2020})}\BibitemShut {NoStop}%
\bibitem [{\citenamefont {Aoki}\ \emph {et~al.}(2021)\citenamefont {Aoki}, \citenamefont {Kimata}, \citenamefont {Sato}, \citenamefont {Knebel}, \citenamefont {Honda}, \citenamefont {Nakamura}, \citenamefont {Li}, \citenamefont {Homma}, \citenamefont {Shimizu}, \citenamefont {Knafo}, \citenamefont {Braithwaite}, \citenamefont {Vali{\v s}ka}, \citenamefont {Pourret}, \citenamefont {Brison},\ and\ \citenamefont {Flouquet}}]{Aoki2021-dz}%
  \BibitemOpen
  \bibfield  {author} {\bibinfo {author} {\bibfnamefont {D.}~\bibnamefont {Aoki}}, \bibinfo {author} {\bibfnamefont {M.}~\bibnamefont {Kimata}}, \bibinfo {author} {\bibfnamefont {Y.~J.}\ \bibnamefont {Sato}}, \bibinfo {author} {\bibfnamefont {G.}~\bibnamefont {Knebel}}, \bibinfo {author} {\bibfnamefont {F.}~\bibnamefont {Honda}}, \bibinfo {author} {\bibfnamefont {A.}~\bibnamefont {Nakamura}}, \bibinfo {author} {\bibfnamefont {D.}~\bibnamefont {Li}}, \bibinfo {author} {\bibfnamefont {Y.}~\bibnamefont {Homma}}, \bibinfo {author} {\bibfnamefont {Y.}~\bibnamefont {Shimizu}}, \bibinfo {author} {\bibfnamefont {W.}~\bibnamefont {Knafo}}, \bibinfo {author} {\bibfnamefont {D.}~\bibnamefont {Braithwaite}}, \bibinfo {author} {\bibfnamefont {M.}~\bibnamefont {Vali{\v s}ka}}, \bibinfo {author} {\bibfnamefont {A.}~\bibnamefont {Pourret}}, \bibinfo {author} {\bibfnamefont {J.-P.}\ \bibnamefont {Brison}},\ and\ \bibinfo {author} {\bibfnamefont {J.}~\bibnamefont {Flouquet}},\ }\bibfield  {title} {\bibinfo {title}
  {{Field-Induced} superconductivity near the superconducting critical pressure in {UTe$_{2}$}},\ }\href {https://doi.org/10.7566/JPSJ.90.074705} {\bibfield  {journal} {\bibinfo  {journal} {J. Phys. Soc. Jpn.}\ }\textbf {\bibinfo {volume} {90}},\ \bibinfo {pages} {074705} (\bibinfo {year} {2021})}\BibitemShut {NoStop}%
\bibitem [{\citenamefont {Aoki}\ \emph {et~al.}(2022{\natexlab{b}})\citenamefont {Aoki}, \citenamefont {Sakai}, \citenamefont {Opletal}, \citenamefont {Tokiwa}, \citenamefont {Ishizuka}, \citenamefont {Yanase}, \citenamefont {Harima}, \citenamefont {Nakamura}, \citenamefont {Li}, \citenamefont {Homma}, \citenamefont {Shimizu}, \citenamefont {Knebel}, \citenamefont {Flouquet},\ and\ \citenamefont {Haga}}]{Aoki2022-dj}%
  \BibitemOpen
  \bibfield  {author} {\bibinfo {author} {\bibfnamefont {D.}~\bibnamefont {Aoki}}, \bibinfo {author} {\bibfnamefont {H.}~\bibnamefont {Sakai}}, \bibinfo {author} {\bibfnamefont {P.}~\bibnamefont {Opletal}}, \bibinfo {author} {\bibfnamefont {Y.}~\bibnamefont {Tokiwa}}, \bibinfo {author} {\bibfnamefont {J.}~\bibnamefont {Ishizuka}}, \bibinfo {author} {\bibfnamefont {Y.}~\bibnamefont {Yanase}}, \bibinfo {author} {\bibfnamefont {H.}~\bibnamefont {Harima}}, \bibinfo {author} {\bibfnamefont {A.}~\bibnamefont {Nakamura}}, \bibinfo {author} {\bibfnamefont {D.}~\bibnamefont {Li}}, \bibinfo {author} {\bibfnamefont {Y.}~\bibnamefont {Homma}}, \bibinfo {author} {\bibfnamefont {Y.}~\bibnamefont {Shimizu}}, \bibinfo {author} {\bibfnamefont {G.}~\bibnamefont {Knebel}}, \bibinfo {author} {\bibfnamefont {J.}~\bibnamefont {Flouquet}},\ and\ \bibinfo {author} {\bibfnamefont {Y.}~\bibnamefont {Haga}},\ }\bibfield  {title} {\bibinfo {title} {First observation of the de haas--van alphen effect and fermi surfaces in the
  unconventional superconductor {UTe$_{2}$}},\ }\href {https://doi.org/10.7566/JPSJ.91.083704} {\bibfield  {journal} {\bibinfo  {journal} {J. Phys. Soc. Jpn.}\ }\textbf {\bibinfo {volume} {91}},\ \bibinfo {pages} {083704} (\bibinfo {year} {2022}{\natexlab{b}})}\BibitemShut {NoStop}%
\bibitem [{\citenamefont {Knebel}\ \emph {et~al.}(2020)\citenamefont {Knebel}, \citenamefont {Kimata}, \citenamefont {Vali{\v s}ka}, \citenamefont {Honda}, \citenamefont {Li}, \citenamefont {Braithwaite}, \citenamefont {Lapertot}, \citenamefont {Knafo}, \citenamefont {Pourret}, \citenamefont {Sato}, \citenamefont {Shimizu}, \citenamefont {Kihara}, \citenamefont {Brison}, \citenamefont {Flouquet},\ and\ \citenamefont {Aoki}}]{Knebel2020-cr}%
  \BibitemOpen
  \bibfield  {author} {\bibinfo {author} {\bibfnamefont {G.}~\bibnamefont {Knebel}}, \bibinfo {author} {\bibfnamefont {M.}~\bibnamefont {Kimata}}, \bibinfo {author} {\bibfnamefont {M.}~\bibnamefont {Vali{\v s}ka}}, \bibinfo {author} {\bibfnamefont {F.}~\bibnamefont {Honda}}, \bibinfo {author} {\bibfnamefont {D.}~\bibnamefont {Li}}, \bibinfo {author} {\bibfnamefont {D.}~\bibnamefont {Braithwaite}}, \bibinfo {author} {\bibfnamefont {G.}~\bibnamefont {Lapertot}}, \bibinfo {author} {\bibfnamefont {W.}~\bibnamefont {Knafo}}, \bibinfo {author} {\bibfnamefont {A.}~\bibnamefont {Pourret}}, \bibinfo {author} {\bibfnamefont {Y.~J.}\ \bibnamefont {Sato}}, \bibinfo {author} {\bibfnamefont {Y.}~\bibnamefont {Shimizu}}, \bibinfo {author} {\bibfnamefont {T.}~\bibnamefont {Kihara}}, \bibinfo {author} {\bibfnamefont {J.-P.}\ \bibnamefont {Brison}}, \bibinfo {author} {\bibfnamefont {J.}~\bibnamefont {Flouquet}},\ and\ \bibinfo {author} {\bibfnamefont {D.}~\bibnamefont {Aoki}},\ }\bibfield  {title} {\bibinfo {title} {Anisotropy
  of the upper critical field in the {Heavy-Fermion} superconductor {UTe$_{2}$} under pressure},\ }\href {https://doi.org/10.7566/JPSJ.89.053707} {\bibfield  {journal} {\bibinfo  {journal} {J. Phys. Soc. Jpn.}\ }\textbf {\bibinfo {volume} {89}},\ \bibinfo {pages} {053707} (\bibinfo {year} {2020})}\BibitemShut {NoStop}%
\bibitem [{\citenamefont {Sakai}\ \emph {et~al.}(2023)\citenamefont {Sakai}, \citenamefont {Tokiwa}, \citenamefont {Opletal}, \citenamefont {Kimata}, \citenamefont {Awaji}, \citenamefont {Sasaki}, \citenamefont {Aoki}, \citenamefont {Kambe}, \citenamefont {Tokunaga},\ and\ \citenamefont {Haga}}]{Sakai2023-qx}%
  \BibitemOpen
  \bibfield  {author} {\bibinfo {author} {\bibfnamefont {H.}~\bibnamefont {Sakai}}, \bibinfo {author} {\bibfnamefont {Y.}~\bibnamefont {Tokiwa}}, \bibinfo {author} {\bibfnamefont {P.}~\bibnamefont {Opletal}}, \bibinfo {author} {\bibfnamefont {M.}~\bibnamefont {Kimata}}, \bibinfo {author} {\bibfnamefont {S.}~\bibnamefont {Awaji}}, \bibinfo {author} {\bibfnamefont {T.}~\bibnamefont {Sasaki}}, \bibinfo {author} {\bibfnamefont {D.}~\bibnamefont {Aoki}}, \bibinfo {author} {\bibfnamefont {S.}~\bibnamefont {Kambe}}, \bibinfo {author} {\bibfnamefont {Y.}~\bibnamefont {Tokunaga}},\ and\ \bibinfo {author} {\bibfnamefont {Y.}~\bibnamefont {Haga}},\ }\bibfield  {title} {\bibinfo {title} {Field induced multiple superconducting phases in {UTe$_{2}$} along hard magnetic axis},\ }\href {https://doi.org/10.1103/PhysRevLett.130.196002} {\bibfield  {journal} {\bibinfo  {journal} {Phys. Rev. Lett.}\ }\textbf {\bibinfo {volume} {130}},\ \bibinfo {pages} {196002} (\bibinfo {year} {2023})}\BibitemShut {NoStop}%
\bibitem [{\citenamefont {Tokiwa}\ \emph {et~al.}(2022)\citenamefont {Tokiwa}, \citenamefont {Opletal}, \citenamefont {Sakai}, \citenamefont {Kubo}, \citenamefont {Yamamoto}, \citenamefont {Kambe}, \citenamefont {Kimata}, \citenamefont {Awaji}, \citenamefont {Sasaki}, \citenamefont {Aoki}, \citenamefont {Tokunaga},\ and\ \citenamefont {Haga}}]{Tokiwa2022-ji}%
  \BibitemOpen
  \bibfield  {author} {\bibinfo {author} {\bibfnamefont {Y.}~\bibnamefont {Tokiwa}}, \bibinfo {author} {\bibfnamefont {P.}~\bibnamefont {Opletal}}, \bibinfo {author} {\bibfnamefont {H.}~\bibnamefont {Sakai}}, \bibinfo {author} {\bibfnamefont {K.}~\bibnamefont {Kubo}}, \bibinfo {author} {\bibfnamefont {E.}~\bibnamefont {Yamamoto}}, \bibinfo {author} {\bibfnamefont {S.}~\bibnamefont {Kambe}}, \bibinfo {author} {\bibfnamefont {M.}~\bibnamefont {Kimata}}, \bibinfo {author} {\bibfnamefont {S.}~\bibnamefont {Awaji}}, \bibinfo {author} {\bibfnamefont {T.}~\bibnamefont {Sasaki}}, \bibinfo {author} {\bibfnamefont {D.}~\bibnamefont {Aoki}}, \bibinfo {author} {\bibfnamefont {Y.}~\bibnamefont {Tokunaga}},\ and\ \bibinfo {author} {\bibfnamefont {Y.}~\bibnamefont {Haga}},\ }\bibfield  {title} {\bibinfo {title} {Stabilization of superconductivity by metamagnetism in an easy-axis magnetic field on {UTe$_{2}$}},\ }\href {https://arxiv.org/abs/2210.11769} {\bibfield  {journal} {\bibinfo  {journal} {arXiv:2210.11769
  [cond-mat.supr-con]}\ } (\bibinfo {year} {2022})}\BibitemShut {NoStop}%
\bibitem [{\citenamefont {Matsumura}\ \emph {et~al.}(2023)\citenamefont {Matsumura}, \citenamefont {Fujibayashi}, \citenamefont {Kinjo}, \citenamefont {Kitagawa}, \citenamefont {Ishida}, \citenamefont {Tokunaga}, \citenamefont {Sakai}, \citenamefont {Kambe}, \citenamefont {Nakamura}, \citenamefont {Shimizu}, \citenamefont {Homma}, \citenamefont {Li}, \citenamefont {Honda},\ and\ \citenamefont {Aoki}}]{Matsumura2023-ia}%
  \BibitemOpen
  \bibfield  {author} {\bibinfo {author} {\bibfnamefont {H.}~\bibnamefont {Matsumura}}, \bibinfo {author} {\bibfnamefont {H.}~\bibnamefont {Fujibayashi}}, \bibinfo {author} {\bibfnamefont {K.}~\bibnamefont {Kinjo}}, \bibinfo {author} {\bibfnamefont {S.}~\bibnamefont {Kitagawa}}, \bibinfo {author} {\bibfnamefont {K.}~\bibnamefont {Ishida}}, \bibinfo {author} {\bibfnamefont {Y.}~\bibnamefont {Tokunaga}}, \bibinfo {author} {\bibfnamefont {H.}~\bibnamefont {Sakai}}, \bibinfo {author} {\bibfnamefont {S.}~\bibnamefont {Kambe}}, \bibinfo {author} {\bibfnamefont {A.}~\bibnamefont {Nakamura}}, \bibinfo {author} {\bibfnamefont {Y.}~\bibnamefont {Shimizu}}, \bibinfo {author} {\bibfnamefont {Y.}~\bibnamefont {Homma}}, \bibinfo {author} {\bibfnamefont {D.}~\bibnamefont {Li}}, \bibinfo {author} {\bibfnamefont {F.}~\bibnamefont {Honda}},\ and\ \bibinfo {author} {\bibfnamefont {D.}~\bibnamefont {Aoki}},\ }\bibfield  {title} {\bibinfo {title} {Large reduction in the a-axis knight shift on {UTe$_{2}$}with {$T_{\rm c}$} = 2.1
  {K}},\ }\href {https://doi.org/10.7566/JPSJ.92.063701} {\bibfield  {journal} {\bibinfo  {journal} {J. Phys. Soc. Jpn.}\ }\textbf {\bibinfo {volume} {92}},\ \bibinfo {pages} {063701} (\bibinfo {year} {2023})}\BibitemShut {NoStop}%
\bibitem [{\citenamefont {Kinjo}\ \emph {et~al.}(2023)\citenamefont {Kinjo}, \citenamefont {Fujibayashi}, \citenamefont {Matsumura}, \citenamefont {Hori}, \citenamefont {Kitagawa}, \citenamefont {Ishida}, \citenamefont {Tokunaga}, \citenamefont {Sakai}, \citenamefont {Kambe}, \citenamefont {Nakamura}, \citenamefont {Shimizu}, \citenamefont {Homma}, \citenamefont {Li}, \citenamefont {Honda},\ and\ \citenamefont {Aoki}}]{Kinjo2023-ri}%
  \BibitemOpen
  \bibfield  {author} {\bibinfo {author} {\bibfnamefont {K.}~\bibnamefont {Kinjo}}, \bibinfo {author} {\bibfnamefont {H.}~\bibnamefont {Fujibayashi}}, \bibinfo {author} {\bibfnamefont {H.}~\bibnamefont {Matsumura}}, \bibinfo {author} {\bibfnamefont {F.}~\bibnamefont {Hori}}, \bibinfo {author} {\bibfnamefont {S.}~\bibnamefont {Kitagawa}}, \bibinfo {author} {\bibfnamefont {K.}~\bibnamefont {Ishida}}, \bibinfo {author} {\bibfnamefont {Y.}~\bibnamefont {Tokunaga}}, \bibinfo {author} {\bibfnamefont {H.}~\bibnamefont {Sakai}}, \bibinfo {author} {\bibfnamefont {S.}~\bibnamefont {Kambe}}, \bibinfo {author} {\bibfnamefont {A.}~\bibnamefont {Nakamura}}, \bibinfo {author} {\bibfnamefont {Y.}~\bibnamefont {Shimizu}}, \bibinfo {author} {\bibfnamefont {Y.}~\bibnamefont {Homma}}, \bibinfo {author} {\bibfnamefont {D.}~\bibnamefont {Li}}, \bibinfo {author} {\bibfnamefont {F.}~\bibnamefont {Honda}},\ and\ \bibinfo {author} {\bibfnamefont {D.}~\bibnamefont {Aoki}},\ }\bibfield  {title} {\bibinfo {title} {Superconducting spin
  reorientation in spin-triplet multiple superconducting phases of {UTe$_{2}$}},\ }\href {https://doi.org/10.1126/sciadv.adg2736} {\bibfield  {journal} {\bibinfo  {journal} {Sci Adv}\ }\textbf {\bibinfo {volume} {9}},\ \bibinfo {pages} {eadg2736} (\bibinfo {year} {2023})}\BibitemShut {NoStop}%
\bibitem [{\citenamefont {Suetsugu}\ \emph {et~al.}(2024)\citenamefont {Suetsugu}, \citenamefont {Shimomura}, \citenamefont {Kamimura}, \citenamefont {Asaba}, \citenamefont {Asaeda}, \citenamefont {Kosuge}, \citenamefont {Sekino}, \citenamefont {Ikemori}, \citenamefont {Kasahara}, \citenamefont {Kohsaka}, \citenamefont {Lee}, \citenamefont {Yanase}, \citenamefont {Sakai}, \citenamefont {Opletal}, \citenamefont {Tokiwa}, \citenamefont {Haga},\ and\ \citenamefont {Matsuda}}]{Suetsugu2024-ua}%
  \BibitemOpen
  \bibfield  {author} {\bibinfo {author} {\bibfnamefont {S.}~\bibnamefont {Suetsugu}}, \bibinfo {author} {\bibfnamefont {M.}~\bibnamefont {Shimomura}}, \bibinfo {author} {\bibfnamefont {M.}~\bibnamefont {Kamimura}}, \bibinfo {author} {\bibfnamefont {T.}~\bibnamefont {Asaba}}, \bibinfo {author} {\bibfnamefont {H.}~\bibnamefont {Asaeda}}, \bibinfo {author} {\bibfnamefont {Y.}~\bibnamefont {Kosuge}}, \bibinfo {author} {\bibfnamefont {Y.}~\bibnamefont {Sekino}}, \bibinfo {author} {\bibfnamefont {S.}~\bibnamefont {Ikemori}}, \bibinfo {author} {\bibfnamefont {Y.}~\bibnamefont {Kasahara}}, \bibinfo {author} {\bibfnamefont {Y.}~\bibnamefont {Kohsaka}}, \bibinfo {author} {\bibfnamefont {M.}~\bibnamefont {Lee}}, \bibinfo {author} {\bibfnamefont {Y.}~\bibnamefont {Yanase}}, \bibinfo {author} {\bibfnamefont {H.}~\bibnamefont {Sakai}}, \bibinfo {author} {\bibfnamefont {P.}~\bibnamefont {Opletal}}, \bibinfo {author} {\bibfnamefont {Y.}~\bibnamefont {Tokiwa}}, \bibinfo {author} {\bibfnamefont {Y.}~\bibnamefont {Haga}},\
  and\ \bibinfo {author} {\bibfnamefont {Y.}~\bibnamefont {Matsuda}},\ }\bibfield  {title} {\bibinfo {title} {Fully gapped pairing state in spin-triplet superconductor {UTe$_{2}$}},\ }\href {https://doi.org/10.1126/sciadv.adk3772} {\bibfield  {journal} {\bibinfo  {journal} {Sci Adv}\ }\textbf {\bibinfo {volume} {10}},\ \bibinfo {pages} {eadk3772} (\bibinfo {year} {2024})}\BibitemShut {NoStop}%
\bibitem [{\citenamefont {Braithwaite}\ \emph {et~al.}(2019)\citenamefont {Braithwaite}, \citenamefont {Vali{\v s}ka}, \citenamefont {Knebel}, \citenamefont {Lapertot}, \citenamefont {Brison}, \citenamefont {Pourret}, \citenamefont {Zhitomirsky}, \citenamefont {Flouquet}, \citenamefont {Honda},\ and\ \citenamefont {Aoki}}]{Braithwaite2019-cr}%
  \BibitemOpen
  \bibfield  {author} {\bibinfo {author} {\bibfnamefont {D.}~\bibnamefont {Braithwaite}}, \bibinfo {author} {\bibfnamefont {M.}~\bibnamefont {Vali{\v s}ka}}, \bibinfo {author} {\bibfnamefont {G.}~\bibnamefont {Knebel}}, \bibinfo {author} {\bibfnamefont {G.}~\bibnamefont {Lapertot}}, \bibinfo {author} {\bibfnamefont {J.-P.}\ \bibnamefont {Brison}}, \bibinfo {author} {\bibfnamefont {A.}~\bibnamefont {Pourret}}, \bibinfo {author} {\bibfnamefont {M.~E.}\ \bibnamefont {Zhitomirsky}}, \bibinfo {author} {\bibfnamefont {J.}~\bibnamefont {Flouquet}}, \bibinfo {author} {\bibfnamefont {F.}~\bibnamefont {Honda}},\ and\ \bibinfo {author} {\bibfnamefont {D.}~\bibnamefont {Aoki}},\ }\bibfield  {title} {\bibinfo {title} {Multiple superconducting phases in a nearly ferromagnetic system},\ }\href {https://doi.org/10.1038/s42005-019-0248-z} {\bibfield  {journal} {\bibinfo  {journal} {Communications Physics}\ }\textbf {\bibinfo {volume} {2}},\ \bibinfo {pages} {1} (\bibinfo {year} {2019})}\BibitemShut {NoStop}%
\bibitem [{\citenamefont {Rosuel}\ \emph {et~al.}(2023)\citenamefont {Rosuel}, \citenamefont {Marcenat}, \citenamefont {Knebel}, \citenamefont {Klein}, \citenamefont {Pourret}, \citenamefont {Marquardt}, \citenamefont {Niu}, \citenamefont {Rousseau}, \citenamefont {Demuer}, \citenamefont {Seyfarth}, \citenamefont {Lapertot}, \citenamefont {Aoki}, \citenamefont {Braithwaite}, \citenamefont {Flouquet},\ and\ \citenamefont {Brison}}]{Rosuel2023-st}%
  \BibitemOpen
  \bibfield  {author} {\bibinfo {author} {\bibfnamefont {A.}~\bibnamefont {Rosuel}}, \bibinfo {author} {\bibfnamefont {C.}~\bibnamefont {Marcenat}}, \bibinfo {author} {\bibfnamefont {G.}~\bibnamefont {Knebel}}, \bibinfo {author} {\bibfnamefont {T.}~\bibnamefont {Klein}}, \bibinfo {author} {\bibfnamefont {A.}~\bibnamefont {Pourret}}, \bibinfo {author} {\bibfnamefont {N.}~\bibnamefont {Marquardt}}, \bibinfo {author} {\bibfnamefont {Q.}~\bibnamefont {Niu}}, \bibinfo {author} {\bibfnamefont {S.}~\bibnamefont {Rousseau}}, \bibinfo {author} {\bibfnamefont {A.}~\bibnamefont {Demuer}}, \bibinfo {author} {\bibfnamefont {G.}~\bibnamefont {Seyfarth}}, \bibinfo {author} {\bibfnamefont {G.}~\bibnamefont {Lapertot}}, \bibinfo {author} {\bibfnamefont {D.}~\bibnamefont {Aoki}}, \bibinfo {author} {\bibfnamefont {D.}~\bibnamefont {Braithwaite}}, \bibinfo {author} {\bibfnamefont {J.}~\bibnamefont {Flouquet}},\ and\ \bibinfo {author} {\bibfnamefont {J.~P.}\ \bibnamefont {Brison}},\ }\bibfield  {title} {\bibinfo {title}
  {{Field-Induced} tuning of the pairing state in a superconductor},\ }\href {https://doi.org/10.1103/PhysRevX.13.011022} {\bibfield  {journal} {\bibinfo  {journal} {Phys. Rev. X}\ }\textbf {\bibinfo {volume} {13}},\ \bibinfo {pages} {011022} (\bibinfo {year} {2023})}\BibitemShut {NoStop}%
\bibitem [{\citenamefont {Thomas}\ \emph {et~al.}(2020)\citenamefont {Thomas}, \citenamefont {Santos}, \citenamefont {Christensen}, \citenamefont {Asaba}, \citenamefont {Ronning}, \citenamefont {Thompson}, \citenamefont {Bauer}, \citenamefont {Fernandes}, \citenamefont {Fabbris},\ and\ \citenamefont {Rosa}}]{Thomas2020-ah}%
  \BibitemOpen
  \bibfield  {author} {\bibinfo {author} {\bibfnamefont {S.~M.}\ \bibnamefont {Thomas}}, \bibinfo {author} {\bibfnamefont {F.~B.}\ \bibnamefont {Santos}}, \bibinfo {author} {\bibfnamefont {M.~H.}\ \bibnamefont {Christensen}}, \bibinfo {author} {\bibfnamefont {T.}~\bibnamefont {Asaba}}, \bibinfo {author} {\bibfnamefont {F.}~\bibnamefont {Ronning}}, \bibinfo {author} {\bibfnamefont {J.~D.}\ \bibnamefont {Thompson}}, \bibinfo {author} {\bibfnamefont {E.~D.}\ \bibnamefont {Bauer}}, \bibinfo {author} {\bibfnamefont {R.~M.}\ \bibnamefont {Fernandes}}, \bibinfo {author} {\bibfnamefont {G.}~\bibnamefont {Fabbris}},\ and\ \bibinfo {author} {\bibfnamefont {P.~F.~S.}\ \bibnamefont {Rosa}},\ }\bibfield  {title} {\bibinfo {title} {Evidence for a pressure-induced antiferromagnetic quantum critical point in intermediate-valence {UTe$_{2}$}},\ }\href {https://doi.org/10.1126/sciadv.abc8709} {\bibfield  {journal} {\bibinfo  {journal} {Science Advances}\ }\textbf {\bibinfo {volume} {6}},\ \bibinfo {pages} {eabc8709} (\bibinfo
  {year} {2020})}\BibitemShut {NoStop}%
\bibitem [{\citenamefont {Rosa}\ \emph {et~al.}(2022)\citenamefont {Rosa}, \citenamefont {Weiland}, \citenamefont {Fender}, \citenamefont {Scott}, \citenamefont {Ronning}, \citenamefont {Thompson}, \citenamefont {Bauer},\ and\ \citenamefont {Thomas}}]{Rosa2022-ud}%
  \BibitemOpen
  \bibfield  {author} {\bibinfo {author} {\bibfnamefont {P.~F.~S.}\ \bibnamefont {Rosa}}, \bibinfo {author} {\bibfnamefont {A.}~\bibnamefont {Weiland}}, \bibinfo {author} {\bibfnamefont {S.~S.}\ \bibnamefont {Fender}}, \bibinfo {author} {\bibfnamefont {B.~L.}\ \bibnamefont {Scott}}, \bibinfo {author} {\bibfnamefont {F.}~\bibnamefont {Ronning}}, \bibinfo {author} {\bibfnamefont {J.~D.}\ \bibnamefont {Thompson}}, \bibinfo {author} {\bibfnamefont {E.~D.}\ \bibnamefont {Bauer}},\ and\ \bibinfo {author} {\bibfnamefont {S.~M.}\ \bibnamefont {Thomas}},\ }\bibfield  {title} {\bibinfo {title} {Single thermodynamic transition at 2 {K} in superconducting {UTe$_{2}$} single crystals},\ }\href {https://doi.org/10.1038/s43246-022-00254-2} {\bibfield  {journal} {\bibinfo  {journal} {Communications Materials}\ }\textbf {\bibinfo {volume} {3}},\ \bibinfo {pages} {1} (\bibinfo {year} {2022})}\BibitemShut {NoStop}%
\bibitem [{\citenamefont {Ishizuka}\ \emph {et~al.}(2019)\citenamefont {Ishizuka}, \citenamefont {Sumita}, \citenamefont {Daido},\ and\ \citenamefont {Yanase}}]{Ishizuka2019-hv}%
  \BibitemOpen
  \bibfield  {author} {\bibinfo {author} {\bibfnamefont {J.}~\bibnamefont {Ishizuka}}, \bibinfo {author} {\bibfnamefont {S.}~\bibnamefont {Sumita}}, \bibinfo {author} {\bibfnamefont {A.}~\bibnamefont {Daido}},\ and\ \bibinfo {author} {\bibfnamefont {Y.}~\bibnamefont {Yanase}},\ }\bibfield  {title} {\bibinfo {title} {{Insulator-Metal} transition and topological superconductivity in {UTe$_{2}$} from a {First-Principles} calculation},\ }\href {https://doi.org/10.1103/PhysRevLett.123.217001} {\bibfield  {journal} {\bibinfo  {journal} {Phys. Rev. Lett.}\ }\textbf {\bibinfo {volume} {123}},\ \bibinfo {pages} {217001} (\bibinfo {year} {2019})}\BibitemShut {NoStop}%
\bibitem [{\citenamefont {Tei}\ \emph {et~al.}(2023)\citenamefont {Tei}, \citenamefont {Mizushima},\ and\ \citenamefont {Fujimoto}}]{Tei2023-nq}%
  \BibitemOpen
  \bibfield  {author} {\bibinfo {author} {\bibfnamefont {J.}~\bibnamefont {Tei}}, \bibinfo {author} {\bibfnamefont {T.}~\bibnamefont {Mizushima}},\ and\ \bibinfo {author} {\bibfnamefont {S.}~\bibnamefont {Fujimoto}},\ }\bibfield  {title} {\bibinfo {title} {Possible realization of topological crystalline superconductivity with time-reversal symmetry in {UTe$_{2}$}},\ }\href {https://doi.org/10.1103/PhysRevB.107.144517} {\bibfield  {journal} {\bibinfo  {journal} {Phys. Rev. B Condens. Matter}\ }\textbf {\bibinfo {volume} {107}},\ \bibinfo {pages} {144517} (\bibinfo {year} {2023})}\BibitemShut {NoStop}%
\bibitem [{\citenamefont {Tei}\ \emph {et~al.}(2024)\citenamefont {Tei}, \citenamefont {Mizushima},\ and\ \citenamefont {Fujimoto}}]{Tei2024-gv}%
  \BibitemOpen
  \bibfield  {author} {\bibinfo {author} {\bibfnamefont {J.}~\bibnamefont {Tei}}, \bibinfo {author} {\bibfnamefont {T.}~\bibnamefont {Mizushima}},\ and\ \bibinfo {author} {\bibfnamefont {S.}~\bibnamefont {Fujimoto}},\ }\bibfield  {title} {\bibinfo {title} {Pairing symmetries of multiple superconducting phases in {UTe$_{2}$}: Competition between ferromagnetic and antiferromagnetic fluctuations},\ }\href {https://doi.org/10.1103/PhysRevB.109.064516} {\bibfield  {journal} {\bibinfo  {journal} {Phys. Rev. B Condens. Matter}\ }\textbf {\bibinfo {volume} {109}},\ \bibinfo {pages} {064516} (\bibinfo {year} {2024})}\BibitemShut {NoStop}%
\bibitem [{\citenamefont {Hakuno}\ \emph {et~al.}(2024)\citenamefont {Hakuno}, \citenamefont {Nogaki},\ and\ \citenamefont {Yanase}}]{Hakuno2024-cx}%
  \BibitemOpen
  \bibfield  {author} {\bibinfo {author} {\bibfnamefont {R.}~\bibnamefont {Hakuno}}, \bibinfo {author} {\bibfnamefont {K.}~\bibnamefont {Nogaki}},\ and\ \bibinfo {author} {\bibfnamefont {Y.}~\bibnamefont {Yanase}},\ }\bibfield  {title} {\bibinfo {title} {Magnetism and superconductivity in mixed-dimensional periodic anderson model for {UTe$_{2}$}},\ }\href {https://doi.org/10.1103/PhysRevB.109.104509} {\bibfield  {journal} {\bibinfo  {journal} {Phys. Rev. B Condens. Matter}\ }\textbf {\bibinfo {volume} {109}},\ \bibinfo {pages} {104509} (\bibinfo {year} {2024})}\BibitemShut {NoStop}%
\bibitem [{\citenamefont {Naritsuka}\ \emph {et~al.}(2021)\citenamefont {Naritsuka}, \citenamefont {Terashima},\ and\ \citenamefont {Matsuda}}]{Naritsuka2021-ux}%
  \BibitemOpen
  \bibfield  {author} {\bibinfo {author} {\bibfnamefont {M.}~\bibnamefont {Naritsuka}}, \bibinfo {author} {\bibfnamefont {T.}~\bibnamefont {Terashima}},\ and\ \bibinfo {author} {\bibfnamefont {Y.}~\bibnamefont {Matsuda}},\ }\bibfield  {title} {\bibinfo {title} {Controlling unconventional superconductivity in artificially engineeredf-electron kondo superlattices},\ }\href {https://doi.org/10.1088/1361-648X/abfdf2} {\bibfield  {journal} {\bibinfo  {journal} {J. Phys. Condens. Matter}\ }\textbf {\bibinfo {volume} {33}},\ \bibinfo {pages} {273001} (\bibinfo {year} {2021})}\BibitemShut {NoStop}%
\bibitem [{\citenamefont {Radzihovsky}(2011)}]{Radzihovsky2011-jx}%
  \BibitemOpen
  \bibfield  {author} {\bibinfo {author} {\bibfnamefont {L.}~\bibnamefont {Radzihovsky}},\ }\bibfield  {title} {\bibinfo {title} {Fluctuations and phase transitions in larkin-ovchinnikov liquid-crystal states of a population-imbalanced resonant fermi gas},\ }\href {https://doi.org/10.1103/PhysRevA.84.023611} {\bibfield  {journal} {\bibinfo  {journal} {Phys. Rev. A}\ }\textbf {\bibinfo {volume} {84}},\ \bibinfo {pages} {023611} (\bibinfo {year} {2011})}\BibitemShut {NoStop}%
\bibitem [{\citenamefont {She}\ and\ \citenamefont {Balatsky}(2012)}]{She2012-lw}%
  \BibitemOpen
  \bibfield  {author} {\bibinfo {author} {\bibfnamefont {J.-H.}\ \bibnamefont {She}}\ and\ \bibinfo {author} {\bibfnamefont {A.~V.}\ \bibnamefont {Balatsky}},\ }\bibfield  {title} {\bibinfo {title} {{Berezinskii-Kosterlitz-Thouless} transition to the superconducting state of heavy-fermion superlattices},\ }\href {https://doi.org/10.1103/PhysRevLett.109.077002} {\bibfield  {journal} {\bibinfo  {journal} {Phys. Rev. Lett.}\ }\textbf {\bibinfo {volume} {109}},\ \bibinfo {pages} {077002} (\bibinfo {year} {2012})}\BibitemShut {NoStop}%
\bibitem [{\citenamefont {Yanase}\ \emph {et~al.}(2022)\citenamefont {Yanase}, \citenamefont {Daido}, \citenamefont {Takasan},\ and\ \citenamefont {Yoshida}}]{Yanase2022-bq}%
  \BibitemOpen
  \bibfield  {author} {\bibinfo {author} {\bibfnamefont {Y.}~\bibnamefont {Yanase}}, \bibinfo {author} {\bibfnamefont {A.}~\bibnamefont {Daido}}, \bibinfo {author} {\bibfnamefont {K.}~\bibnamefont {Takasan}},\ and\ \bibinfo {author} {\bibfnamefont {T.}~\bibnamefont {Yoshida}},\ }\bibfield  {title} {\bibinfo {title} {Topological d-wave superconductivity in two dimensions},\ }\href {https://doi.org/10.1016/j.physe.2022.115143} {\bibfield  {journal} {\bibinfo  {journal} {Physica E}\ }\textbf {\bibinfo {volume} {140}},\ \bibinfo {pages} {115143} (\bibinfo {year} {2022})}\BibitemShut {NoStop}%
\bibitem [{\citenamefont {Asaba}\ \emph {et~al.}(2024)\citenamefont {Asaba}, \citenamefont {Naritsuka}, \citenamefont {Asaeda}, \citenamefont {Kosuge}, \citenamefont {Ikemori}, \citenamefont {Suetsugu}, \citenamefont {Kasahara}, \citenamefont {Kohsaka}, \citenamefont {Terashima}, \citenamefont {Daido}, \citenamefont {Yanase},\ and\ \citenamefont {Matsuda}}]{Asaba2024-jp}%
  \BibitemOpen
  \bibfield  {author} {\bibinfo {author} {\bibfnamefont {T.}~\bibnamefont {Asaba}}, \bibinfo {author} {\bibfnamefont {M.}~\bibnamefont {Naritsuka}}, \bibinfo {author} {\bibfnamefont {H.}~\bibnamefont {Asaeda}}, \bibinfo {author} {\bibfnamefont {Y.}~\bibnamefont {Kosuge}}, \bibinfo {author} {\bibfnamefont {S.}~\bibnamefont {Ikemori}}, \bibinfo {author} {\bibfnamefont {S.}~\bibnamefont {Suetsugu}}, \bibinfo {author} {\bibfnamefont {Y.}~\bibnamefont {Kasahara}}, \bibinfo {author} {\bibfnamefont {Y.}~\bibnamefont {Kohsaka}}, \bibinfo {author} {\bibfnamefont {T.}~\bibnamefont {Terashima}}, \bibinfo {author} {\bibfnamefont {A.}~\bibnamefont {Daido}}, \bibinfo {author} {\bibfnamefont {Y.}~\bibnamefont {Yanase}},\ and\ \bibinfo {author} {\bibfnamefont {Y.}~\bibnamefont {Matsuda}},\ }\bibfield  {title} {\bibinfo {title} {Evidence for a finite-momentum cooper pair in tricolor d-wave superconducting superlattices},\ }\href {https://doi.org/10.1038/s41467-024-47875-4} {\bibfield  {journal} {\bibinfo  {journal} {Nat.
  Commun.}\ }\textbf {\bibinfo {volume} {15}},\ \bibinfo {pages} {3861} (\bibinfo {year} {2024})}\BibitemShut {NoStop}%
\bibitem [{\citenamefont {Fulde}\ and\ \citenamefont {Ferrell}(1964)}]{Fulde1964-bx}%
  \BibitemOpen
  \bibfield  {author} {\bibinfo {author} {\bibfnamefont {P.}~\bibnamefont {Fulde}}\ and\ \bibinfo {author} {\bibfnamefont {R.~A.}\ \bibnamefont {Ferrell}},\ }\bibfield  {title} {\bibinfo {title} {Superconductivity in a strong spin-exchange field},\ }\href {https://doi.org/10.1103/physrev.135.a550} {\bibfield  {journal} {\bibinfo  {journal} {Phys. Rev.}\ }\textbf {\bibinfo {volume} {135}},\ \bibinfo {pages} {A550} (\bibinfo {year} {1964})}\BibitemShut {NoStop}%
\bibitem [{\citenamefont {Larkin}\ and\ \citenamefont {Ovchinnikov}(1964)}]{Larkin1964-aa}%
  \BibitemOpen
  \bibfield  {author} {\bibinfo {author} {\bibfnamefont {A.~I.}\ \bibnamefont {Larkin}}\ and\ \bibinfo {author} {\bibfnamefont {Y.~N.}\ \bibnamefont {Ovchinnikov}},\ }\bibfield  {title} {\bibinfo {title} {Nonuniform state of superconductors},\ }\href@noop {} {\bibfield  {journal} {\bibinfo  {journal} {Zh. Eksp. Teor. Fiz.}\ }\textbf {\bibinfo {volume} {47}},\ \bibinfo {pages} {1136} (\bibinfo {year} {1964})},\ \bibinfo {note} {[Sov. Phys. JETP 20, 762 (1965)]}\BibitemShut {NoStop}%
\bibitem [{\citenamefont {Yin}\ \emph {et~al.}(2014)\citenamefont {Yin}, \citenamefont {Martikainen},\ and\ \citenamefont {Törmä}}]{Yin2014-wo}%
  \BibitemOpen
  \bibfield  {author} {\bibinfo {author} {\bibfnamefont {S.}~\bibnamefont {Yin}}, \bibinfo {author} {\bibfnamefont {J.-P.}\ \bibnamefont {Martikainen}},\ and\ \bibinfo {author} {\bibfnamefont {P.}~\bibnamefont {Törmä}},\ }\bibfield  {title} {\bibinfo {title} {Fulde-ferrell states and berezinskii-kosterlitz-thouless phase transition in two-dimensional imbalanced fermi gases},\ }\href {https://doi.org/10.1103/PhysRevB.89.014507} {\bibfield  {journal} {\bibinfo  {journal} {Phys. Rev. B}\ }\textbf {\bibinfo {volume} {89}},\ \bibinfo {pages} {014507} (\bibinfo {year} {2014})}\BibitemShut {NoStop}%
\bibitem [{\citenamefont {Xu}\ and\ \citenamefont {Zhang}(2015)}]{Xu2015-ql}%
  \BibitemOpen
  \bibfield  {author} {\bibinfo {author} {\bibfnamefont {Y.}~\bibnamefont {Xu}}\ and\ \bibinfo {author} {\bibfnamefont {C.}~\bibnamefont {Zhang}},\ }\bibfield  {title} {\bibinfo {title} {Berezinskii-kosterlitz-thouless phase transition in {2D} spin-orbit-coupled fulde-ferrell superfluids},\ }\href {https://doi.org/10.1103/PhysRevLett.114.110401} {\bibfield  {journal} {\bibinfo  {journal} {Phys. Rev. Lett.}\ }\textbf {\bibinfo {volume} {114}},\ \bibinfo {pages} {110401} (\bibinfo {year} {2015})}\BibitemShut {NoStop}%
\bibitem [{\citenamefont {Nagata}\ \emph {et~al.}(1992)\citenamefont {Nagata}, \citenamefont {Aochi}, \citenamefont {Abe}, \citenamefont {Ebisu}, \citenamefont {Hagino}, \citenamefont {Seki},\ and\ \citenamefont {Tsutsumi}}]{Nagata1992-bt}%
  \BibitemOpen
  \bibfield  {author} {\bibinfo {author} {\bibfnamefont {S.}~\bibnamefont {Nagata}}, \bibinfo {author} {\bibfnamefont {T.}~\bibnamefont {Aochi}}, \bibinfo {author} {\bibfnamefont {T.}~\bibnamefont {Abe}}, \bibinfo {author} {\bibfnamefont {S.}~\bibnamefont {Ebisu}}, \bibinfo {author} {\bibfnamefont {T.}~\bibnamefont {Hagino}}, \bibinfo {author} {\bibfnamefont {Y.}~\bibnamefont {Seki}},\ and\ \bibinfo {author} {\bibfnamefont {K.}~\bibnamefont {Tsutsumi}},\ }\bibfield  {title} {\bibinfo {title} {Superconductivity in the layered compound {2H-TaS$_{2}$}},\ }\href {https://doi.org/10.1016/0022-3697(92)90242-6} {\bibfield  {journal} {\bibinfo  {journal} {J. Phys. Chem. Solids}\ }\textbf {\bibinfo {volume} {53}},\ \bibinfo {pages} {1259} (\bibinfo {year} {1992})}\BibitemShut {NoStop}%
\bibitem [{\citenamefont {Navarro-Moratalla}\ \emph {et~al.}(2016)\citenamefont {Navarro-Moratalla}, \citenamefont {Island}, \citenamefont {Ma{\~n}as-Valero}, \citenamefont {Pinilla-Cienfuegos}, \citenamefont {Castellanos-Gomez}, \citenamefont {Quereda}, \citenamefont {Rubio-Bollinger}, \citenamefont {Chirolli}, \citenamefont {Silva-Guill{\'e}n}, \citenamefont {Agra{\"\i}t}, \citenamefont {Steele}, \citenamefont {Guinea}, \citenamefont {van~der Zant},\ and\ \citenamefont {Coronado}}]{Navarro-Moratalla2016-mg}%
  \BibitemOpen
  \bibfield  {author} {\bibinfo {author} {\bibfnamefont {E.}~\bibnamefont {Navarro-Moratalla}}, \bibinfo {author} {\bibfnamefont {J.~O.}\ \bibnamefont {Island}}, \bibinfo {author} {\bibfnamefont {S.}~\bibnamefont {Ma{\~n}as-Valero}}, \bibinfo {author} {\bibfnamefont {E.}~\bibnamefont {Pinilla-Cienfuegos}}, \bibinfo {author} {\bibfnamefont {A.}~\bibnamefont {Castellanos-Gomez}}, \bibinfo {author} {\bibfnamefont {J.}~\bibnamefont {Quereda}}, \bibinfo {author} {\bibfnamefont {G.}~\bibnamefont {Rubio-Bollinger}}, \bibinfo {author} {\bibfnamefont {L.}~\bibnamefont {Chirolli}}, \bibinfo {author} {\bibfnamefont {J.~A.}\ \bibnamefont {Silva-Guill{\'e}n}}, \bibinfo {author} {\bibfnamefont {N.}~\bibnamefont {Agra{\"\i}t}}, \bibinfo {author} {\bibfnamefont {G.~A.}\ \bibnamefont {Steele}}, \bibinfo {author} {\bibfnamefont {F.}~\bibnamefont {Guinea}}, \bibinfo {author} {\bibfnamefont {H.~S.~J.}\ \bibnamefont {van~der Zant}},\ and\ \bibinfo {author} {\bibfnamefont {E.}~\bibnamefont {Coronado}},\ }\bibfield  {title}
  {\bibinfo {title} {Enhanced superconductivity in atomically thin {TaS$_{2}$}},\ }\href {https://doi.org/10.1038/ncomms11043} {\bibfield  {journal} {\bibinfo  {journal} {Nat. Commun.}\ }\textbf {\bibinfo {volume} {7}},\ \bibinfo {pages} {11043} (\bibinfo {year} {2016})}\BibitemShut {NoStop}%
\bibitem [{\citenamefont {Yang}\ \emph {et~al.}(2018)\citenamefont {Yang}, \citenamefont {Fang}, \citenamefont {Fatemi}, \citenamefont {Ruhman}, \citenamefont {Navarro-Moratalla}, \citenamefont {Watanabe}, \citenamefont {Taniguchi}, \citenamefont {Kaxiras},\ and\ \citenamefont {Jarillo-Herrero}}]{Yang2018-ck}%
  \BibitemOpen
  \bibfield  {author} {\bibinfo {author} {\bibfnamefont {Y.}~\bibnamefont {Yang}}, \bibinfo {author} {\bibfnamefont {S.}~\bibnamefont {Fang}}, \bibinfo {author} {\bibfnamefont {V.}~\bibnamefont {Fatemi}}, \bibinfo {author} {\bibfnamefont {J.}~\bibnamefont {Ruhman}}, \bibinfo {author} {\bibfnamefont {E.}~\bibnamefont {Navarro-Moratalla}}, \bibinfo {author} {\bibfnamefont {K.}~\bibnamefont {Watanabe}}, \bibinfo {author} {\bibfnamefont {T.}~\bibnamefont {Taniguchi}}, \bibinfo {author} {\bibfnamefont {E.}~\bibnamefont {Kaxiras}},\ and\ \bibinfo {author} {\bibfnamefont {P.}~\bibnamefont {Jarillo-Herrero}},\ }\bibfield  {title} {\bibinfo {title} {Enhanced superconductivity upon weakening of charge density wave transport in {-TaS$_{2}$} in the two-dimensional limit},\ }\href {https://doi.org/10.1103/PhysRevB.98.035203} {\bibfield  {journal} {\bibinfo  {journal} {Phys. Rev. B Condens. Matter}\ }\textbf {\bibinfo {volume} {98}},\ \bibinfo {pages} {035203} (\bibinfo {year} {2018})}\BibitemShut {NoStop}%
\bibitem [{\citenamefont {Bekaert}\ \emph {et~al.}(2020)\citenamefont {Bekaert}, \citenamefont {Khestanova}, \citenamefont {Hopkinson}, \citenamefont {Birkbeck}, \citenamefont {Clark}, \citenamefont {Zhu}, \citenamefont {Bandurin}, \citenamefont {Gorbachev}, \citenamefont {Fairclough}, \citenamefont {Zou}, \citenamefont {Hamer}, \citenamefont {Terry}, \citenamefont {Peters}, \citenamefont {Sanchez}, \citenamefont {Partoens}, \citenamefont {Haigh}, \citenamefont {Milo{\v s}evi{\'c}},\ and\ \citenamefont {Grigorieva}}]{Bekaert2020-ph}%
  \BibitemOpen
  \bibfield  {author} {\bibinfo {author} {\bibfnamefont {J.}~\bibnamefont {Bekaert}}, \bibinfo {author} {\bibfnamefont {E.}~\bibnamefont {Khestanova}}, \bibinfo {author} {\bibfnamefont {D.~G.}\ \bibnamefont {Hopkinson}}, \bibinfo {author} {\bibfnamefont {J.}~\bibnamefont {Birkbeck}}, \bibinfo {author} {\bibfnamefont {N.}~\bibnamefont {Clark}}, \bibinfo {author} {\bibfnamefont {M.}~\bibnamefont {Zhu}}, \bibinfo {author} {\bibfnamefont {D.~A.}\ \bibnamefont {Bandurin}}, \bibinfo {author} {\bibfnamefont {R.}~\bibnamefont {Gorbachev}}, \bibinfo {author} {\bibfnamefont {S.}~\bibnamefont {Fairclough}}, \bibinfo {author} {\bibfnamefont {Y.}~\bibnamefont {Zou}}, \bibinfo {author} {\bibfnamefont {M.}~\bibnamefont {Hamer}}, \bibinfo {author} {\bibfnamefont {D.~J.}\ \bibnamefont {Terry}}, \bibinfo {author} {\bibfnamefont {J.~J.~P.}\ \bibnamefont {Peters}}, \bibinfo {author} {\bibfnamefont {A.~M.}\ \bibnamefont {Sanchez}}, \bibinfo {author} {\bibfnamefont {B.}~\bibnamefont {Partoens}}, \bibinfo {author} {\bibfnamefont
  {S.~J.}\ \bibnamefont {Haigh}}, \bibinfo {author} {\bibfnamefont {M.~V.}\ \bibnamefont {Milo{\v s}evi{\'c}}},\ and\ \bibinfo {author} {\bibfnamefont {I.~V.}\ \bibnamefont {Grigorieva}},\ }\bibfield  {title} {\bibinfo {title} {Enhanced superconductivity in {Few-Layer} {TaS$_{2}$} due to healing by oxygenation},\ }\href {https://doi.org/10.1021/acs.nanolett.0c00871} {\bibfield  {journal} {\bibinfo  {journal} {Nano Lett.}\ }\textbf {\bibinfo {volume} {20}},\ \bibinfo {pages} {3808} (\bibinfo {year} {2020})}\BibitemShut {NoStop}%
\bibitem [{\citenamefont {Almoalem}\ \emph {et~al.}(2024)\citenamefont {Almoalem}, \citenamefont {Feldman}, \citenamefont {Mangel}, \citenamefont {Shlafman}, \citenamefont {Yaish}, \citenamefont {Fischer}, \citenamefont {Moshe}, \citenamefont {Ruhman},\ and\ \citenamefont {Kanigel}}]{Almoalem2024-cm}%
  \BibitemOpen
  \bibfield  {author} {\bibinfo {author} {\bibfnamefont {A.}~\bibnamefont {Almoalem}}, \bibinfo {author} {\bibfnamefont {I.}~\bibnamefont {Feldman}}, \bibinfo {author} {\bibfnamefont {I.}~\bibnamefont {Mangel}}, \bibinfo {author} {\bibfnamefont {M.}~\bibnamefont {Shlafman}}, \bibinfo {author} {\bibfnamefont {Y.~E.}\ \bibnamefont {Yaish}}, \bibinfo {author} {\bibfnamefont {M.~H.}\ \bibnamefont {Fischer}}, \bibinfo {author} {\bibfnamefont {M.}~\bibnamefont {Moshe}}, \bibinfo {author} {\bibfnamefont {J.}~\bibnamefont {Ruhman}},\ and\ \bibinfo {author} {\bibfnamefont {A.}~\bibnamefont {Kanigel}},\ }\bibfield  {title} {\bibinfo {title} {The observation of $\pi$-shifts in the little-parks effect in $\rm{4H_{b}}$-$\rm{TaS_{2}}$},\ }\href {https://doi.org/10.1038/s41467-024-48260-x} {\bibfield  {journal} {\bibinfo  {journal} {Nat. Commun.}\ }\textbf {\bibinfo {volume} {15}},\ \bibinfo {pages} {4623} (\bibinfo {year} {2024})}\BibitemShut {NoStop}%
\bibitem [{\citenamefont {Wan}\ \emph {et~al.}(2024)\citenamefont {Wan}, \citenamefont {Qiu}, \citenamefont {Ren}, \citenamefont {Qian}, \citenamefont {Li}, \citenamefont {Xu}, \citenamefont {Zhou}, \citenamefont {Zhou}, \citenamefont {Zhou}, \citenamefont {Wang}, \citenamefont {Yang}, \citenamefont {Sofer}, \citenamefont {Huang}, \citenamefont {Wang},\ and\ \citenamefont {Duan}}]{Wan2024-wx}%
  \BibitemOpen
  \bibfield  {author} {\bibinfo {author} {\bibfnamefont {Z.}~\bibnamefont {Wan}}, \bibinfo {author} {\bibfnamefont {G.}~\bibnamefont {Qiu}}, \bibinfo {author} {\bibfnamefont {H.}~\bibnamefont {Ren}}, \bibinfo {author} {\bibfnamefont {Q.}~\bibnamefont {Qian}}, \bibinfo {author} {\bibfnamefont {Y.}~\bibnamefont {Li}}, \bibinfo {author} {\bibfnamefont {D.}~\bibnamefont {Xu}}, \bibinfo {author} {\bibfnamefont {J.}~\bibnamefont {Zhou}}, \bibinfo {author} {\bibfnamefont {J.}~\bibnamefont {Zhou}}, \bibinfo {author} {\bibfnamefont {B.}~\bibnamefont {Zhou}}, \bibinfo {author} {\bibfnamefont {L.}~\bibnamefont {Wang}}, \bibinfo {author} {\bibfnamefont {T.-H.}\ \bibnamefont {Yang}}, \bibinfo {author} {\bibfnamefont {Z.}~\bibnamefont {Sofer}}, \bibinfo {author} {\bibfnamefont {Y.}~\bibnamefont {Huang}}, \bibinfo {author} {\bibfnamefont {K.~L.}\ \bibnamefont {Wang}},\ and\ \bibinfo {author} {\bibfnamefont {X.}~\bibnamefont {Duan}},\ }\bibfield  {title} {\bibinfo {title} {Unconventional superconductivity in chiral
  molecule-{TaS2} hybrid superlattices},\ }\href {https://doi.org/10.1038/s41586-024-07625-4} {\bibfield  {journal} {\bibinfo  {journal} {Nature}\ }\textbf {\bibinfo {volume} {632}},\ \bibinfo {pages} {69} (\bibinfo {year} {2024})}\BibitemShut {NoStop}%
\bibitem [{\citenamefont {Fischer}\ \emph {et~al.}(2023{\natexlab{a}})\citenamefont {Fischer}, \citenamefont {Lee},\ and\ \citenamefont {Ruhman}}]{Fischer2023-yq}%
  \BibitemOpen
  \bibfield  {author} {\bibinfo {author} {\bibfnamefont {M.~H.}\ \bibnamefont {Fischer}}, \bibinfo {author} {\bibfnamefont {P.~A.}\ \bibnamefont {Lee}},\ and\ \bibinfo {author} {\bibfnamefont {J.}~\bibnamefont {Ruhman}},\ }\bibfield  {title} {\bibinfo {title} {Mechanism for {$\pi$} phase shifts in {Little-Parks} experiments: Application to {Hb-TaS$_{2}$} and to {-TaS$_{2}$} intercalated with chiral molecules},\ }\href {https://doi.org/10.1103/PhysRevB.108.L180505} {\bibfield  {journal} {\bibinfo  {journal} {Phys. Rev. B Condens. Matter}\ }\textbf {\bibinfo {volume} {108}},\ \bibinfo {pages} {L180505} (\bibinfo {year} {2023}{\natexlab{a}})}\BibitemShut {NoStop}%
\bibitem [{\citenamefont {Wan}\ \emph {et~al.}(2023)\citenamefont {Wan}, \citenamefont {Zheliuk}, \citenamefont {Yuan}, \citenamefont {Peng}, \citenamefont {Zhang}, \citenamefont {Liang}, \citenamefont {Zeitler}, \citenamefont {Wiedmann}, \citenamefont {Hussey}, \citenamefont {Palstra},\ and\ \citenamefont {Ye}}]{Wan2023-rh}%
  \BibitemOpen
  \bibfield  {author} {\bibinfo {author} {\bibfnamefont {P.}~\bibnamefont {Wan}}, \bibinfo {author} {\bibfnamefont {O.}~\bibnamefont {Zheliuk}}, \bibinfo {author} {\bibfnamefont {N.~F.~Q.}\ \bibnamefont {Yuan}}, \bibinfo {author} {\bibfnamefont {X.}~\bibnamefont {Peng}}, \bibinfo {author} {\bibfnamefont {L.}~\bibnamefont {Zhang}}, \bibinfo {author} {\bibfnamefont {M.}~\bibnamefont {Liang}}, \bibinfo {author} {\bibfnamefont {U.}~\bibnamefont {Zeitler}}, \bibinfo {author} {\bibfnamefont {S.}~\bibnamefont {Wiedmann}}, \bibinfo {author} {\bibfnamefont {N.~E.}\ \bibnamefont {Hussey}}, \bibinfo {author} {\bibfnamefont {T.~T.~M.}\ \bibnamefont {Palstra}},\ and\ \bibinfo {author} {\bibfnamefont {J.}~\bibnamefont {Ye}},\ }\bibfield  {title} {\bibinfo {title} {Orbital fulde-ferrell-larkin-ovchinnikov state in an ising superconductor},\ }\href {https://doi.org/10.1038/s41586-023-05967-z} {\bibfield  {journal} {\bibinfo  {journal} {Nature}\ }\textbf {\bibinfo {volume} {619}},\ \bibinfo {pages} {46} (\bibinfo {year}
  {2023})}\BibitemShut {NoStop}%
\bibitem [{\citenamefont {Tokura}\ and\ \citenamefont {Nagaosa}(2018)}]{Tokura2018-yp}%
  \BibitemOpen
  \bibfield  {author} {\bibinfo {author} {\bibfnamefont {Y.}~\bibnamefont {Tokura}}\ and\ \bibinfo {author} {\bibfnamefont {N.}~\bibnamefont {Nagaosa}},\ }\bibfield  {title} {\bibinfo {title} {Nonreciprocal responses from non-centrosymmetric quantum materials},\ }\href {https://doi.org/10.1038/s41467-018-05759-4} {\bibfield  {journal} {\bibinfo  {journal} {Nat. Commun.}\ }\textbf {\bibinfo {volume} {9}},\ \bibinfo {pages} {3740} (\bibinfo {year} {2018})}\BibitemShut {NoStop}%
\bibitem [{\citenamefont {Ideue}\ and\ \citenamefont {Iwasa}(2021)}]{Ideue2021-yo}%
  \BibitemOpen
  \bibfield  {author} {\bibinfo {author} {\bibfnamefont {T.}~\bibnamefont {Ideue}}\ and\ \bibinfo {author} {\bibfnamefont {Y.}~\bibnamefont {Iwasa}},\ }\bibfield  {title} {\bibinfo {title} {Symmetry breaking and nonlinear electric transport in van der waals nanostructures},\ }\href {https://doi.org/10.1146/annurev-conmatphys-060220-100347} {\bibfield  {journal} {\bibinfo  {journal} {Annu. Rev. Condens. Matter Phys.}\ }\textbf {\bibinfo {volume} {12}},\ \bibinfo {pages} {201} (\bibinfo {year} {2021})}\BibitemShut {NoStop}%
\bibitem [{\citenamefont {Hoshino}\ \emph {et~al.}(2018)\citenamefont {Hoshino}, \citenamefont {Wakatsuki}, \citenamefont {Hamamoto},\ and\ \citenamefont {Nagaosa}}]{Hoshino2018-qr}%
  \BibitemOpen
  \bibfield  {author} {\bibinfo {author} {\bibfnamefont {S.}~\bibnamefont {Hoshino}}, \bibinfo {author} {\bibfnamefont {R.}~\bibnamefont {Wakatsuki}}, \bibinfo {author} {\bibfnamefont {K.}~\bibnamefont {Hamamoto}},\ and\ \bibinfo {author} {\bibfnamefont {N.}~\bibnamefont {Nagaosa}},\ }\bibfield  {title} {\bibinfo {title} {Nonreciprocal charge transport in two-dimensional noncentrosymmetric superconductors},\ }\href {https://doi.org/10.1103/PhysRevB.98.054510} {\bibfield  {journal} {\bibinfo  {journal} {Phys. Rev. B Condens. Matter}\ }\textbf {\bibinfo {volume} {98}},\ \bibinfo {pages} {054510} (\bibinfo {year} {2018})}\BibitemShut {NoStop}%
\bibitem [{\citenamefont {Nagaosa}\ and\ \citenamefont {Yanase}(2024)}]{Nagaosa2024-uk}%
  \BibitemOpen
  \bibfield  {author} {\bibinfo {author} {\bibfnamefont {N.}~\bibnamefont {Nagaosa}}\ and\ \bibinfo {author} {\bibfnamefont {Y.}~\bibnamefont {Yanase}},\ }\bibfield  {title} {\bibinfo {title} {Nonreciprocal transport and optical phenomena in quantum materials},\ }\href {https://doi.org/https://doi.org/10.1146/annurev-conmatphys-032822-033734} {\bibfield  {journal} {\bibinfo  {journal} {Ann. Rev. Condens. Matter Phys.}\ }\textbf {\bibinfo {volume} {15}},\ \bibinfo {pages} {63} (\bibinfo {year} {2024})}\BibitemShut {NoStop}%
\bibitem [{\citenamefont {Varlamov}\ and\ \citenamefont {Reggiani}(1992)}]{Varlamov1992-eu}%
  \BibitemOpen
  \bibfield  {author} {\bibinfo {author} {\bibfnamefont {A.~A.}\ \bibnamefont {Varlamov}}\ and\ \bibinfo {author} {\bibfnamefont {L.}~\bibnamefont {Reggiani}},\ }\bibfield  {title} {\bibinfo {title} {Nonlinear fluctuation conductivity of a layered superconductor: Crossover in strong electric fields},\ }\href {https://doi.org/10.1103/physrevb.45.1060} {\bibfield  {journal} {\bibinfo  {journal} {Phys. Rev. B Condens. Matter}\ }\textbf {\bibinfo {volume} {45}},\ \bibinfo {pages} {1060} (\bibinfo {year} {1992})}\BibitemShut {NoStop}%
\bibitem [{\citenamefont {Mishonov}\ \emph {et~al.}(2002)\citenamefont {Mishonov}, \citenamefont {Posazhennikova},\ and\ \citenamefont {Indekeu}}]{Mishonov2002-ud}%
  \BibitemOpen
  \bibfield  {author} {\bibinfo {author} {\bibfnamefont {T.}~\bibnamefont {Mishonov}}, \bibinfo {author} {\bibfnamefont {A.}~\bibnamefont {Posazhennikova}},\ and\ \bibinfo {author} {\bibfnamefont {J.}~\bibnamefont {Indekeu}},\ }\bibfield  {title} {\bibinfo {title} {Fluctuation conductivity in superconductors in strong electric fields},\ }\href {https://doi.org/10.1103/PhysRevB.65.064519} {\bibfield  {journal} {\bibinfo  {journal} {Phys. Rev. B Condens. Matter}\ }\textbf {\bibinfo {volume} {65}},\ \bibinfo {pages} {064519} (\bibinfo {year} {2002})}\BibitemShut {NoStop}%
\bibitem [{\citenamefont {Puica}\ and\ \citenamefont {Lang}(2006)}]{Puica2006-gp}%
  \BibitemOpen
  \bibfield  {author} {\bibinfo {author} {\bibfnamefont {I.}~\bibnamefont {Puica}}\ and\ \bibinfo {author} {\bibfnamefont {W.}~\bibnamefont {Lang}},\ }\bibfield  {title} {\bibinfo {title} {Out-of-plane fluctuation conductivity of layered superconductors in strong electric fields},\ }\href {https://doi.org/10.1103/PhysRevB.73.024502} {\bibfield  {journal} {\bibinfo  {journal} {Phys. Rev. B}\ }\textbf {\bibinfo {volume} {73}},\ \bibinfo {pages} {024502} (\bibinfo {year} {2006})}\BibitemShut {NoStop}%
\bibitem [{\citenamefont {Larkin}\ and\ \citenamefont {Varlamov}(2005)}]{larkin2005theory}%
  \BibitemOpen
  \bibfield  {author} {\bibinfo {author} {\bibfnamefont {A.}~\bibnamefont {Larkin}}\ and\ \bibinfo {author} {\bibfnamefont {A.}~\bibnamefont {Varlamov}},\ }\href@noop {} {\emph {\bibinfo {title} {Theory of Fluctuations in Superconductors}}},\ Vol.\ \bibinfo {volume} {127}\ (\bibinfo  {publisher} {Oxford University Press},\ \bibinfo {address} {Oxford},\ \bibinfo {year} {2005})\BibitemShut {NoStop}%
\bibitem [{\citenamefont {Konschelle}\ \emph {et~al.}(2007)\citenamefont {Konschelle}, \citenamefont {Cayssol},\ and\ \citenamefont {Buzdin}}]{Konschelle2007-me}%
  \BibitemOpen
  \bibfield  {author} {\bibinfo {author} {\bibfnamefont {F.}~\bibnamefont {Konschelle}}, \bibinfo {author} {\bibfnamefont {J.}~\bibnamefont {Cayssol}},\ and\ \bibinfo {author} {\bibfnamefont {A.~I.}\ \bibnamefont {Buzdin}},\ }\bibfield  {title} {\bibinfo {title} {Anomalous fluctuation regimes at {FFLO} transition},\ }\href {https://doi.org/10.1209/0295-5075/79/67001} {\bibfield  {journal} {\bibinfo  {journal} {EPL}\ }\textbf {\bibinfo {volume} {79}},\ \bibinfo {pages} {67001} (\bibinfo {year} {2007})}\BibitemShut {NoStop}%
\bibitem [{\citenamefont {Konschelle}\ \emph {et~al.}(2009)\citenamefont {Konschelle}, \citenamefont {Cayssol},\ and\ \citenamefont {Buzdin}}]{Konschelle2009-mj}%
  \BibitemOpen
  \bibfield  {author} {\bibinfo {author} {\bibfnamefont {F.}~\bibnamefont {Konschelle}}, \bibinfo {author} {\bibfnamefont {J.}~\bibnamefont {Cayssol}},\ and\ \bibinfo {author} {\bibfnamefont {A.~I.}\ \bibnamefont {Buzdin}},\ }\bibfield  {title} {\bibinfo {title} {Oscillations of magnetization and conductivity in anisotropic {Fulde-Ferrell-Larkin-Ovchinnikov} superconductors},\ }\href {https://doi.org/10.1103/PhysRevB.79.224526} {\bibfield  {journal} {\bibinfo  {journal} {Phys. Rev. B Condens. Matter}\ }\textbf {\bibinfo {volume} {79}},\ \bibinfo {pages} {224526} (\bibinfo {year} {2009})}\BibitemShut {NoStop}%
\bibitem [{\citenamefont {Wakatsuki}\ and\ \citenamefont {Nagaosa}(2018)}]{Wakatsuki2018-vd}%
  \BibitemOpen
  \bibfield  {author} {\bibinfo {author} {\bibfnamefont {R.}~\bibnamefont {Wakatsuki}}\ and\ \bibinfo {author} {\bibfnamefont {N.}~\bibnamefont {Nagaosa}},\ }\bibfield  {title} {\bibinfo {title} {Nonreciprocal current in noncentrosymmetric rashba superconductors},\ }\href {https://doi.org/10.1103/PhysRevLett.121.026601} {\bibfield  {journal} {\bibinfo  {journal} {Phys. Rev. Lett.}\ }\textbf {\bibinfo {volume} {121}},\ \bibinfo {pages} {026601} (\bibinfo {year} {2018})}\BibitemShut {NoStop}%
\bibitem [{\citenamefont {Daido}\ and\ \citenamefont {Yanase}(2024)}]{Daido2024-is}%
  \BibitemOpen
  \bibfield  {author} {\bibinfo {author} {\bibfnamefont {A.}~\bibnamefont {Daido}}\ and\ \bibinfo {author} {\bibfnamefont {Y.}~\bibnamefont {Yanase}},\ }\bibfield  {title} {\bibinfo {title} {Rectification and nonlinear hall effect by fluctuating finite-momentum cooper pairs},\ }\href {https://doi.org/10.1103/PhysRevResearch.6.L022009} {\bibfield  {journal} {\bibinfo  {journal} {Phys. Rev. Res.}\ }\textbf {\bibinfo {volume} {6}},\ \bibinfo {pages} {L022009} (\bibinfo {year} {2024})}\BibitemShut {NoStop}%
\bibitem [{\citenamefont {Nunchot}\ \emph {et~al.}(2022)\citenamefont {Nunchot}, \citenamefont {Nakashima},\ and\ \citenamefont {Ikeda}}]{Nunchot2022-mc}%
  \BibitemOpen
  \bibfield  {author} {\bibinfo {author} {\bibfnamefont {N.}~\bibnamefont {Nunchot}}, \bibinfo {author} {\bibfnamefont {D.}~\bibnamefont {Nakashima}},\ and\ \bibinfo {author} {\bibfnamefont {R.}~\bibnamefont {Ikeda}},\ }\bibfield  {title} {\bibinfo {title} {Fluctuation conductivity and vortex state in a superconductor with strong paramagnetic pair breaking},\ }\href {https://doi.org/10.1103/PhysRevB.105.174510} {\bibfield  {journal} {\bibinfo  {journal} {Phys. Rev. B Condens. Matter}\ }\textbf {\bibinfo {volume} {105}},\ \bibinfo {pages} {174510} (\bibinfo {year} {2022})}\BibitemShut {NoStop}%
\bibitem [{\citenamefont {Yoshida}\ \emph {et~al.}(2012)\citenamefont {Yoshida}, \citenamefont {Sigrist},\ and\ \citenamefont {Yanase}}]{Yoshida2012-wa}%
  \BibitemOpen
  \bibfield  {author} {\bibinfo {author} {\bibfnamefont {T.}~\bibnamefont {Yoshida}}, \bibinfo {author} {\bibfnamefont {M.}~\bibnamefont {Sigrist}},\ and\ \bibinfo {author} {\bibfnamefont {Y.}~\bibnamefont {Yanase}},\ }\bibfield  {title} {\bibinfo {title} {Pair-density wave states through spin-orbit coupling in multilayer superconductors},\ }\href {https://doi.org/10.1103/PhysRevB.86.134514} {\bibfield  {journal} {\bibinfo  {journal} {Phys. Rev. B Condens. Matter}\ }\textbf {\bibinfo {volume} {86}},\ \bibinfo {pages} {134514} (\bibinfo {year} {2012})}\BibitemShut {NoStop}%
\bibitem [{\citenamefont {Maruyama}\ \emph {et~al.}(2012)\citenamefont {Maruyama}, \citenamefont {Sigrist},\ and\ \citenamefont {Yanase}}]{Maruyama2012-cb}%
  \BibitemOpen
  \bibfield  {author} {\bibinfo {author} {\bibfnamefont {D.}~\bibnamefont {Maruyama}}, \bibinfo {author} {\bibfnamefont {M.}~\bibnamefont {Sigrist}},\ and\ \bibinfo {author} {\bibfnamefont {Y.}~\bibnamefont {Yanase}},\ }\bibfield  {title} {\bibinfo {title} {Locally non-centrosymmetric superconductivity in multilayer systems},\ }\href {https://doi.org/10.1143/JPSJ.81.034702} {\bibfield  {journal} {\bibinfo  {journal} {J. Phys. Soc. Jpn.}\ }\textbf {\bibinfo {volume} {81}},\ \bibinfo {pages} {034702} (\bibinfo {year} {2012})}\BibitemShut {NoStop}%
\bibitem [{\citenamefont {Sigrist}\ \emph {et~al.}(2014)\citenamefont {Sigrist}, \citenamefont {Agterberg}, \citenamefont {Fischer}, \citenamefont {Goryo}, \citenamefont {Loder}, \citenamefont {Rhim}, \citenamefont {Maruyama}, \citenamefont {Yanase}, \citenamefont {Yoshida},\ and\ \citenamefont {Youn}}]{Sigrist2014-rc}%
  \BibitemOpen
  \bibfield  {author} {\bibinfo {author} {\bibfnamefont {M.}~\bibnamefont {Sigrist}}, \bibinfo {author} {\bibfnamefont {D.~F.}\ \bibnamefont {Agterberg}}, \bibinfo {author} {\bibfnamefont {M.~H.}\ \bibnamefont {Fischer}}, \bibinfo {author} {\bibfnamefont {J.}~\bibnamefont {Goryo}}, \bibinfo {author} {\bibfnamefont {F.}~\bibnamefont {Loder}}, \bibinfo {author} {\bibfnamefont {S.-H.}\ \bibnamefont {Rhim}}, \bibinfo {author} {\bibfnamefont {D.}~\bibnamefont {Maruyama}}, \bibinfo {author} {\bibfnamefont {Y.}~\bibnamefont {Yanase}}, \bibinfo {author} {\bibfnamefont {T.}~\bibnamefont {Yoshida}},\ and\ \bibinfo {author} {\bibfnamefont {S.~J.}\ \bibnamefont {Youn}},\ }\bibfield  {title} {\bibinfo {title} {Superconductors with staggered non-centrosymmetricity},\ }\href {https://doi.org/10.7566/JPSJ.83.061014} {\bibfield  {journal} {\bibinfo  {journal} {J. Phys. Soc. Jpn.}\ }\textbf {\bibinfo {volume} {83}},\ \bibinfo {pages} {061014} (\bibinfo {year} {2014})}\BibitemShut {NoStop}%
\bibitem [{\citenamefont {Nakamura}\ and\ \citenamefont {Yanase}(2017)}]{Nakamura2017-vb}%
  \BibitemOpen
  \bibfield  {author} {\bibinfo {author} {\bibfnamefont {Y.}~\bibnamefont {Nakamura}}\ and\ \bibinfo {author} {\bibfnamefont {Y.}~\bibnamefont {Yanase}},\ }\bibfield  {title} {\bibinfo {title} {Odd-parity superconductivity in bilayer transition metal dichalcogenides},\ }\href {https://doi.org/10.1103/PhysRevB.96.054501} {\bibfield  {journal} {\bibinfo  {journal} {Phys. Rev. B Condens. Matter}\ }\textbf {\bibinfo {volume} {96}},\ \bibinfo {pages} {054501} (\bibinfo {year} {2017})}\BibitemShut {NoStop}%
\bibitem [{\citenamefont {M{\"o}ckli}\ \emph {et~al.}(2018)\citenamefont {M{\"o}ckli}, \citenamefont {Yanase},\ and\ \citenamefont {Sigrist}}]{Mockli2018-ci}%
  \BibitemOpen
  \bibfield  {author} {\bibinfo {author} {\bibfnamefont {D.}~\bibnamefont {M{\"o}ckli}}, \bibinfo {author} {\bibfnamefont {Y.}~\bibnamefont {Yanase}},\ and\ \bibinfo {author} {\bibfnamefont {M.}~\bibnamefont {Sigrist}},\ }\bibfield  {title} {\bibinfo {title} {Orbitally limited pair-density-wave phase of multilayer superconductors},\ }\href {https://doi.org/10.1103/PhysRevB.97.144508} {\bibfield  {journal} {\bibinfo  {journal} {Phys. Rev. B Condens. Matter}\ }\textbf {\bibinfo {volume} {97}},\ \bibinfo {pages} {144508} (\bibinfo {year} {2018})}\BibitemShut {NoStop}%
\bibitem [{\citenamefont {Wakatsuki}\ \emph {et~al.}(2017)\citenamefont {Wakatsuki}, \citenamefont {Saito}, \citenamefont {Hoshino}, \citenamefont {Itahashi}, \citenamefont {Ideue}, \citenamefont {Ezawa}, \citenamefont {Iwasa},\ and\ \citenamefont {Nagaosa}}]{Wakatsuki2017-hw}%
  \BibitemOpen
  \bibfield  {author} {\bibinfo {author} {\bibfnamefont {R.}~\bibnamefont {Wakatsuki}}, \bibinfo {author} {\bibfnamefont {Y.}~\bibnamefont {Saito}}, \bibinfo {author} {\bibfnamefont {S.}~\bibnamefont {Hoshino}}, \bibinfo {author} {\bibfnamefont {Y.~M.}\ \bibnamefont {Itahashi}}, \bibinfo {author} {\bibfnamefont {T.}~\bibnamefont {Ideue}}, \bibinfo {author} {\bibfnamefont {M.}~\bibnamefont {Ezawa}}, \bibinfo {author} {\bibfnamefont {Y.}~\bibnamefont {Iwasa}},\ and\ \bibinfo {author} {\bibfnamefont {N.}~\bibnamefont {Nagaosa}},\ }\bibfield  {title} {\bibinfo {title} {Nonreciprocal charge transport in noncentrosymmetric superconductors},\ }\href {https://doi.org/10.1126/sciadv.1602390} {\bibfield  {journal} {\bibinfo  {journal} {Sci Adv}\ }\textbf {\bibinfo {volume} {3}},\ \bibinfo {pages} {e1602390} (\bibinfo {year} {2017})}\BibitemShut {NoStop}%
\bibitem [{\citenamefont {Yoshida}\ \emph {et~al.}(2013)\citenamefont {Yoshida}, \citenamefont {Sigrist},\ and\ \citenamefont {Yanase}}]{Yoshida2013}%
  \BibitemOpen
  \bibfield  {author} {\bibinfo {author} {\bibfnamefont {T.}~\bibnamefont {Yoshida}}, \bibinfo {author} {\bibfnamefont {M.}~\bibnamefont {Sigrist}},\ and\ \bibinfo {author} {\bibfnamefont {Y.}~\bibnamefont {Yanase}},\ }\bibfield  {title} {\bibinfo {title} {Complex-stripe phases induced by staggered rashba spin–orbit coupling},\ }\href {https://doi.org/10.7566/JPSJ.82.074714} {\bibfield  {journal} {\bibinfo  {journal} {Journal of the Physical Society of Japan}\ }\textbf {\bibinfo {volume} {82}},\ \bibinfo {pages} {074714} (\bibinfo {year} {2013})}\BibitemShut {NoStop}%
\bibitem [{\citenamefont {Agterberg}\ \emph {et~al.}(2020)\citenamefont {Agterberg}, \citenamefont {Davis}, \citenamefont {Edkins}, \citenamefont {Fradkin}, \citenamefont {Van~Harlingen}, \citenamefont {Kivelson}, \citenamefont {Lee}, \citenamefont {Radzihovsky}, \citenamefont {Tranquada},\ and\ \citenamefont {Wang}}]{Agterberg2020-gs}%
  \BibitemOpen
  \bibfield  {author} {\bibinfo {author} {\bibfnamefont {D.~F.}\ \bibnamefont {Agterberg}}, \bibinfo {author} {\bibfnamefont {J.~C.~S.}\ \bibnamefont {Davis}}, \bibinfo {author} {\bibfnamefont {S.~D.}\ \bibnamefont {Edkins}}, \bibinfo {author} {\bibfnamefont {E.}~\bibnamefont {Fradkin}}, \bibinfo {author} {\bibfnamefont {D.~J.}\ \bibnamefont {Van~Harlingen}}, \bibinfo {author} {\bibfnamefont {S.~A.}\ \bibnamefont {Kivelson}}, \bibinfo {author} {\bibfnamefont {P.~A.}\ \bibnamefont {Lee}}, \bibinfo {author} {\bibfnamefont {L.}~\bibnamefont {Radzihovsky}}, \bibinfo {author} {\bibfnamefont {J.~M.}\ \bibnamefont {Tranquada}},\ and\ \bibinfo {author} {\bibfnamefont {Y.}~\bibnamefont {Wang}},\ }\bibfield  {title} {\bibinfo {title} {The physics of {Pair-Density} waves: Cuprate superconductors and beyond},\ }\href {https://doi.org/10.1146/annurev-conmatphys-031119-050711} {\bibfield  {journal} {\bibinfo  {journal} {Annu. Rev. Condens. Matter Phys.}\ }\textbf {\bibinfo {volume} {11}},\ \bibinfo {pages} {231} (\bibinfo
  {year} {2020})}\BibitemShut {NoStop}%
\bibitem [{\citenamefont {Matsuda}\ and\ \citenamefont {Shimahara}(2007)}]{Matsuda2007-xw}%
  \BibitemOpen
  \bibfield  {author} {\bibinfo {author} {\bibfnamefont {Y.}~\bibnamefont {Matsuda}}\ and\ \bibinfo {author} {\bibfnamefont {H.}~\bibnamefont {Shimahara}},\ }\bibfield  {title} {\bibinfo {title} {{Fulde--Ferrell--Larkin--Ovchinnikov} state in heavy fermion superconductors},\ }\href {https://doi.org/10.1143/JPSJ.76.051005} {\bibfield  {journal} {\bibinfo  {journal} {J. Phys. Soc. Jpn.}\ }\textbf {\bibinfo {volume} {76}},\ \bibinfo {pages} {051005} (\bibinfo {year} {2007})}\BibitemShut {NoStop}%
\bibitem [{\citenamefont {Fischer}\ \emph {et~al.}(2023{\natexlab{b}})\citenamefont {Fischer}, \citenamefont {Sigrist}, \citenamefont {Agterberg},\ and\ \citenamefont {Yanase}}]{Fischer2023-kd}%
  \BibitemOpen
  \bibfield  {author} {\bibinfo {author} {\bibfnamefont {M.~H.}\ \bibnamefont {Fischer}}, \bibinfo {author} {\bibfnamefont {M.}~\bibnamefont {Sigrist}}, \bibinfo {author} {\bibfnamefont {D.~F.}\ \bibnamefont {Agterberg}},\ and\ \bibinfo {author} {\bibfnamefont {Y.}~\bibnamefont {Yanase}},\ }\bibfield  {title} {\bibinfo {title} {Superconductivity and local {Inversion-Symmetry} breaking},\ }\href {https://doi.org/10.1146/annurev-conmatphys-040521-042511} {\bibfield  {journal} {\bibinfo  {journal} {Annu. Rev. Condens. Matter Phys.}\ }\textbf {\bibinfo {volume} {14}},\ \bibinfo {pages} {153} (\bibinfo {year} {2023}{\natexlab{b}})}\BibitemShut {NoStop}%
\bibitem [{\citenamefont {Bohm}(1949)}]{Bohm1949-pd}%
  \BibitemOpen
  \bibfield  {author} {\bibinfo {author} {\bibfnamefont {D.}~\bibnamefont {Bohm}},\ }\bibfield  {title} {\bibinfo {title} {Note on a theorem of bloch concerning possible causes of superconductivity},\ }\href {https://doi.org/10.1103/physrev.75.502} {\bibfield  {journal} {\bibinfo  {journal} {Phys. Rev.}\ }\textbf {\bibinfo {volume} {75}},\ \bibinfo {pages} {502} (\bibinfo {year} {1949})}\BibitemShut {NoStop}%
\bibitem [{\citenamefont {Barzykin}\ and\ \citenamefont {Gor'kov}(2002)}]{Barzykin2002-fo}%
  \BibitemOpen
  \bibfield  {author} {\bibinfo {author} {\bibfnamefont {V.}~\bibnamefont {Barzykin}}\ and\ \bibinfo {author} {\bibfnamefont {L.~P.}\ \bibnamefont {Gor'kov}},\ }\bibfield  {title} {\bibinfo {title} {Inhomogeneous stripe phase revisited for surface superconductivity},\ }\href {https://doi.org/10.1103/PhysRevLett.89.227002} {\bibfield  {journal} {\bibinfo  {journal} {Phys. Rev. Lett.}\ }\textbf {\bibinfo {volume} {89}},\ \bibinfo {pages} {227002} (\bibinfo {year} {2002})}\BibitemShut {NoStop}%
\bibitem [{\citenamefont {Agterberg}(2003)}]{Agterberg2003-jx}%
  \BibitemOpen
  \bibfield  {author} {\bibinfo {author} {\bibfnamefont {D.~F.}\ \bibnamefont {Agterberg}},\ }\bibfield  {title} {\bibinfo {title} {Novel magnetic field effects in unconventional superconductors},\ }\href {https://doi.org/10.1016/S0921-4534(03)00634-8} {\bibfield  {journal} {\bibinfo  {journal} {Physica C Supercond.}\ }\textbf {\bibinfo {volume} {387}},\ \bibinfo {pages} {13} (\bibinfo {year} {2003})}\BibitemShut {NoStop}%
\bibitem [{\citenamefont {Dimitrova}\ and\ \citenamefont {Feigel'man}(2003)}]{Dimitrova2003-vp}%
  \BibitemOpen
  \bibfield  {author} {\bibinfo {author} {\bibfnamefont {O.~V.}\ \bibnamefont {Dimitrova}}\ and\ \bibinfo {author} {\bibfnamefont {M.~V.}\ \bibnamefont {Feigel'man}},\ }\bibfield  {title} {\bibinfo {title} {Phase diagram of a surface superconductor in parallel magnetic field},\ }\href {https://doi.org/10.1134/1.1644308} {\bibfield  {journal} {\bibinfo  {journal} {Journal of Experimental and Theoretical Physics Letters}\ }\textbf {\bibinfo {volume} {78}},\ \bibinfo {pages} {637} (\bibinfo {year} {2003})}\BibitemShut {NoStop}%
\bibitem [{\citenamefont {Kaur}\ \emph {et~al.}(2005)\citenamefont {Kaur}, \citenamefont {Agterberg},\ and\ \citenamefont {Sigrist}}]{Kaur2005-vk}%
  \BibitemOpen
  \bibfield  {author} {\bibinfo {author} {\bibfnamefont {R.~P.}\ \bibnamefont {Kaur}}, \bibinfo {author} {\bibfnamefont {D.~F.}\ \bibnamefont {Agterberg}},\ and\ \bibinfo {author} {\bibfnamefont {M.}~\bibnamefont {Sigrist}},\ }\bibfield  {title} {\bibinfo {title} {Helical vortex phase in the noncentrosymmetric {CePt3Si}},\ }\href {https://doi.org/10.1103/PhysRevLett.94.137002} {\bibfield  {journal} {\bibinfo  {journal} {Phys. Rev. Lett.}\ }\textbf {\bibinfo {volume} {94}},\ \bibinfo {pages} {137002} (\bibinfo {year} {2005})}\BibitemShut {NoStop}%
\bibitem [{\citenamefont {Agterberg}\ and\ \citenamefont {Kaur}(2007)}]{Agterberg2007-zi}%
  \BibitemOpen
  \bibfield  {author} {\bibinfo {author} {\bibfnamefont {D.~F.}\ \bibnamefont {Agterberg}}\ and\ \bibinfo {author} {\bibfnamefont {R.~P.}\ \bibnamefont {Kaur}},\ }\bibfield  {title} {\bibinfo {title} {Magnetic-field-induced helical and stripe phases in rashba superconductors},\ }\href {https://doi.org/10.1103/PhysRevB.75.064511} {\bibfield  {journal} {\bibinfo  {journal} {Phys. Rev. B Condens. Matter}\ }\textbf {\bibinfo {volume} {75}},\ \bibinfo {pages} {064511} (\bibinfo {year} {2007})}\BibitemShut {NoStop}%
\bibitem [{\citenamefont {Dimitrova}\ and\ \citenamefont {Feigel'man}(2007)}]{Dimitrova2007-fx}%
  \BibitemOpen
  \bibfield  {author} {\bibinfo {author} {\bibfnamefont {O.}~\bibnamefont {Dimitrova}}\ and\ \bibinfo {author} {\bibfnamefont {M.~V.}\ \bibnamefont {Feigel'man}},\ }\bibfield  {title} {\bibinfo {title} {Theory of a two-dimensional superconductor with broken inversion symmetry},\ }\href {https://doi.org/10.1103/PhysRevB.76.014522} {\bibfield  {journal} {\bibinfo  {journal} {Phys. Rev. B Condens. Matter}\ }\textbf {\bibinfo {volume} {76}},\ \bibinfo {pages} {014522} (\bibinfo {year} {2007})}\BibitemShut {NoStop}%
\bibitem [{\citenamefont {Yanase}\ and\ \citenamefont {Sigrist}(2008)}]{Yanase2008-mk}%
  \BibitemOpen
  \bibfield  {author} {\bibinfo {author} {\bibfnamefont {Y.}~\bibnamefont {Yanase}}\ and\ \bibinfo {author} {\bibfnamefont {M.}~\bibnamefont {Sigrist}},\ }\bibfield  {title} {\bibinfo {title} {Helical superconductivity in non-centrosymmetric superconductors with dominantly spin triplet pairing},\ }\href {https://doi.org/10.1143/JPSJS.77SA.342} {\bibfield  {journal} {\bibinfo  {journal} {J. Phys. Soc. Jpn.}\ }\textbf {\bibinfo {volume} {77}},\ \bibinfo {pages} {342} (\bibinfo {year} {2008})}\BibitemShut {NoStop}%
\bibitem [{\citenamefont {Samokhin}(2008)}]{Samokhin2008-oa}%
  \BibitemOpen
  \bibfield  {author} {\bibinfo {author} {\bibfnamefont {K.~V.}\ \bibnamefont {Samokhin}},\ }\bibfield  {title} {\bibinfo {title} {Upper critical field in noncentrosymmetric superconductors},\ }\href {https://doi.org/10.1103/PhysRevB.78.224520} {\bibfield  {journal} {\bibinfo  {journal} {Phys. Rev. B Condens. Matter}\ }\textbf {\bibinfo {volume} {78}},\ \bibinfo {pages} {224520} (\bibinfo {year} {2008})}\BibitemShut {NoStop}%
\bibitem [{\citenamefont {Michaeli}\ \emph {et~al.}(2012)\citenamefont {Michaeli}, \citenamefont {Potter},\ and\ \citenamefont {Lee}}]{Michaeli2012-ks}%
  \BibitemOpen
  \bibfield  {author} {\bibinfo {author} {\bibfnamefont {K.}~\bibnamefont {Michaeli}}, \bibinfo {author} {\bibfnamefont {A.~C.}\ \bibnamefont {Potter}},\ and\ \bibinfo {author} {\bibfnamefont {P.~A.}\ \bibnamefont {Lee}},\ }\bibfield  {title} {\bibinfo {title} {Superconducting and ferromagnetic phases in {SrTiO3/LaAlO3} oxide interface structures: possibility of finite momentum pairing},\ }\href {https://doi.org/10.1103/PhysRevLett.108.117003} {\bibfield  {journal} {\bibinfo  {journal} {Phys. Rev. Lett.}\ }\textbf {\bibinfo {volume} {108}},\ \bibinfo {pages} {117003} (\bibinfo {year} {2012})}\BibitemShut {NoStop}%
\bibitem [{\citenamefont {Houzet}\ and\ \citenamefont {Meyer}(2015)}]{Houzet2015-ie}%
  \BibitemOpen
  \bibfield  {author} {\bibinfo {author} {\bibfnamefont {M.}~\bibnamefont {Houzet}}\ and\ \bibinfo {author} {\bibfnamefont {J.~S.}\ \bibnamefont {Meyer}},\ }\bibfield  {title} {\bibinfo {title} {Quasiclassical theory of disordered rashba superconductors},\ }\href {https://doi.org/10.1103/PhysRevB.92.014509} {\bibfield  {journal} {\bibinfo  {journal} {Phys. Rev. B Condens. Matter}\ }\textbf {\bibinfo {volume} {92}},\ \bibinfo {pages} {014509} (\bibinfo {year} {2015})}\BibitemShut {NoStop}%
\bibitem [{\citenamefont {Bauer}\ and\ \citenamefont {Sigrist}(2012)}]{bauer2012non}%
  \BibitemOpen
  \bibfield  {author} {\bibinfo {author} {\bibfnamefont {E.}~\bibnamefont {Bauer}}\ and\ \bibinfo {author} {\bibfnamefont {M.}~\bibnamefont {Sigrist}},\ }\href@noop {} {\emph {\bibinfo {title} {Non-centrosymmetric Superconductors: Introduction and Overview}}},\ Vol.\ \bibinfo {volume} {847}\ (\bibinfo  {publisher} {Springer},\ \bibinfo {address} {New York},\ \bibinfo {year} {2012})\BibitemShut {NoStop}%
\bibitem [{\citenamefont {Yuan}(2023)}]{Yuan2023-eg}%
  \BibitemOpen
  \bibfield  {author} {\bibinfo {author} {\bibfnamefont {N.~F.~Q.}\ \bibnamefont {Yuan}},\ }\bibfield  {title} {\bibinfo {title} {Orbital fulde-ferrell-larkin-ovchinnikov state in an ising superconductor},\ }\href {https://doi.org/10.1103/physrevresearch.5.043122} {\bibfield  {journal} {\bibinfo  {journal} {Phys. Rev. Res.}\ }\textbf {\bibinfo {volume} {5}},\ \bibinfo {pages} {043122} (\bibinfo {year} {2023})}\BibitemShut {NoStop}%
\bibitem [{\citenamefont {Xie}\ and\ \citenamefont {Law}(2023)}]{Xie2023-gs}%
  \BibitemOpen
  \bibfield  {author} {\bibinfo {author} {\bibfnamefont {Y.-M.}\ \bibnamefont {Xie}}\ and\ \bibinfo {author} {\bibfnamefont {K.~T.}\ \bibnamefont {Law}},\ }\bibfield  {title} {\bibinfo {title} {Orbital {Fulde-Ferrell} pairing state in moir{\'e} ising superconductors},\ }\href {https://doi.org/10.1103/PhysRevLett.131.016001} {\bibfield  {journal} {\bibinfo  {journal} {Phys. Rev. Lett.}\ }\textbf {\bibinfo {volume} {131}},\ \bibinfo {pages} {016001} (\bibinfo {year} {2023})}\BibitemShut {NoStop}%
\bibitem [{\citenamefont {Nakamura}\ \emph {et~al.}(2024)\citenamefont {Nakamura}, \citenamefont {Daido},\ and\ \citenamefont {Yanase}}]{Nakamura2024-qe}%
  \BibitemOpen
  \bibfield  {author} {\bibinfo {author} {\bibfnamefont {K.}~\bibnamefont {Nakamura}}, \bibinfo {author} {\bibfnamefont {A.}~\bibnamefont {Daido}},\ and\ \bibinfo {author} {\bibfnamefont {Y.}~\bibnamefont {Yanase}},\ }\bibfield  {title} {\bibinfo {title} {Orbital effect on the intrinsic superconducting diode effect},\ }\href {https://doi.org/10.1103/PhysRevB.109.094501} {\bibfield  {journal} {\bibinfo  {journal} {Phys. Rev. B Condens. Matter}\ }\textbf {\bibinfo {volume} {109}},\ \bibinfo {pages} {094501} (\bibinfo {year} {2024})}\BibitemShut {NoStop}%
\end{thebibliography}%

\appendix
\section{Derivation of GL factor $\hat{\alpha}$}
\label{app:GL_alpha}
In this appendix, we demonstrate the derivation of the GL factor $\hat{\alpha}$ from GL free energy,
\begin{equation}
    F = -\frac{1}{\beta}\tr\log e^{-\beta\hat{H}}.
\end{equation}
Here, we can rewrite the Hamiltonian as
\begin{equation}
    \hat{H} = \frac{1}{2}\sum_{\bm{k}}\Psi(\bm{k})^{\dagger}\hat{H}_{\mathrm{BdG}}(\bm{k})\Psi(\bm{k})+2\sum_{\bm{k}}\xi(\bm{k})+\sum_{m,\bm{k}}\frac{\abs{\Delta}^{2}}{U}
\end{equation}
with the BdG Hamiltoninan $\hat{H}_{\mathrm{BdG}}$,
\begin{equation}
  \hat{H}_{\mathrm{N}}(\bm{k}) =
  \mqty(
  H_{\mathrm{N}}(\bm{k}) & 0 \\
  0 & -H_{\mathrm{N}}(\bm{-k})^{T}
  ),
\end{equation}
\begin{equation}
    \Delta = \Delta_{1}\varphi_{1}+\Delta_{2}\varphi_{2}
\end{equation}
\begin{equation}
  \hat{\Delta} =
  \mqty(
  0 & \Delta \\
  \Delta^{\dagger} & 0
  ),
\end{equation}
\begin{equation}
    \hat{H}_{\mathrm{BdG}}(\bm{k}) = \hat{H}_{\mathrm{N}}(\bm{k}) + \hat{\Delta},
\end{equation}
and the Nambu spinor,
\begin{equation}
    \Psi(\bm{k})^{\dagger} = (c^{\dagger}_{\bm{k}\uparrow 1},c^{\dagger}_{\bm{k}\downarrow 1},c^{\dagger}_{\bm{k}\uparrow 2},c^{\dagger}_{\bm{k}\downarrow 2},c_{\bm{-k},\uparrow 1},c_{\bm{-k}\downarrow 1},c_{\bm{-k},\uparrow 2},c_{\bm{-k}\downarrow 2}).
\end{equation}
Then, the GL free energy is
\begin{align}
    F 
    &= -\frac{1}{2\beta}\sum_{\bm{k}}\tr\log\qty(1+e^{-\beta E_{n}(\bm{k})})+2\sum_{\bm{k}}\xi(\bm{k})+\sum_{m,\bm{k}}\frac{\abs{\Delta}^{2}}{U} \\
    &= F|_{\Delta=0}+\delta F,
\end{align}
where $E_{n}(\bm{k})$ represents eigenvalues of $\hat{H}_{\mathrm{BdG}}(\bm{k})$.
By expanding $\delta F$ by $\Delta$, we can obtain the GL factor $\hat{\alpha}$,
\begin{widetext}
    \begin{align}
        \delta F/V
        =& \sum_{m}\frac{\abs{\Delta}^{2}}{U} + \frac{1}{2\beta V}\sum_{\bm{k},\omega_{n}}\tr_{\mathrm{N}}\qty(\frac{1}{i\omega_{n}-H_{\mathrm{N}}(\bm{k})}\Delta(\bm{k})\frac{1}{i\omega_{n}+H_{\mathrm{N}}(-\bm{k})^{T}}\Delta(\bm{k})^{\dagger})+\order{\Delta^{4}} \nonumber\\
        =& \sum_{m}\frac{\abs{\Delta_{m}}^{2}}{U}
        + \frac{1}{2V}\sum_{\bm{k},n,m}F_{nm}(\bm{k})\left[
        \Delta_{1}^{*}\Delta_{1}\abs{\bra{u_{n}(\bm{k})}\varphi_{1}\ket{u_{m}^{*}(\bm{-k})}}^{2}\right. +\Delta_{2}^{*}\Delta_{2}\abs{\bra{u_{n}(\bm{k})}\varphi_{2}\ket{u_{m}^{*}(\bm{-k})}}^{2} \nonumber\\
        &+\Delta_{1}^{*}\Delta_{2}\bra{u_{m}^{*}(\bm{-k})}\varphi_{1}^{\dagger}
        \ket{u_{n}(\bm{k})}\bra{u_{n}(\bm{k})}\varphi_{2}\ket{u_{m}^{*}(\bm{-k})}\left.+\Delta_{2}^{*}\Delta_{1}\bra{u_{m}^{*}(\bm{-k})}\varphi_{2}^{\dagger}
        \ket{u_{n}(\bm{k})}\bra{u_{n}(\bm{k})}\varphi_{1}\ket{u_{m}^{*}(\bm{-k})}
        \right]+\order{\Delta^{4}} \nonumber\\
        =& \mqty(\Delta_{1}^{*} & \Delta_{2}^{*})\hat{\alpha}\mqty(\Delta_{1} \\ \Delta_{2}) + \order{\Delta^{4}}
    \end{align}
\end{widetext}
with $F_{mn}(\bm{k})$ in Eq.~\eqref{eq:F_nm}.
Then we obtain the expression of $\hat{\alpha}$ given in Eq.~\eqref{eq:alpha}.

We prove that $\sigma_{\mathrm{BCS-PDW}}$ vanishes in the system in the perpendicular Zeeman field.
In this case, the system has both $C_{2z}$ symmetry and $M_{z}$ symmetry,
\begin{equation}
    \hat{\alpha}(\bm{q}) = \hat{\alpha}(-\bm{q}),
\end{equation}
\begin{equation}
    \sigma_{x}\hat{\alpha}(\bm{q})\sigma^{\dagger}_{x} = \hat{\alpha}(\bm{q}).
\end{equation}
Thus, we can express $\hat{\alpha}(\bm{q})$ by,
\begin{equation}
    \hat{\alpha}(\bm{q}) =
    \mqty(
    a(\bm{q}) & b(\bm{q}) \\
    b(\bm{q}) & a(\bm{q})
    ),
\end{equation}
with real numbers $a(\bm{q})$ and $b(\bm{q})$.
Accordingly, eigenvectors of $\hat{\alpha}$ are given by
\begin{align}
    \ket{\mathrm{BCS}} &=\frac{1}{\sqrt{2}} \mqty(1 \\ 1), \\
    \ket{\mathrm{PDW}} &=\frac{1}{\sqrt{2}} \mqty(1 \\ -1),
\end{align}
regardless of $\bm{q}$.
This forces $\sigma_{\mathrm{BCS-PDW}}$ to vanish, since
\begin{align}
\bra{\mathrm{BCS}}j_{i}\ket{\mathrm{PDW}}&=\frac{1}{2}\mqty(1 & 1)
    \partial_{i}
    \mqty(
    a(\bm{q}) & b(\bm{q}) \\
    b(\bm{q}) & a(\bm{q})
    )
    \mqty(1 \\ -1) \nonumber\\
    &= \frac{1}{2}\partial_{i}[a(\bm{q})+b(\bm{q})-a(\bm{q})-b(\bm{q})]\nonumber\\
    &= 0.
\end{align}
Thus, the reciprocal paraconductivity is given by summing up the BCS and PDW contributions, $\sigma_{\mathrm{BCS}}$ and $\sigma_{\mathrm{PDW}}$, in the system in the perpendicular Zeeman field.
\end{document}